%% file: vluks.tex
%
%
%

\input mn


\input psfig

\def\scuba{SCUBA}
\def\blast{BLAST}
\def\shades{{\sl SHADES}}

\pageoffset{-2.5pc}{0pc}

\loadboldmathnames


\pagerange{0--0}    
\pubyear{2004}
\volume{000}

\begintopmatter  

\title{The extragalactic sub-mm population: predictions for the \scuba\ Half-Degree
Extragalactic Survey (\shades)}
\author{Eelco van Kampen$^{1,2}$, Will J. Percival$^1$, Miller Crawford$^1$,
James S. Dunlop$^1$, Susie E. Scott$^1$,
Neil Bevis$^3$, Seb Oliver$^3$, Frazer Pearce$^4$, Scott T. Kay$^3$,
Enrique Gazta$\tilde {\rm n}$aga$^{5,6}$, David H. Hughes$^5$,
Itziar Aretxaga$^5$ }
\smallskip
\affiliation{$^1$ Institute for Astronomy, University of Edinburgh,
Royal Observatory, Blackford Hill, Edinburgh EH9 3HJ}
\smallskip
\affiliation{$^2$ Institute for Astrophysics, University of Innsbruck,
Technikerstr. 25, A-6020 Innsbruck, Austria}
\smallskip
\affiliation{$^3$ Astronomy Centre, University of Sussex, Falmer, Brighton BN1 9QH} 
\smallskip
\affiliation{$^4$ School of Physics \& Astronomy, University of Nottingham, 
  University Park, Nottingham, NG7 2RD} 
\smallskip
\affiliation{$^5$ Instituto Nacional de Astrof\'isica, Optica y Electr\'onica,
Apt. Postal 51 y 216, Puebla, Pue, Mexico}
\smallskip
\affiliation{$^6$ Institut d'Estudis Espacials de Catalunya, Edifici Nexus,
Gran Capita 2-4, desp. 201, 08034 Barcelona, Spain}

\shortauthor{van Kampen et al.}
\shorttitle{predictions for {\it SHADES}}
\acceptedline{Accepted ... Received ...; in original form ...}

\abstract {We present predictions for the angular correlation function
and redshift distribution for \shades, the \scuba\ HAlf-Degree
Extragalactic Survey, which will yield a sample of around 300 sub-mm
sources in the 850 micron waveband in two separate fields.
Complete and unbiased 
photometric redshift information on these sub-mm sources will be derived
by combining the SCUBA data with i) deep radio imaging already obtained 
with the VLA, ii) guaranteed-time Spitzer data at mid-infrared 
wavelengths, and iii) far-infrared maps to be produced by \blast,
the Balloon-borne Large-Aperture Sub-millimeter Telescope.
Predictions for the redshift distribution and clustering properties
of the final anticipated \shades\ sample have been computed for a wide variety 
of models, each constrained to fit the observed number counts.
Since we are dealing with around 150 sources per field, we use the
sky-averaged angular correlation function to produce a more robust fit
of a power-law shape $w(\theta)=(\theta/A)^{-\delta}$ to the model data.
Comparing the predicted distributions of redshift and of the clustering
amplitude $A$ and slope $\delta$, we find that models can be constrained
from the combined SHADES data with the expected photometric redshift
information.}

\maketitle   

\section{Introduction}

The understanding of galaxy formation and evolution is making rapid
progress because of a combination of a wealth of new observational data
at a wide range of wavelengths and redshifts, and an increased
understanding of which physical processes underlie the formation and
evolution of galaxies. However,
an important problem with most current galaxy formation models is that
it is difficult to establish whether a set of model parameters that produces
a good match to observations is unique. The main reason for this is that, for
most models, there are degeneracies amongst the various free
parameters. As a significant fraction of observational data used to
constrain the model parameters are obtained from our local universe
the `uniqueness problem' can be resolved by comparing model predictions
and observations at high redshift, which in many respects is
independent from a comparison at low redshift.

One major difference between low and high redshifts is the frequency
and intensity of major mergers: at high redshifts they occur far more often and
are more intense as the participating galaxies are likely to be more
gas-rich than their low-redshift counterparts. They are also expected to be
dust-enshrouded at the peak of their burst of star formation, which means
that they are most easily detected in the sub-mm or far-infrared wavebands.

For this purpose, highly valuable observational data will be provided
by the \scuba\ HAlf-Degree Extragalactic Survey (\shades, see
{\tt http://www.roe.ac.uk/ifa/shades} and Mortier et al. 2005 for details).
This survey, which commenced in December 2002, has been designed to cover
0.5 sq.\ degrees to a 3.5-$\sigma$ detection limit of $S_{850\mu m}$ = 8\,mJy,
split between two 0.25 sq.\ degree fields. The two survey areas,  
the Lockman Hole East, and the Subaru-XMM Deep Field (SXDF),  
have been selected on the basis of low galactic confusion 
at sub-mm wavelengths, and the wealth of existing or anticipated 
supporting multi-frequency data from radio to X-ray wavelengths.

In addition the \scuba\ data will be combined with data from the
VLA, the Spitzer telescope, and
\blast, a Balloon-borne Large-Aperture Sub-millimeter Telescope
(see {\tt http://chile1.physics.upenn.edu/blastpublic} and Devlin 2001
for details), which will undertake a series of nested extragalactic surveys
at 250, 350 and 500\,$\mu$m. This experiment will significantly extend
the wavelength range, sensitivity, and area of existing ground-based
extragalactic sub-mm surveys (Hughes et al.\ 2002).

It is anticipated that spectroscopic redshifts will ultimately be
obtained for a substantial fraction of the \shades\ sources
(e.g. Chapman et al. 2003, 2004). However, the key point for the work
presented here is that, even where optical/near-infrared
spectroscopy is impossible, the long-wavelength data provided
by the combined \scuba\ + VLA + Spitzer + \blast\ dataset
will yield photometric redshifts for {\it all} sources with
uncertainties of $\delta z<0.5$ (Aretxaga et al.\ 2004).
This offers a unique powerful way of providing the complete and
unbiased redshift and SED information required to measure the
clustering properties of sub-mm sources, and the cosmic history
of dust-enshrouded star formation that takes place in very massive
star-bursts with inferred star-formation rates of order
$1000\,{\rm M_{\odot} yr^{-1}}$ (Scott et al. 2002).

These massive star-bursts could be associated with the formation of the
progenitors of massive ellipticals if sustained for a significant
amount of time (up to 1\,Gyr).  However, \scuba\ sources could also be
associated with bright, but short-lived bursts of intense
star-formation occurring in more modest galaxies drawn from the
high-redshift galaxy population already discovered at optical/UV
wavelengths (Adelberger \& Steidel 2000 and many others).
If the bright \scuba\ sources are indeed the progenitors of massive
ellipticals then they are likely to be more strongly clustered than when
drawn from the population of less massive galaxies. This is an inevitable
result of gravitational collapse from Gaussian initial density fluctuations:
the rare high-mass peaks are strongly biased with respect to the mass
(Kaiser 1984).

There is abundant evidence that this bias does occur at high redshift:
the correlations of Lyman-break galaxies at $z\simeq 3$ are almost
identical to those of present-day field galaxies, even though the mass
must be much more uniform at early times. Moreover, the correlations
increase with UV luminosity (Giavalisco \& Dickinson 2001), reaching
scale-lengths of $r_0 \simeq 7.5 \, h^{-1}{\rm Mpc}$, 1.5 times the
present-day value. Daddi et al.\ (2000) find a trend of clustering
with colour for EROs, reaching $r_0 \simeq 11 \, h^{-1}{\rm Mpc}$ for
$R-K > 5$, which corresponds to fluctuations in projected number
density that are $\sim$unity on the scale of the \scuba field-of-view,
falling to 10\% rms on 1-degree scales. For \shades\ to detect the
clustering properties of bright sub-mm sources over co-moving scales
reaching $\geq$10\,Mpc, the survey needs to cover a significant
fraction of a square degree. At the time of writing the survey is
set to reach half a square degree within three years, and is making
good progress towards achieving that goal.

The direct predecessor of \shades\ was the 8-mJy survey of Scott et
al. (2002; see also Ivison et al. 2002). 
The correlation function for \scuba\ sources derived from
this survey alone did not yield a significant detection of clustering,
even though the large uncertainties meant it was still consistent
with the strong clustering displayed by EROs. There are
nevertheless good reasons for believing the \scuba\ source population to
be highly clustered, and some observational evidence for this is now
found (Blain et al.\ 2004). In particular, cross-correlation with X-ray
sources (Almaini et al. 2003) and Lyman-break galaxies (Webb et
al. 2003) yield clearly-significant detections of clustering. 

Scott et al. (2005) have recently performed a combined
clustering analysis on the three main existing blank-field \scuba\
surveys (the 8-mJy survey, the CUDSS survey by Webb et al.\ 2003, and the
Hawaii survey by Barger et al.\ 1999) to determine whether the existing data
are capable of revealing significant clustering within the sub-mm population
alone. Even though this analysis is based on combining
data from several small fields, it has yielded the first significant
(5-$\sigma$) measurement of sub-mm source clustering on scales $\simeq
0.5-2$\,arcmin, of a strength that does indeed
appear comparable to that found by Daddi et al. (2000) for EROs.
Interestingly, if the integral constraint (see Section 4.2 for its
definition) is varied as a free parameter, the inferred clustering
in fact becomes stronger than that displayed by the ERO population.

Finally, Blain et al. (2004) have found tentative evidence for a
clustering length of $(6.9\pm 2.1) h^{-1}$ Mpc (comoving) for those
submm galaxies for which they could obtain spectroscopic redshifts.
As this was only possible for sources that are also detected at
radio wavelengths, their sample is incomplete and likely to be biased.
Furthermore, Adelberger (2004) argued that the method used is prone to
systematic errors and is unnecessarily noisy.


The paper is organized as follows: Section 2 provides an overview of
\shades\ and its main aims, Section 3 describes the various models
used to make predictions, which are compared to each other 
in Section 4. This section also presents the actual predictions for 
\shades, and we discuss these in Section 5. 

\section{\shades: a wide-area sub-mm survey with redshift information}

The science goals of \shades\ are to help answer three fundamental
questions about galaxy formation: What is the cosmic history of
massive dust-enshrouded star-formation activity? Are \scuba\ sources
the progenitors of present-day massive ellipticals?  What fraction of
\scuba\ sources harbour a dust-obscured AGN? The aim of this
paper is to review and compare the predictions of various existing models
for the bright sub-mm population, and to consider how they can be best
tested and constrained by the final \shades\ dataset, and thus help
answer the first two questions. The third question it not addressed
in this paper, as it involves a detailed analysis of the combined radio,
mid-infrared and X-ray properties of the \shades\ sources.

An important property of \shades\ is having meaningful redshift estimates,
which provides vital information for estimating the bolometric luminosity
of the sources, and hence the cosmic history of energy output from
dust-enshrouded star-formation activity.
Redshift information also holds the key to measuring
the clustering properties of the sub-mm source population. Although
precise spectroscopic measurements of the redshift of a sample of
\shades\ sources will be possible if reliable radio/optical/IR
counterparts can be identified and readily followed-up with 10-m class optical 
telescopes (e.g. Chapman et al. 2003, 2004), in practice one will not be able
to derive this information for the majority of sources in the
survey. However, combination of the \shades\ and \blast\ data will allow the use of
sub-mm photometric redshift techniques, yielding
crude estimates ($\delta z < 0.5$) for individual sources detected in both surveys. 

A Monte-Carlo based photometric-redshift technique has been designed by
Hughes et al. (2002) and tested by Aretxaga et al. (2003, 2004). Here sub-mm
photometric information is combined with prior information on the population,
such as the number counts and the likely evolution of the luminosity
function of dust-enshrouded galaxies, to weight the output redshifts
provided by a large sample of template SEDs. These SEDs represent the wide
range of temperatures, dust emissivities and luminosities found in
nearby IR-bright galaxies.

Even though the redshift distributions are relatively wide, the
detailed information on the shape of the distribution, combined with a
large number of sources, provides a powerful statistical measurement
of population properties such as the parent redshift-distribution and the
global star-formation rate. While, naively, these measurements might seem
insufficiently crude, the combination of the redshift distributions 
of hundreds of sources can indeed measure the history of star
formation of the galaxies detected in more than two sub-mm bands 
($L_{\rm FIR}>2\times 10^{13}$ L$_\odot$) with an accuracy of $\sim$20\% 
(Hughes et al. 2002).

While simulations show that photometric redshift estimates detected only
from sub-mm data have errors of order 0.5 (Hughes et al. 2002), it has been
shown empirically that the inclusion of additional photometric information
provided by detections or upper limits at 1.4 GHz (from the VLA), and
at 170--70 $\mu$m (from the Spitzer Space Telescope),  
increases the accuracy of the photometric
redshifts to $\pm$0.3 (Aretxaga et al. 2004).


\section{Four alternative models of the extragalactic sub-mm population}

Four different models for the clustering of \scuba\ galaxies are presented:
a 'simple merger' model, a 'hydrodynamical' model, a 'stable clustering' model
and a 'phenomenological' model. Some of these are designed especially
with \shades\ in mind, while for other models the
\scuba\ predictions are just part of a range of predictions. The models
also vary in the level of complexity, and in the underlying assumptions,
including the choices for the cosmological parameters, even though
differences in the latter are minor compared to the fundamental
differences between the models.

The aim of this paper is
not to perform a detailed comparison between these models, or between
models and data, but simply to present predictions for a diverse range of
realistic models. The goal
is to study the ability of SHADES to measure clustering, and establish
its capability to distinguish between models.

\subsection{A simple merger model} 
 
The simple merger model is included in order to help
determine the important processes at work in the creation of \scuba\ galaxies.
The underlying premise of this model is that 8-mJy \scuba\ galaxies are 
formed by obscured star formation driven by the violent merger between 
two galaxy sized haloes. The emission is assumed to be above the 8-mJy 
detection threshold for a lifetime $t_{\rm life}$ after the galaxy 
haloes have merged. No direct link is made between the luminosity of 
the SCUBA galaxy and the properties of the merger except that a lower 
limit is placed on the final mass of haloes that contain a detectable 
8-mJy SCUBA source. In other words, a Poisson sampling of massive halo
mergers is assumed to form bright SCUBA galaxies. We have
adopted a mass limit of
$10^{13}\,{\rm M}_\odot$, corresponding to `radio-galaxy' mass haloes. 

Halo mergers were found in a $256^3$ N-body simulation run within a
co-moving $(100 h^{-1}{\rm Mpc})^3$ box using {\sc gadget}, a
publicly available parallel tree code (Springel, Yoshida
\& White 2001). Cosmological parameters were
assumed to have their concordance values ($\Omega_{\rm m}=0.3$,
$\Omega_\Lambda=0.7$, $h=0.70$~\&~$n_s=1$), and the power spectrum
normalization was set at $\sigma_8=0.9$.  Outputs from the simulation
were obtained at 434 epochs, separated approximately uniformly in
time, and halos were found at each epoch using a standard
friends-of-friends routine with linking length $b=0.2$. New halos were
defined to be halos with $>50\%$ of the constituent particles not
having previously been recorded in a halo of equal, or greater
mass. Of these, the halo was said to have been created by major merger
if there were two progenitors at the previous time output that had
mass between $25\%$ and $75\%$ of the final mass.

Obtaining the right number density of \scuba\ sources is limited by the
definition of merging used, the lifetime of emission above the
detection threshold, and the proportion of mergers that result in
SCUBA sources. We therefore simply assume that all of the mergers,
defined as above, result in a luminous \scuba\ source, and allow
$t_{\rm life}$ to vary to give $\sim$300 sources in
$0.5$\,deg$^2$. Because the density of high-mass
($>10^{13}\,{\rm M}_\odot$) mergers is low,
obtaining the correct number density of
\scuba\ sources required a relatively long lifetime
$t_{\rm life}=8\times10^8$\,years.

Mock \scuba\ catalogues for a $0.5$\,deg$^2$ survey were calculated by 
placing a (co-moving) light cone through an array of simulation 
boxes. This is done by selecting output time-steps such that
corresponding redshifts are separated by a box length in co-moving
coordinates. Boxes are reflected, rotated and translated randomly to reduce
the artificial correlation between neighbouring boxes inherent in using a
single simulation, this necessarily reduces real correlations due to
structures that would cross boxes.
Mergers that occurred less than the model lifetime 
before the time corresponding to their luminosity distance were 
flagged as potential \scuba\ sources, and their angular positions and 
redshifts were recorded in order to create mock catalogues. 
 
Obviously, while this model does predict both the spatial distribution 
and redshift space distribution of the \scuba\ sources, it does
not predict the luminosity function. In fact, we note that
following successful comparison between analytic theory and numerical
simulations, both the redshift space distribution and the spatial
distribution of \scuba\ galaxies in this model could have been
accurately estimated analytically (Percival, Miller \& Peacock 2000;
Percival et al. 2003).

\subsection{A hydrodynamical model} 

At the heart of the model is a simulation from Muanwong
et al. (2002) that is an adaptive particle-particle, particle-mesh
code incorporating smoothed particle hydrodynamics (SPH).  The
underlying code is HYDRA (Couchman, Thomas \& Pearce 1995) with the
addition of a standard pair-wise artificial viscosity (Thacker et
al. 2000).  The cosmological model is $\Omega_{\rm m}=0.35$,
$\Omega_\Lambda=0.65$, $h=0.71$, $\sigma_8=0.9$, $\Omega_{\rm b}
=0.019 h^{-2}$.  The simulation used here employs a box of co-moving
size $(100 h^{-1}{\rm Mpc})^3$ with $160^3$ dark matter
particles and $160^3$ gas particles, and is evolved between
$50>z>0$ in approximately 2000 time-steps.  The simulations have various
components: non-interacting dark matter; gas; ``star-like'' (same as gas,
but forming stars); ``galaxy-fragments'', which are collisionless.
Evolution of the various components is as follows: all particles evolve
under gravity; gas can adiabatically heat and cool; gas can also
radiatively cool; at $\rho/\bar\rho >500,\; T<12,000$K gas particles
become ``star-like'' (at this point all the mass is deemed to have been
converted into stars).  An aggregation of 13 or more close
``star-like'' particles become a ``fragment''.  Fragments may accrete
more star-like particles but do not merge.

As a complete model of galaxy formation this simulation has a number
of strengths and weaknesses. It provides a self consistent treatment
of large-scale-structure and galaxy evolution. However, the limited
resolution and the arbitrary solution to ``cooling catastrophe''
necessitated by this, limit its validity.  For the present purpose the
full power of the simulation is not used; it serves as an ingredient
to a more phenomenological model.
 
The first step is to construct a ``galaxy-fragment'' light-cone in the
usual way (as described in Section 3.1).
With the \shades\ sample area it is not necessary
to use more than a single box transverse to the line-of-sight.

The redshift distribution of the fragments in this cone $(dN/dz)_{frag}$
is measured using redshift bins of uniform width $\Delta z = 0.7$,
however there are many more fragments in each bin than would be
detectable. Each fragment is treated as the possible location of a
SCUBA source and is selected based upon its star-formation rate
(SFR), which is measured as the mass per unit time of "star-like"
particles accreted to this fragment (averaged over the last output
time-step). The required number of fragments with the greatest SFR are
selected from each bin such that the redshift distribution matches
a particular model: $(dN/dz)_{SCUBA}$. For this paper, the
analytical form of Baugh et al. (1996) was adopted with a median
redshift of 2.3 and the normalization such to give 300 sources in the
full $0.5$\,deg$^2$ sample size. Hence, the redshift distribution produced
is not derived from the hydrodynamical simulations and this model
merely makes a reasonable choice as to which fragments SHADES will
include, and for these, encodes the positional information from
the simulations.

\subsection{A stable clustering model} 

This is the model of Hughes \& Gazta\~naga (2000), in which a single
output from a N-body simulation that fits well the local spatial
correlation function as measured by APM (Gazta\~naga \& Baugh 1998) is
used to generate a population of SCUBA galaxies in a lightcone. This
corresponds to assuming stable clustering, i.e. a constant spatial
correlation function in co-moving space. Fixing the spatial correlation
function does not imply that we also fix the angular correlation, as that
depends on lightcone geometry, luminosity evolution, and the redshift
selection function.  The prescription for galaxy formation corresponds
to the assumption that the probability for finding a galaxy somewhere within
the lightcone is simply proportional to the local dark matter density,
with the total number of galaxies normalized to the surface-density required
for a given flux limit. Although the redshift distribution is (exponentially)
cut-off beyond $z=6$, this model contains the highest redshift SCUBA galaxies
of all models considered in this paper.

This model has been used in the photometric
redshift estimation technique of Aretxaga et al. (2003), to constrain
sample size and depth given the correlation length, and to test
correlation function measurements from surveys with relatively small
sky coverage (Gazta\~naga \& Hughes 2001).

\beginfigure{1} 
{\psfig{file=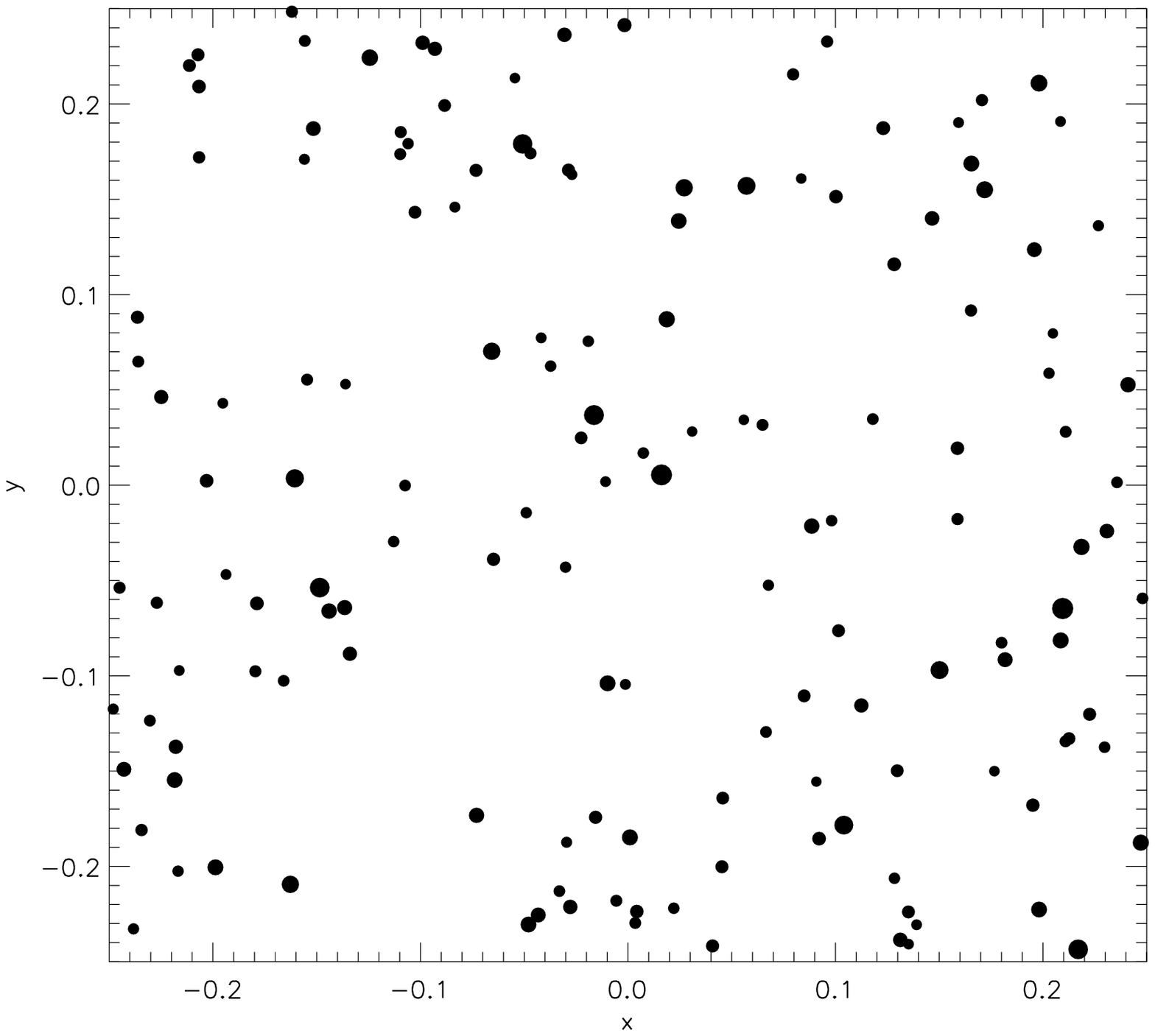,width=8.7cm,silent=1}} 
\caption{{\bf Figure 1.} A simulated distribution of 8-mJy sources
for the phenomenological model described in Section 3.4, with no
redshift selection. The size of the field is equal to each of the
two fields that will make up \shades: a quarter square degree.
The diameter of each dot is proportional to the sub-mm magnitude of
the source it represents.} 
\endfigure 

\subsection{A phenomenological model} 

The phenomenological galaxy formation model of van Kampen, Rimes \&
Peacock (2005), a revised version of the model of van Kampen, Jimenez
\& Peacock (1999), is semi-numerical, in the sense that the merging
history of galaxy haloes is taken directly from N-body simulations
that include special techniques to prevent galaxy-scale haloes
undergoing `overmerging' owing to inadequate numerical resolution.
When haloes merge, a criterion based on dynamical friction is used to
decide how many galaxies exist in the newly merged halo. The most
massive of those galaxies becomes the single central galaxy to which
gas can cool, while the others become its satellites.
 
When a halo first forms, it is assumed to have an isothermal-sphere
density profile. A fraction $\Omega_{\rm b}/\Omega_{\rm m}$ of this is
in the form of gas at the virial temperature, which can cool to form stars
within a single galaxy at the centre of the halo.  Application of the
standard radiative cooling curve shows the rate at which this hot gas
cools and is able to form stars.  Energy output from
supernovae reheats some of the cooled gas back to the hot phase. When
haloes merge, all hot gas is stripped and ends up in the new halo.
Thus, each halo maintains an internal account of the amounts of gas
being transferred between the two phases, and consumed by the
formation of stars.

The model includes two modes of star formation: quiescent star
formation in disks, and star-bursts during major merger events.  Having
formed stars, in order to predict the appearance of the resulting
galaxy it is necessary to assume an IMF, which is generally taken to
be Salpeter's, and to have a spectral synthesis code, for which we use
the spectral models of Bruzual \& Charlot (1993).  The evolution of
the metals is followed, because the cooling of the hot gas depends on
metal content, and a stellar population of high metallicity will be
much redder than a low metallicity one of the same age. It is taken as
established that the population of brown dwarfs makes a negligible
contribution to the total stellar mass density, and the model does not
allow an adjustable $M/L$ ratio for the stellar population.  The
cosmological model adopted is $\Omega_{\rm m}=0.3$,
$\Omega_\Lambda=0.7$, $h=0.7$, $\sigma_8=0.93$, $\Omega_{\rm b}=0.02
h^{-2}$.  The 850$\mu$m flux is assumed to be proportional to the star
formation rate (with 8mJy corresponding to 1000 $M_\odot$/yr, as found
by Scott et al. (2002), with a random term of order 50 per cent added or
subtracted to mimic the uncertainty in dust temperature, grain sizes,
and other properties that are not yet included in the modelling.

The model used in this paper has a
mixture of bursting and quiescent star formation, with most of the
recent star formation occurring in discs, following the Schmidt law
with a threshold according to the Kennicutt criterion, and most of the
high-redshift star formation resulting from merger-driven star-bursts.
The model is similar in philosophy to that of Hatton et al. (2003) and
Baugh et al. (2004), but with many differences in the details and parameters
adopted.

\subsection{Model comparison}

In this section we compare the models in order to get a qualitative
description of where the differences lie.

What drives the flux in each of the models? In the simple merger model,
an actual flux is not calculated; rather, a merger mass threshold is
related to a flux threshold. The hydrodynamical and stable clustering
models have some of the properties of the galaxies fixed, but not the flux,
which is assigned statistically. The phenomenological model is the only one
that generates a flux from the actual physical properties of the galaxies,
taking into account the approximations and assumptions made.

What drives the clustering signal in each of the models? In the stable
clustering model the {\it spatial} correlation function
is fixed in comoving space, which means that the angular correlation function
is built up along the lightcone in a way that depends on the selection of galaxies
as a function of redshift. The simple merger, hydrodynamical and phenomenological
models produce galaxies first, and built up the angular correlation function
along the line of sight. In the simple merger a one-to-one correspondence between
galaxies and haloes is assumed, whereas the other two models have more complex
relations between mass and light.

What determines the redshift distribution in each of the models? In the
hydrodynamical model it is simply taken from Baugh et al. (1996), whereas
for the other models it is actually an outcome of the models, although in
the stable clustering model an exponential cut-off at $z\approx 6$ is applied.
The main determining factor for the simple merger and phenomenological models
for the redshift distribution is the merger rate as a function of redshift.
For the simple merger model this is obvious, but for the phenomenological
model this follows from the dominant contribution of merger-driven starbursts
to the sub-mm flux. 

Besides the models used in this paper, other models exist in the
literature, which are similar in philosophy to those included here, but
still produce different predictions for the sub-mm population.
Models similar to the phenomenological model are those of Hatton et al.
(2003) and Baugh et al. (2004), which differ mainly in the details of
the physics implemented, and the choice of parameters. 

\beginfigure{2}
{\psfig{file=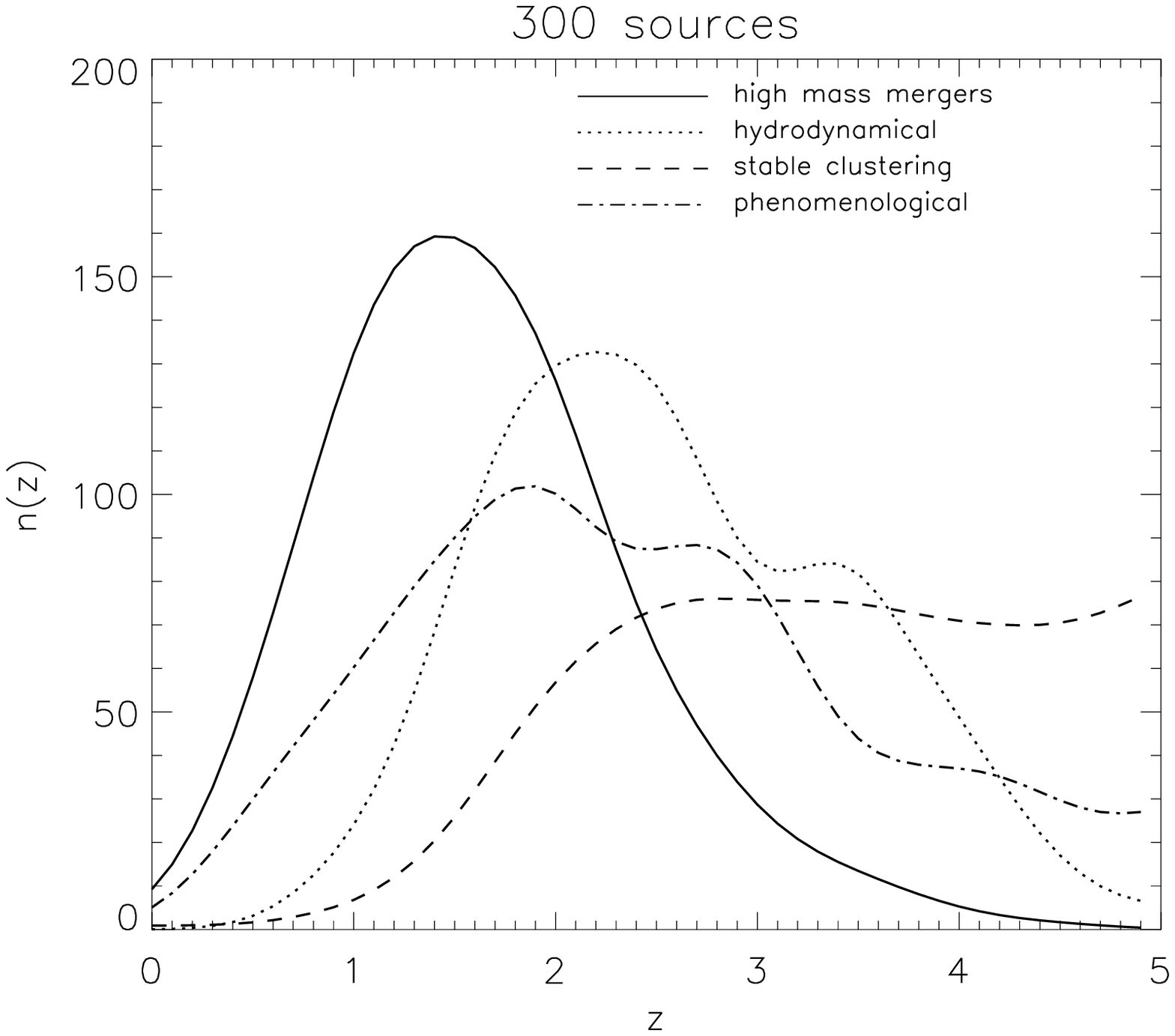,width=9cm}}    
\caption{{\bf Figure 2.} Redshift distributions for all models with 300 sources,
averaged over 25 mock \shades\ datasets for each model. The model distributions
were convolved with a Gaussian of radius 0.4, roughly mimicking the statistical
uncertainty in the photometric redshift errors.} 
\endfigure

\section{Model predictions compared} 

In Fig.\ 1 we show a complete simulation of one of the two 850\,$\mu m$ datasets
that will comprise \shades, produced using the
phenomenological model of van Kampen et al. (2004).
A simple square geometry was chosen, although the actual survey geometry
of each of the two \shades\ fields could be of a somewhat different shape.
All sources with fluxes larger than 8 mJy are shown, where the symbol size
is proportional to the logarithm of the flux, i.e. a sub-mm magnitude.

We now consider simulated \shades\ datasets as predicted from the
various alternative models of the sub-mm source population
presented in Section 3. We do not consider the effects of noise and 
sidelobes, but we do take into account the effects of the 15 arcsec SCUBA
beam by merging into single sub-mm sources anything closer together
than 7 arcsec. This reflects the 
resolution expected for the final source extraction from the \shades\
images. After the removal of close pairs we assume the model sub-mm
sources to reflect the final \shades\ source list.

The final survey is expected to contain around 300 sources, i.e.
150 sources per field. We produce, for each model, 50 realizations of
individual fields, i.e. 25 mock \shades\ datasets of 300 sources each.

\subsection{Predictions for the redshift distribution} 

For the four models, we show in Fig.\ 2 the redshift distributions expected
after smoothing with a Gaussian filter of radius 0.4, which reflects, very
crudely, the resolution achievable with photometric redshifts. The
distributions are obtained by averaging over all realisations for each
model, and are normalized to the total source count of 300.

Even after relatively heavy smoothing, we see that the redshift distributions
are rather different, and are clearly distinguishable from each other. 
This means that even crude but complete redshift information will be of
enormous benefit in differentiating between and constraining models.
Obviously, obtaining more accurate redshifts should help to tune the
models that survive this first test even further.

For the purpose of comparing clustering properties between the models,
note from Fig.\ 2 that the redshift range $2<z<3$ is the only range
where all models have a reasonable number of sources to attempt a
correlation function analysis.
The differences in the distribution stem from the different assumptions
for each of the models: the simple merger model assumes that only high-mass
mergers can form \scuba\ sources, which, in a hierarchical structure formation
scenarion, necessarily places them at lower redshifts as compared to the other
models, while the simple, unbiased galaxy formation prescription of the stable
clustering model places \scuba\ sources at relatively high redshifts.

Current spectroscopic redshift measurements for sub-mm selected galaxies
are incomplete and only available for small samples, so any redshift distribution
derived from such measurements is tentative. Chapman et al. (2004) claim that
a Gaussian distribution with $\bar z=2.4$ and $\sigma_z=0.65$ fits their
available data well, but the incompleteness of their dataset imposed through their
radio selection hinder a comparison to models.


\subsection{Clustering measures} 

The estimated redshifts have a predicted accuracy of $\delta z \sim \pm 0.4$,
which means that we cannot directly measure the 3D spatial correlation
function $\xi(r)$. However, we do not have to restrict ourselves to measuring
angular clustering, as photometric redshifts can be used to boost the angular
clustering signal-to-noise by splitting the sample in redshift bins,
or by only considering pairs of galaxies that lie at similar redshifts.
Even so, the measured correlation function
will be noisy, so we use integrals of this function, as considered in
the early days of optical galaxy surveys when total source counts were
much lower than today (e.g. Davis \& Peebles 1983).

\subsubsection{Estimating the angular correlation function}

The method for modelling the clustering of sources proceeds
as follows. From the data, the Landy \& Szalay (1993) estimator
$1+w_{LS}=1+(DD-2DR+RR)/RR$ is calculated, where $DD$, $DR$ and $RR$ are
the (normalized) galaxy-galaxy, galaxy-random and random-random
pair counts at separation $\theta$, calculated from the galaxy
sample and a large random catalogue containing 10000
points that Poisson samples the survey region.
This estimator is then fitted by its expected value
$$1+\langle w_{LS} \rangle = [1+w(\theta)]/(1+w_\Omega)\ , \eqno\stepeq$$ 
where $w_\Omega$ is the integral of the model two-point correlation
function over the sampling geometry:
$$ w_\Omega = \int_\Omega G_p(\theta)w(\theta)d\Omega\ . \eqno\stepeq$$
The function $G_p(\theta)$ is the probability density function
of finding two randomly placed points in the survey at a
distance $\theta$. This ``integral constraint'' corrects for the
effect of not knowing the true density of objects
(Groth \& Peebles 1977; Landy \& Szalay 1993) and stops the
recovered correlation function being biased to low values
compared with the true function.
Note that eq.\ (1) implies that the true correlation function is biased
low by a {\it factor} $1+w_\Omega$, whereas often this is approximated
by $w(\theta) = w_{LS} - w_\Omega$, i.e. ignoring the term
$w_{LS} w_\Omega$.

We also introduce an alternative to the standard angular correlation function
that takes redshift information into account in an unorthodox way.
In the counting of $DD$ pairs, we just consider those pairs that
have a redshift separation of at most 0.4, whereas the $DR$ and $RR$
counts are still obtained for all galaxy pairs. This is
equivalent to removing distant pairs that are expected to be
unclustered from an analysis of the angular correlation function of
all of the objects in the survey. It is clear that this approach must
increase the signal-to-noise of the recovered correlation function.

\subsubsection{The sky-averaged angular correlation function}

For the relatively small number of sources being detected in \shades,
we measure an integral of the Landy \& Szalay estimator.
Such an approach has previously been used to analyse clustering within
early galaxy redshift surveys (eg. the CfA survey; Davis \& Peebles 1983).
The statistic that was often obtained in these analyses was
the integrated quantity $J_3$, defined as
$$J_3(r) \equiv \int_0^r \xi(y) y^2 dy \ .\eqno\stepeq$$
The dimensionless analogue of $J_3$ is called the
volume-averaged correlation function:
$$\bar\xi(r) \equiv {3\over r^3} \int_0^r \xi(y) y^2 dy
   = 3 {J_3(r)\over r^3}\ .\eqno\stepeq$$
This measures the fluctuation power up to the scale $r$, and is
therefore a useful measure for a survey that is limited
in object numbers. For reference,
$\bar\xi(10 h^{-1} {\rm Mpc})=0.83$ was found for the
optical CfA survey (Davis \& Peebles 1983; no error given).

As we cannot measure the spatial clustering function $\xi(r)$, as the
redshift determinations are very uncertain, we use the two-dimensional
version of $\bar\xi$, the sky-averaged angular correlation function
$$\bar w(\theta) \equiv {2\over\theta^2} \int_0^\theta w(\phi) \phi d\phi \
  ,\eqno\stepeq$$
where $w(\theta)$ is the angular correlation function, which is the
projection of $\xi(r)$ along the line-of-sight. Our estimator of this
statistic was calculated by numerically integrating the angular
correlation function (calculated using the Landy \& Szalay 1993
estimator), in the form of
logarithmically binned estimates $w_i$, up to the angle $\theta_i$
using eq. (5):
$$\bar w_i = {2\over\theta_i^2} \sum_{j\leq i} w_j \theta_j^2 \Delta
                    \ , \eqno\stepeq$$
where $\Delta$ is the logarithmic binsize.
The errors on $\bar w_i$ are obtained by propagating the errors on $w_i$
through this summation.
This estimate for the true sky-averaged angular correlation function
is also biased, and has its own integral constraint, similar to the
one for $w(\theta)$. For a power-law
correlation function $w_{\rm pl}(\theta)=(\theta/A)^{-\delta}$
$$\bar w_{\rm pl}(\theta) = {2\over 2-\delta}
             \Bigl({\theta \over A}\Bigr)^{-\delta} .\eqno\stepeq$$ 
Thus, the sky-averaged correlation function is also a power-law, with
the same slope and different amplitude (except for $\delta=2$),
and the integral constraint $\bar w_\Omega$ is scaled by the same
factor $2/(2-\delta)$ with respect to $w_\Omega$ (see eq. 2).
We can therefore fit power-law models to either $\bar w(\theta)$ or
$w(\theta)$, and constrain the same parameters.

\subsubsection{Fitting to the model correlation functions}

Traditionally, $\chi^2$ minimization is used to fit a power-law function
to the estimated correlation function. This type of minimization is, strictly
speaking, only valid for binned data that is uncorrelated and has Gaussian
errors. Our estimates of $w(\theta)$ and $\bar w(\theta)$ are correlated
for different maximum separations, and have errors that are
non-Gaussian (it is easy to see that, in the absence of clustering,
the errors are strictly Poisson, as shown by Landy \& Szalay (1993)
for the estimator of $w(\theta)$). In order to
constrain models of the correlation function using our binned
estimates, we should therefore perform a full likelihood calculation
taking into account the potentially complex shape of the likelihood.
Both $w(\theta)$ and $\bar w(\theta)$ obviously contain
the same information, and should therefore result in the same
likelihood surface for a given model.

The reason for fitting $\bar w(\theta)$ rather than $w(\theta)$ lies
in the approximations that are made in estimating the likelihood. For
the sky-averaged angular correlation function, the bins are dependent
on more pairs of galaxies than the corresponding direct estimator of
the correlation function, so that the sky-averaged statistic will have
a distribution closer to a Gaussian form. Switching from a
direct estimate to a sky-averaged estimate increases correlations
between data points. However, this can be taken into account properly
in the fitting procedure by calculating
the full covariance matrix for the binned data, which is then 
diagonalized by a unitary transformation to produce an alternative
$\chi^2$-statistic (e.g. Fisher et al. 1994).
This statistic is subsequently employed to fit models to the data.


A single parameter fit to the correlation function is often adopted
assuming that $w(\theta) = (\theta/A)^{-0.8}$ (e.g. Roche et al. 1993;
Daddi et al. 2000). It is not at all clear what the slope for
high-redshift sub-mm sources is going to be, but it is easier
to fit a one-parameter function for a small number of galaxies.
In the following we consider both a one-parameter model with
constrained power-law slope $\delta=0.8$, and a two-parameter fit for
the generic power-law $w(\theta) = (\theta/A)^{-\delta}$ to both
$w(\theta)$ and $\bar w(\theta)$. We employ non-linear $\chi^2$-fitting
for both functions, using the Levenberg-Marquardt method (Press et al.
1988), which allows us to
easily take into account the multiplicative integral constraint.
For each fit the $\chi^2$ probability $Q$ is calculated using the
incomplete gamma function, and any fits with $Q<0.1$ are discarded.

\beginfigure{3}
{\psfig{file=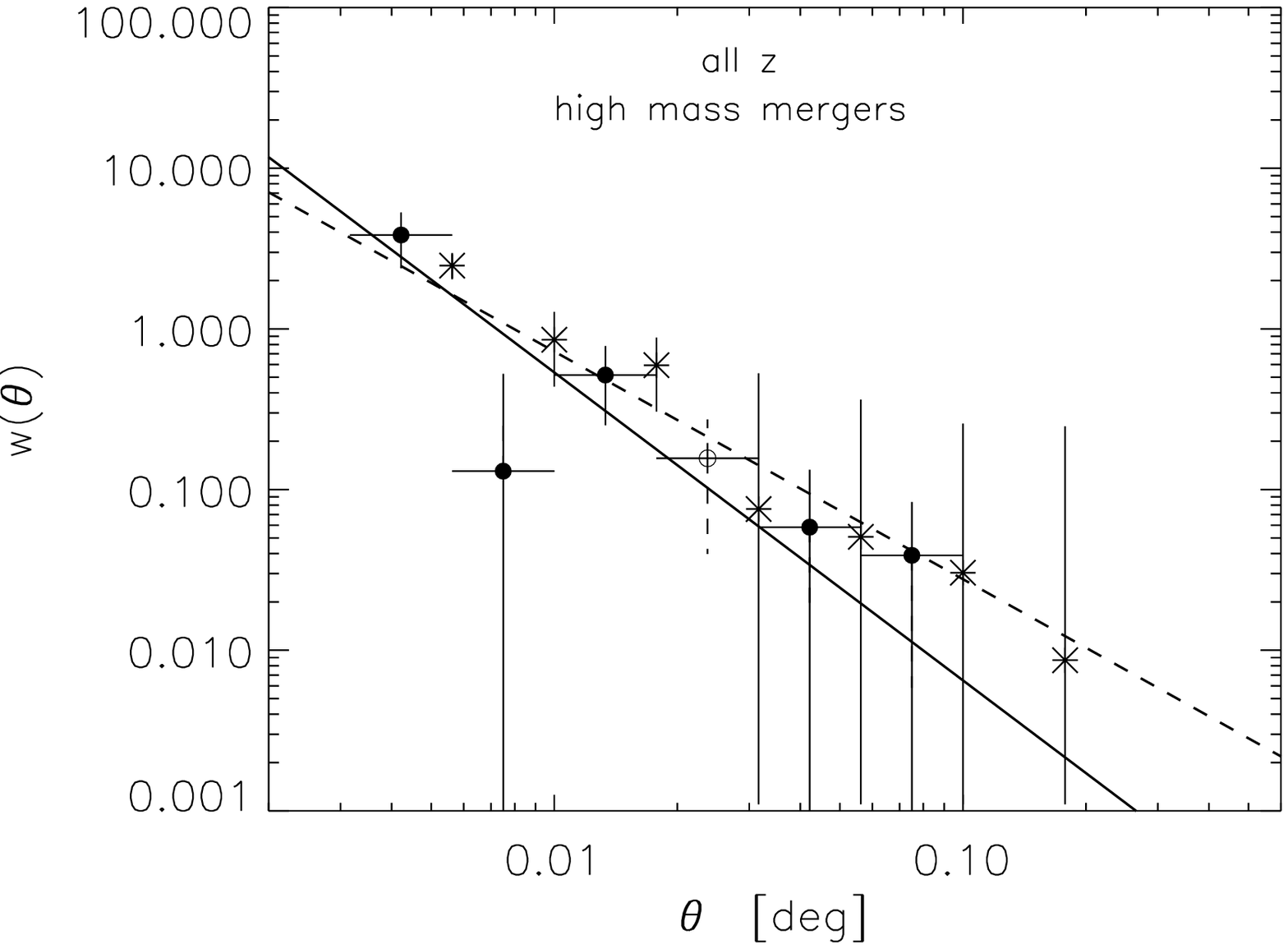,width=7.1cm}}
\vskip -0.5cm
{\psfig{file=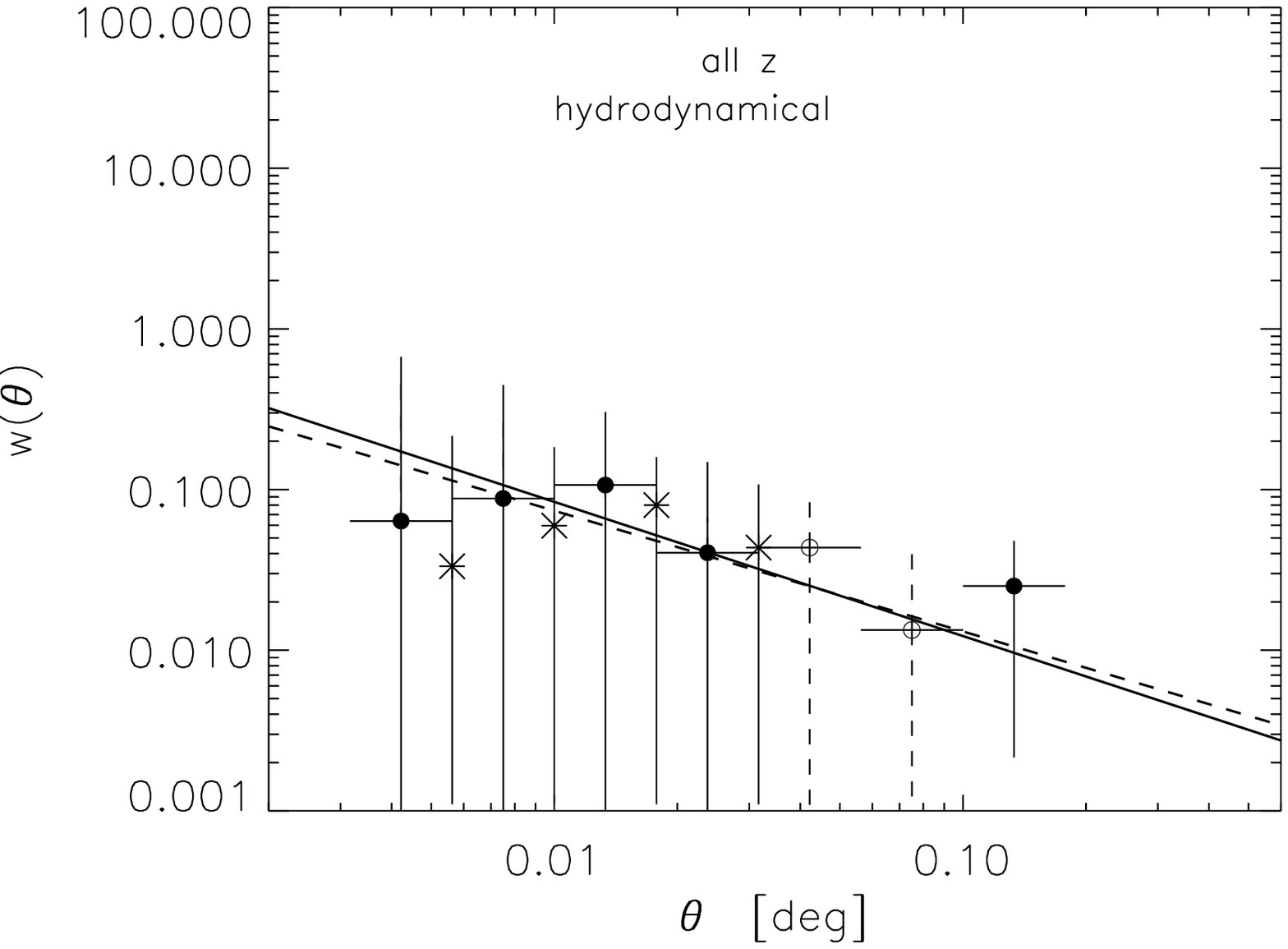,width=7.1cm}} 
\vskip -0.5cm
{\psfig{file=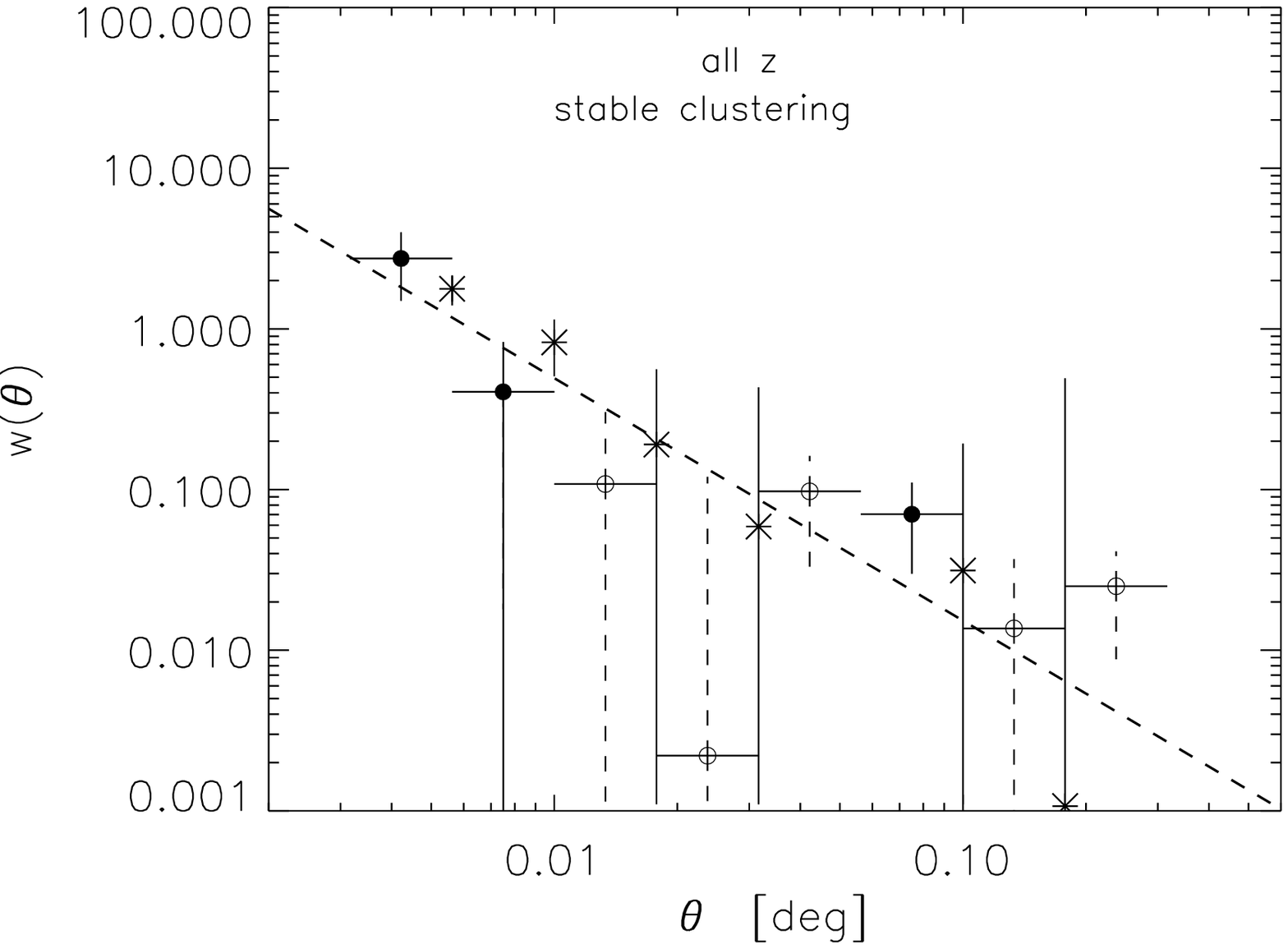,width=7.1cm}} 
\vskip -0.5cm
{\psfig{file=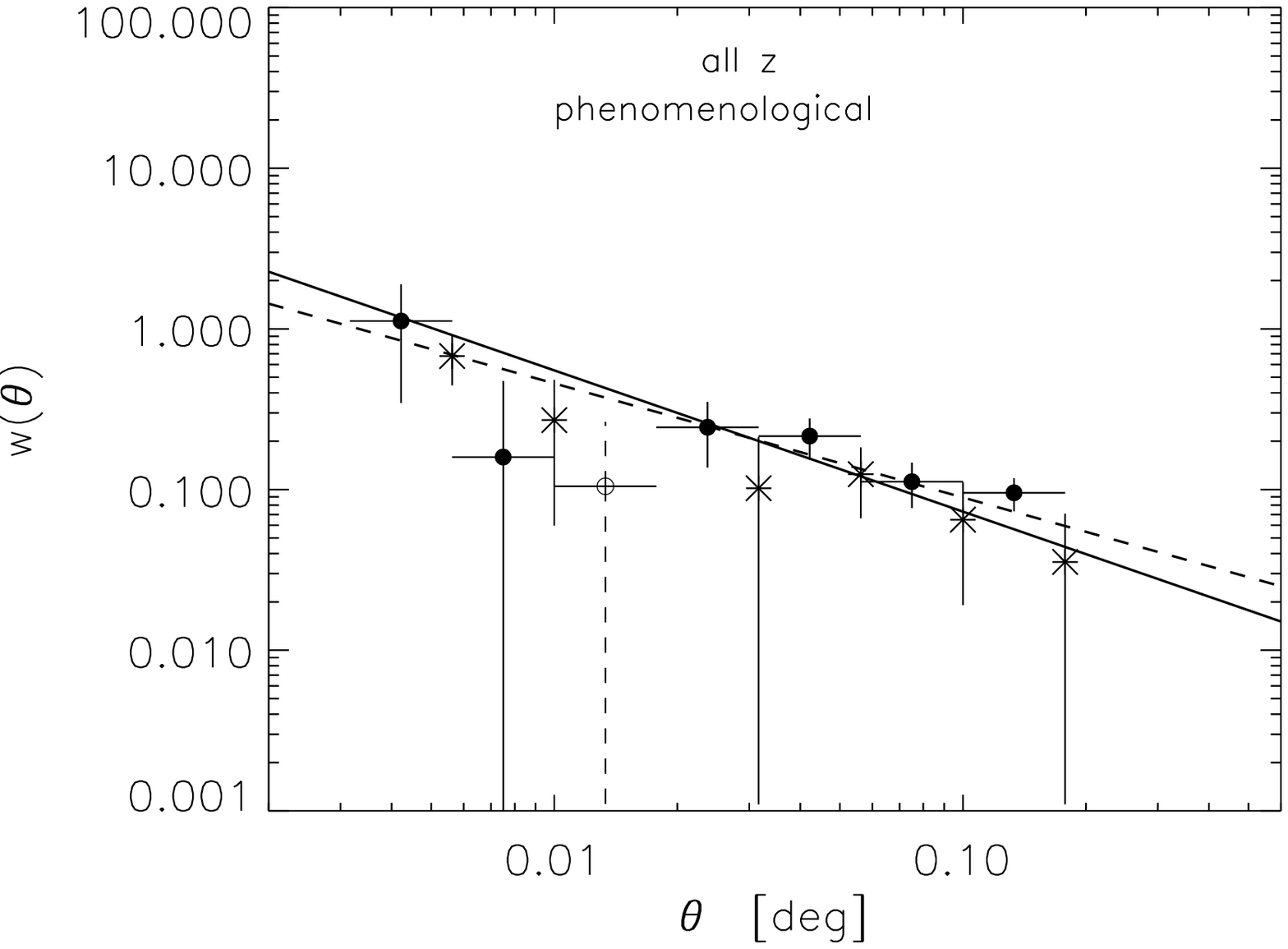,width=7.1cm}} 
\caption{{\bf Figure 3.} Bias-corrected angular correlation function
$w(\theta)$ (round symbols, open meaning negative) and its sky-averaged
counterpart $\bar w(\theta)$ (stars, negative values not plotted)
for a single realisation of each model, with best fit power-law functions
overplotted for both, and no redshift selection. See text for full details.}
\endfigure 

\beginfigure{4} 
{\psfig{file=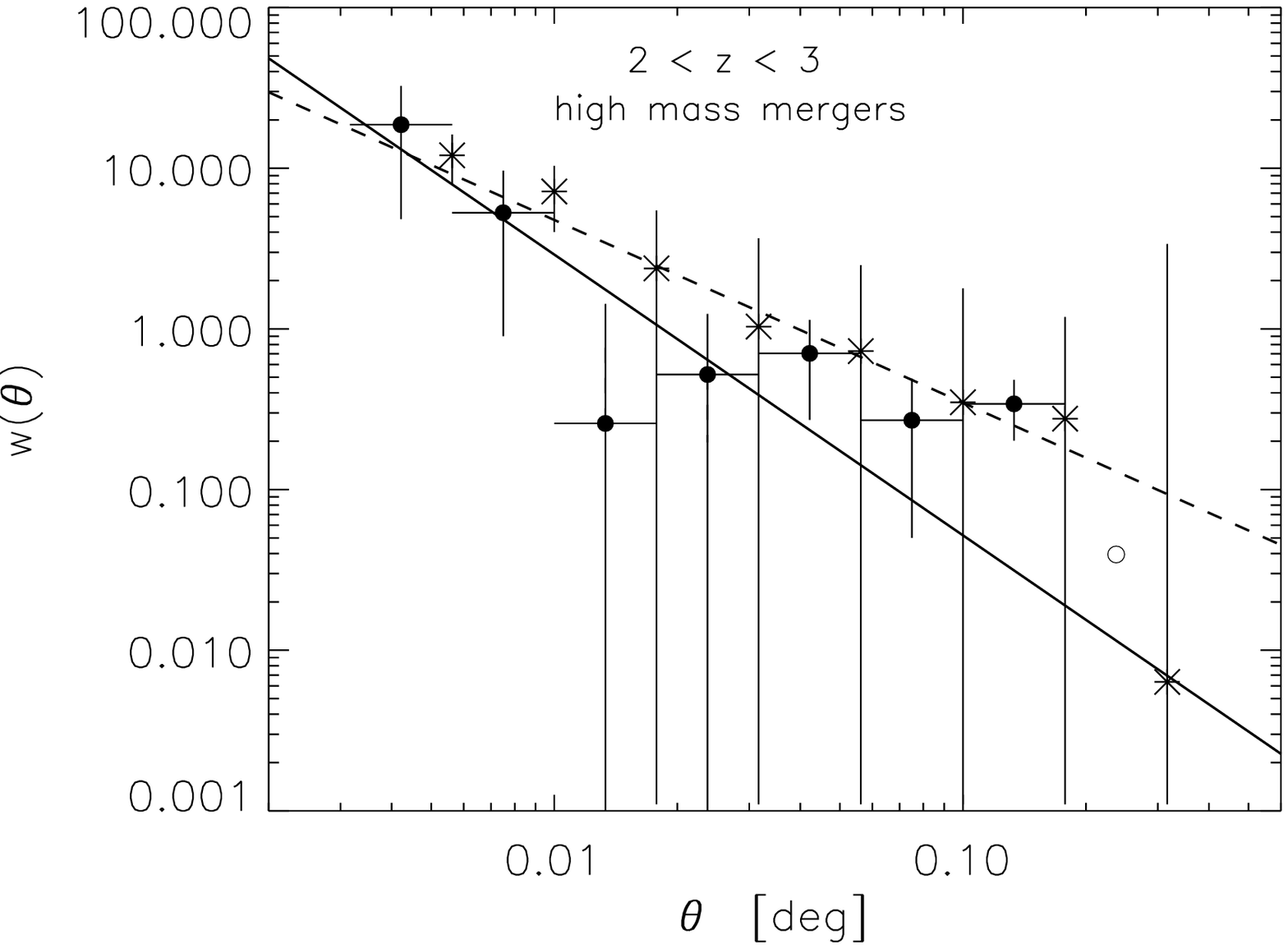,width=7.1cm}}
\vskip -0.5cm
{\psfig{file=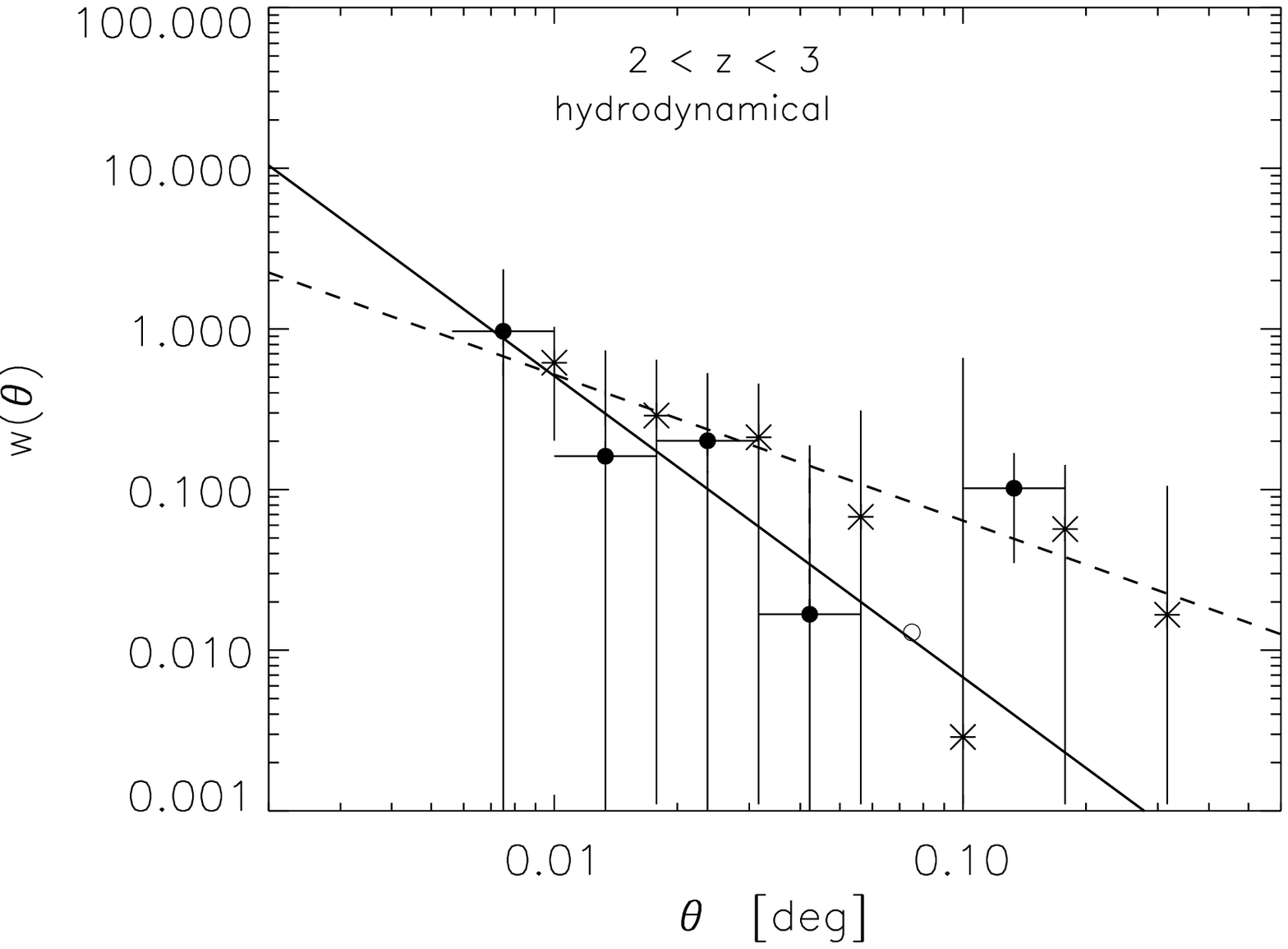,width=7.1cm}} 
\vskip -0.5cm
{\psfig{file=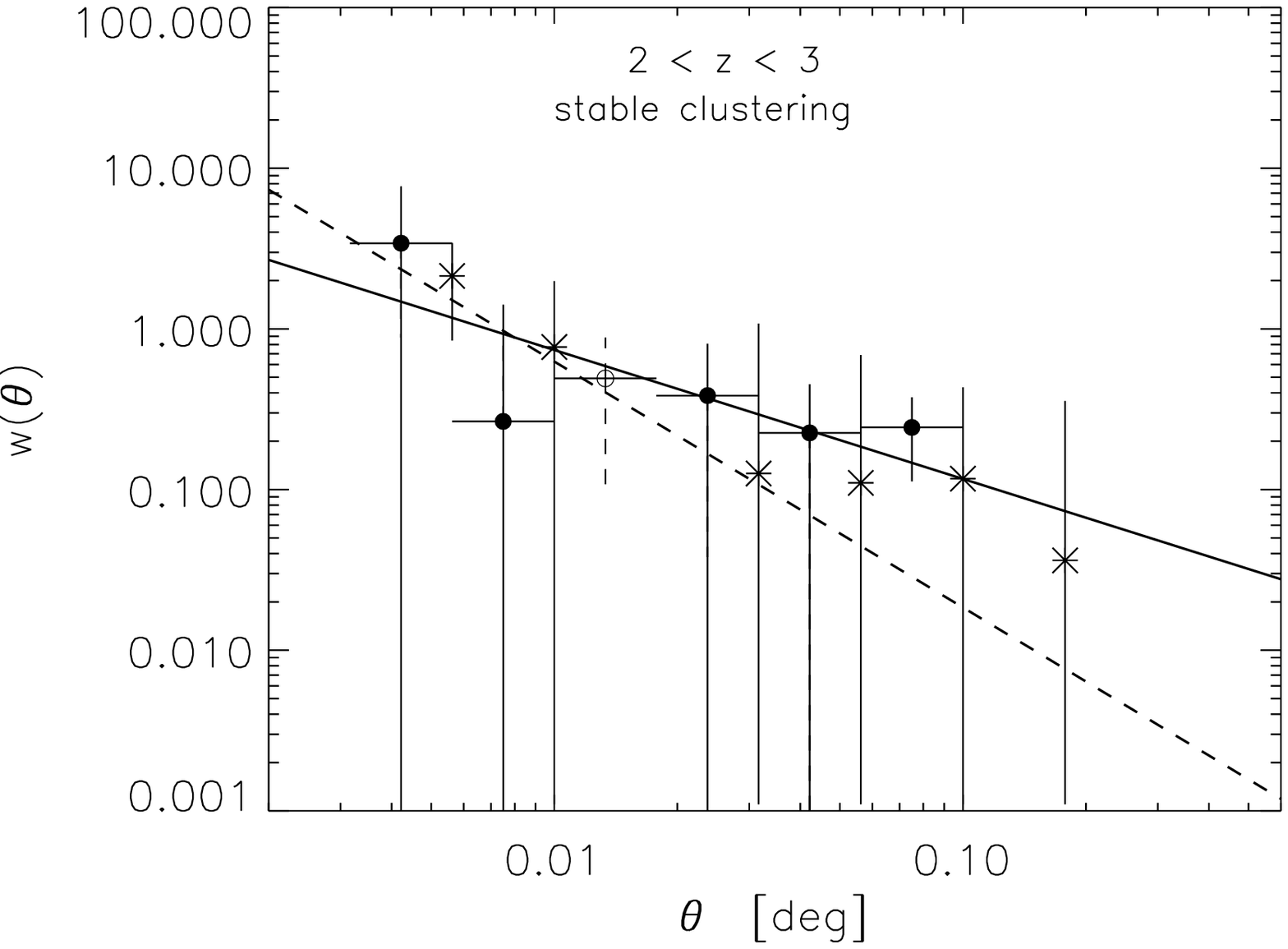,width=7.1cm}} 
\vskip -0.5cm
{\psfig{file=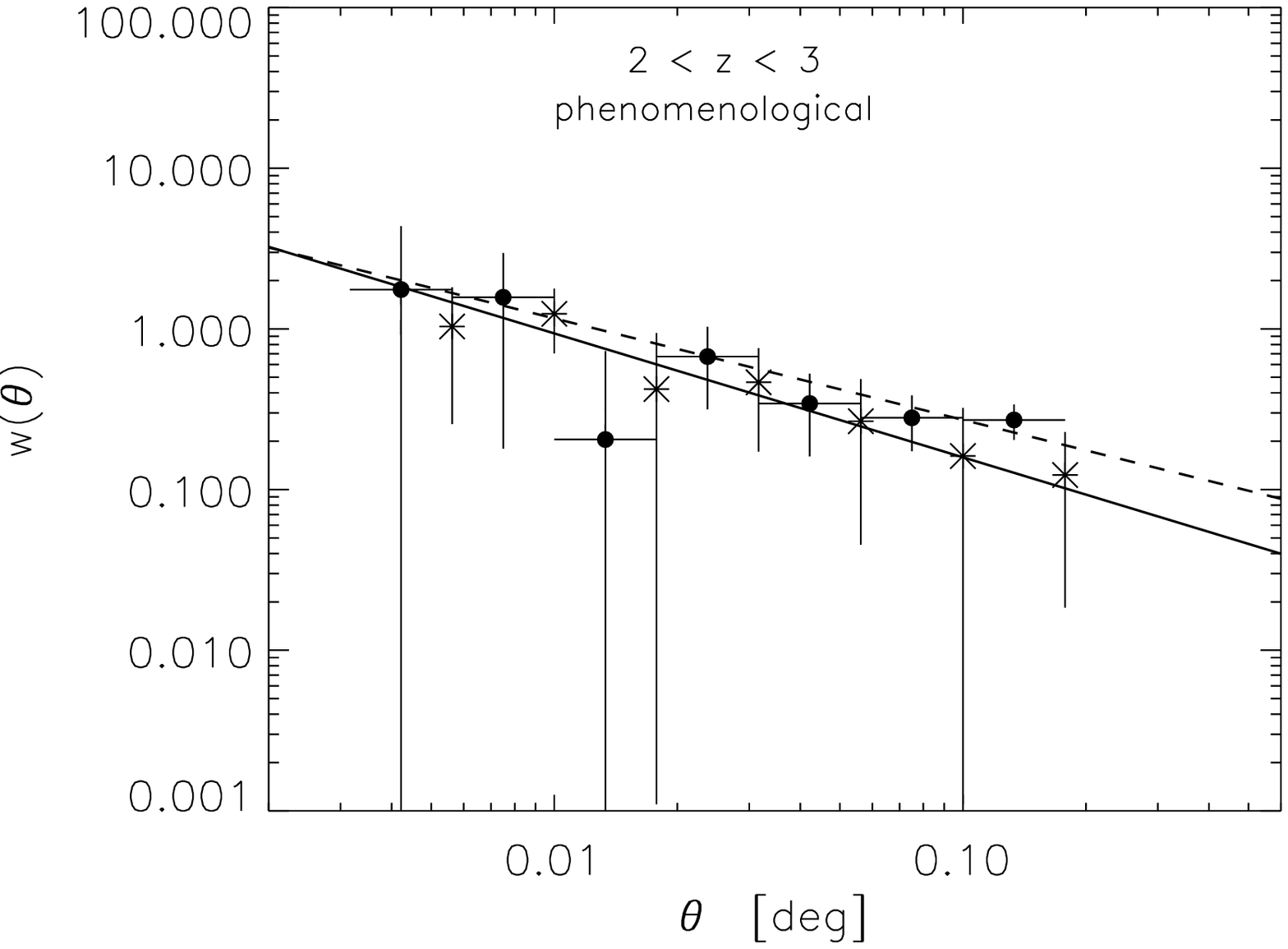,width=7.1cm}}
\caption{{\bf Figure 4.} Same as Fig.\ 3,
but with sources selected to be in the redshift range $2<z<3$,
which is the range where all models have a reasonable amount
of sources in this redshift bin (see Fig. 2). All realisations
are exactly the same as in Fig. 3, in order to demonstrate
what redshift availability can achieve.} 
\endfigure 

\beginfigure{5} 
{\psfig{file=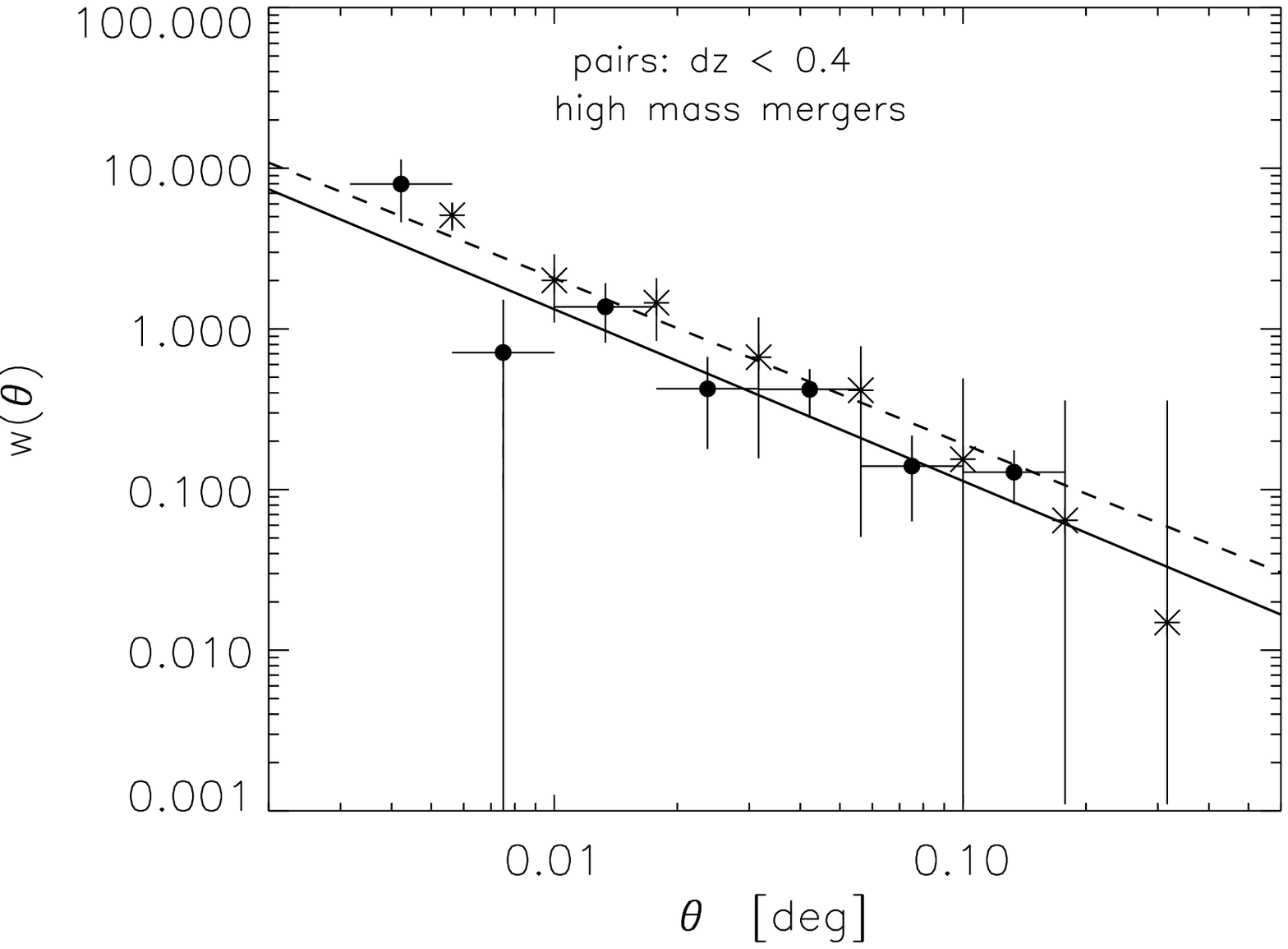,width=7.1cm}}
\vskip -0.5cm
{\psfig{file=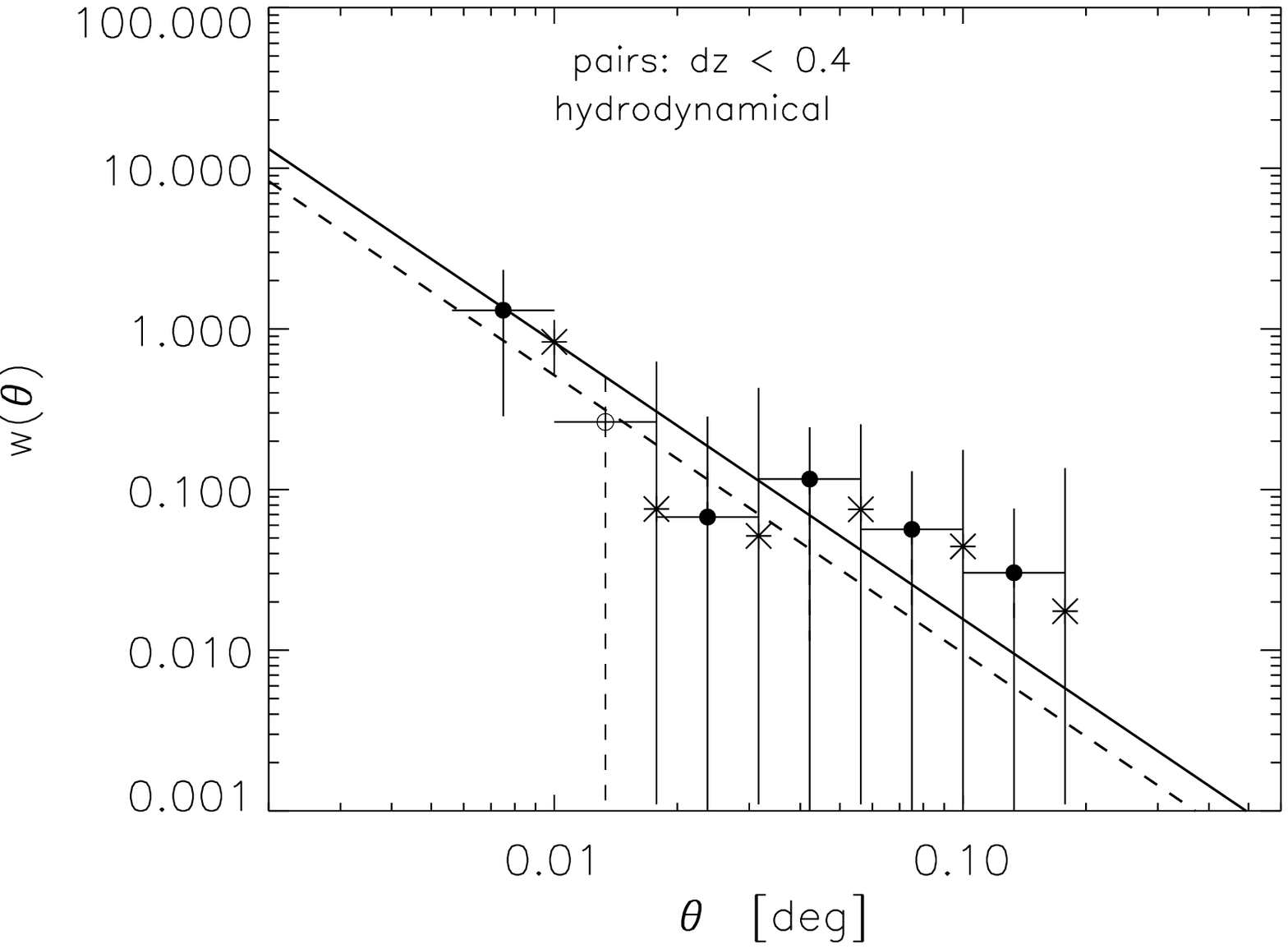,width=7.1cm}} 
\vskip -0.5cm
{\psfig{file=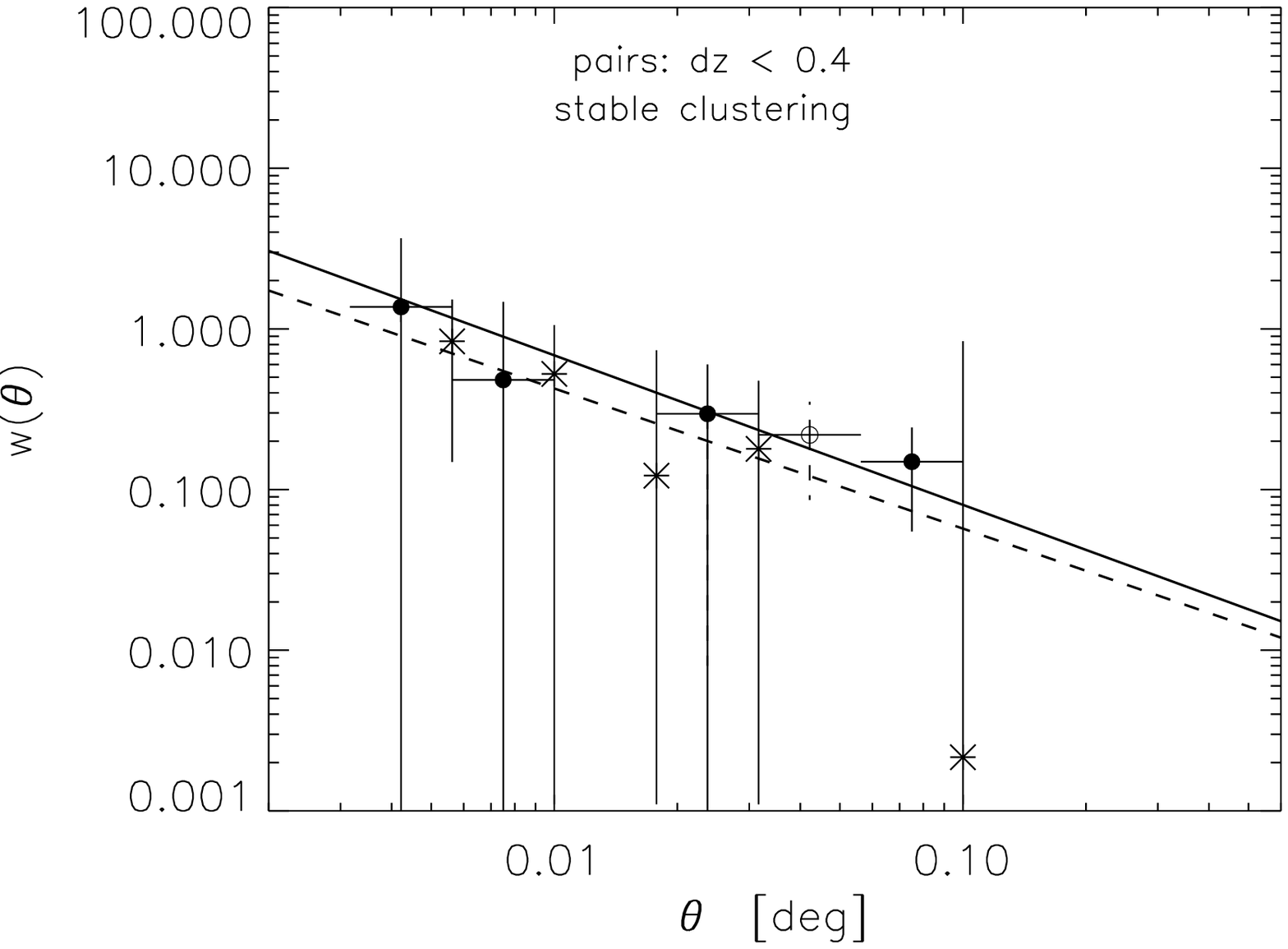,width=7.1cm}} 
\vskip -0.5cm
{\psfig{file=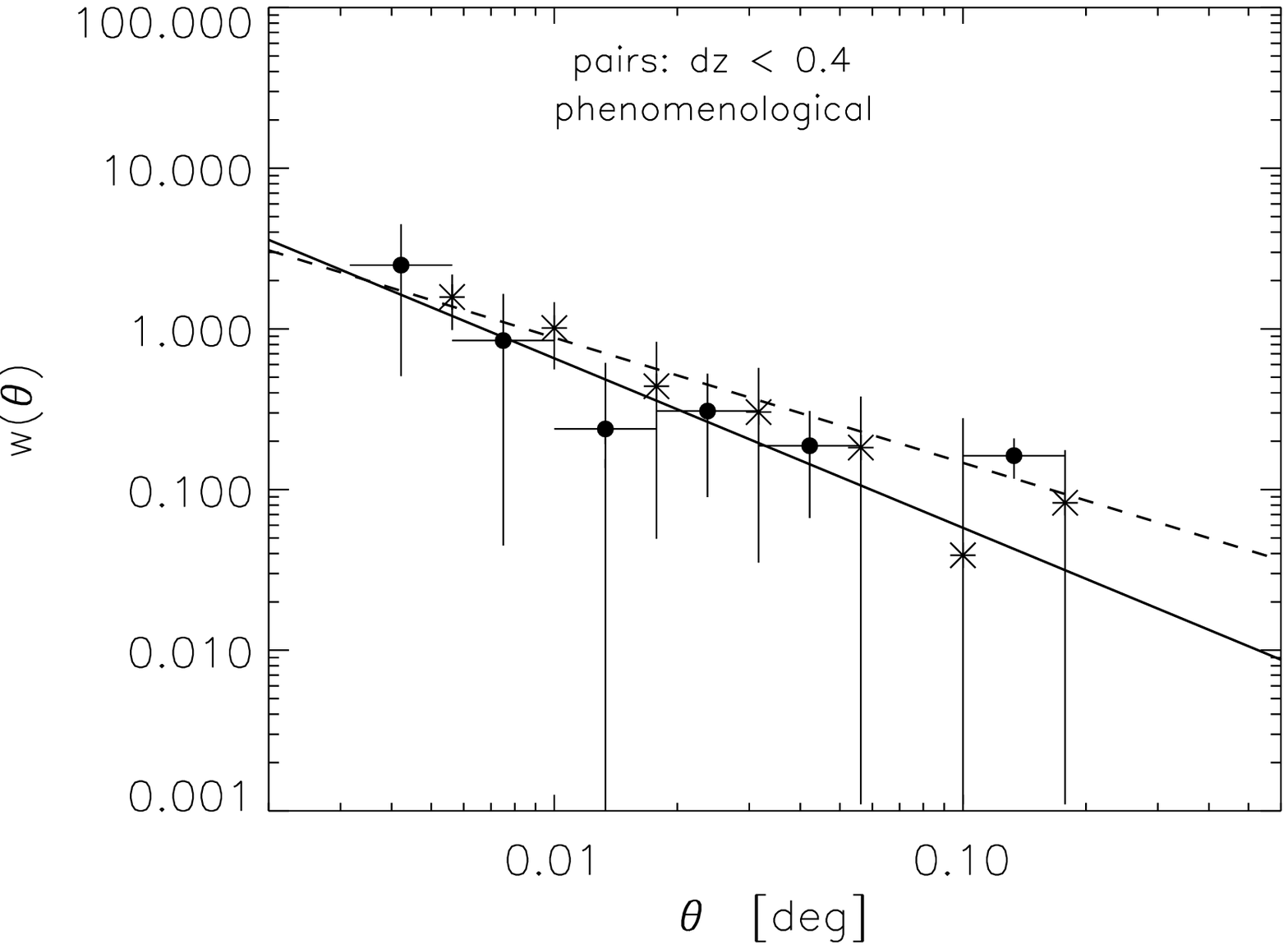,width=7.1cm}}
\caption{{\bf Figure 5.} Same as Fig.\ 3,
but with sources selected to be in redshift pairs with
$\delta z < 0.4$. All realisations are exactly the same as
in Figs. 3 and 4.} 
\endfigure

\beginfigure{6}
{\psfig{file=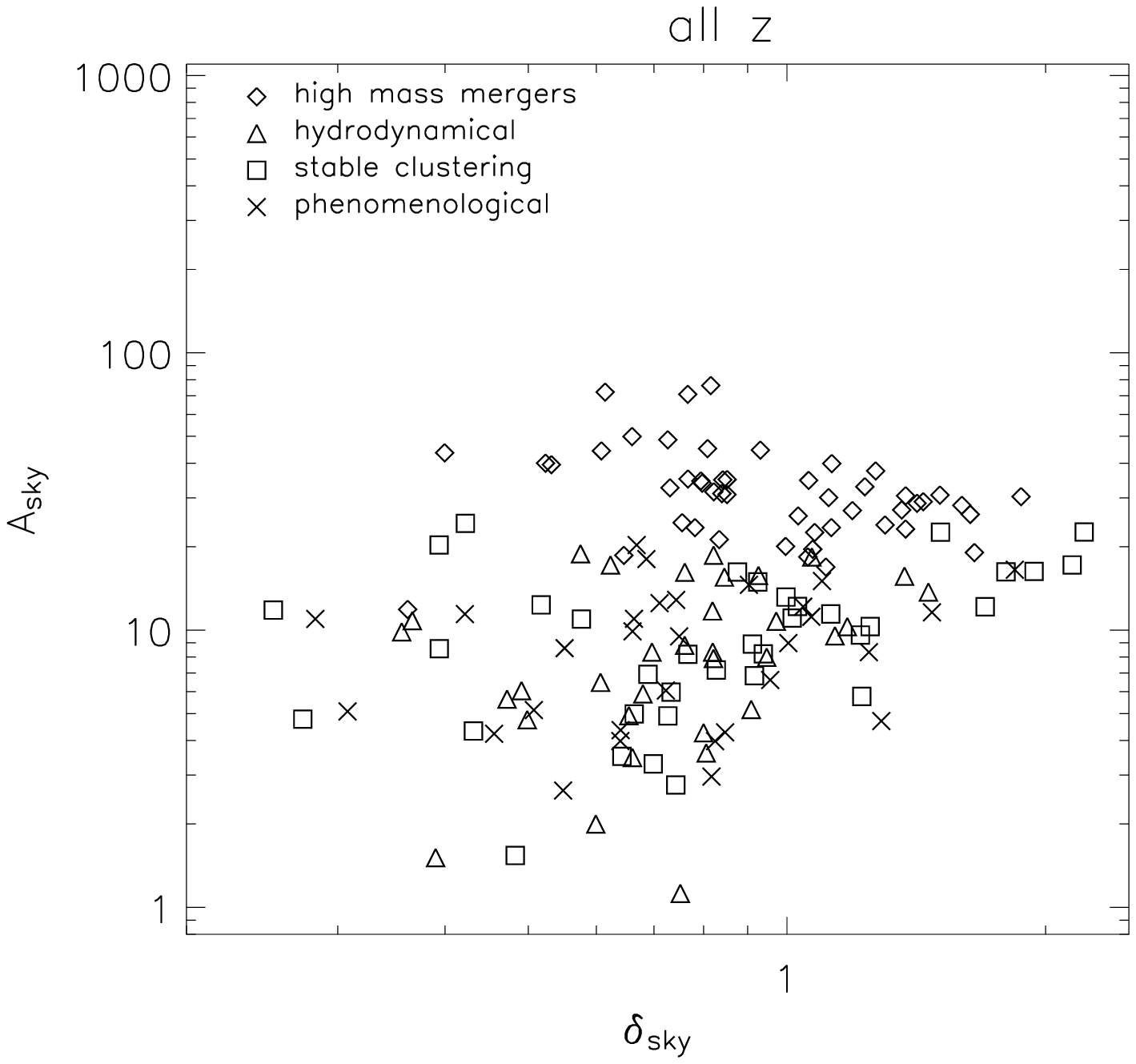,width=8.3cm}} 
\caption{{\bf Figure 6a.} Scatter plot for the two fitting parameters
of the sky-averaged angular correlation function, $A_{\rm sky}$ and
$\delta_{\rm sky}$, for 50 fields of 150 sources. Only fits of sufficient
quality, i.e. those with a $\chi^2$ probability larger than 0.1,
are included (see text for details).}
\vskip 0.5cm
\endfigure 

\beginfigure{7}
{\psfig{file=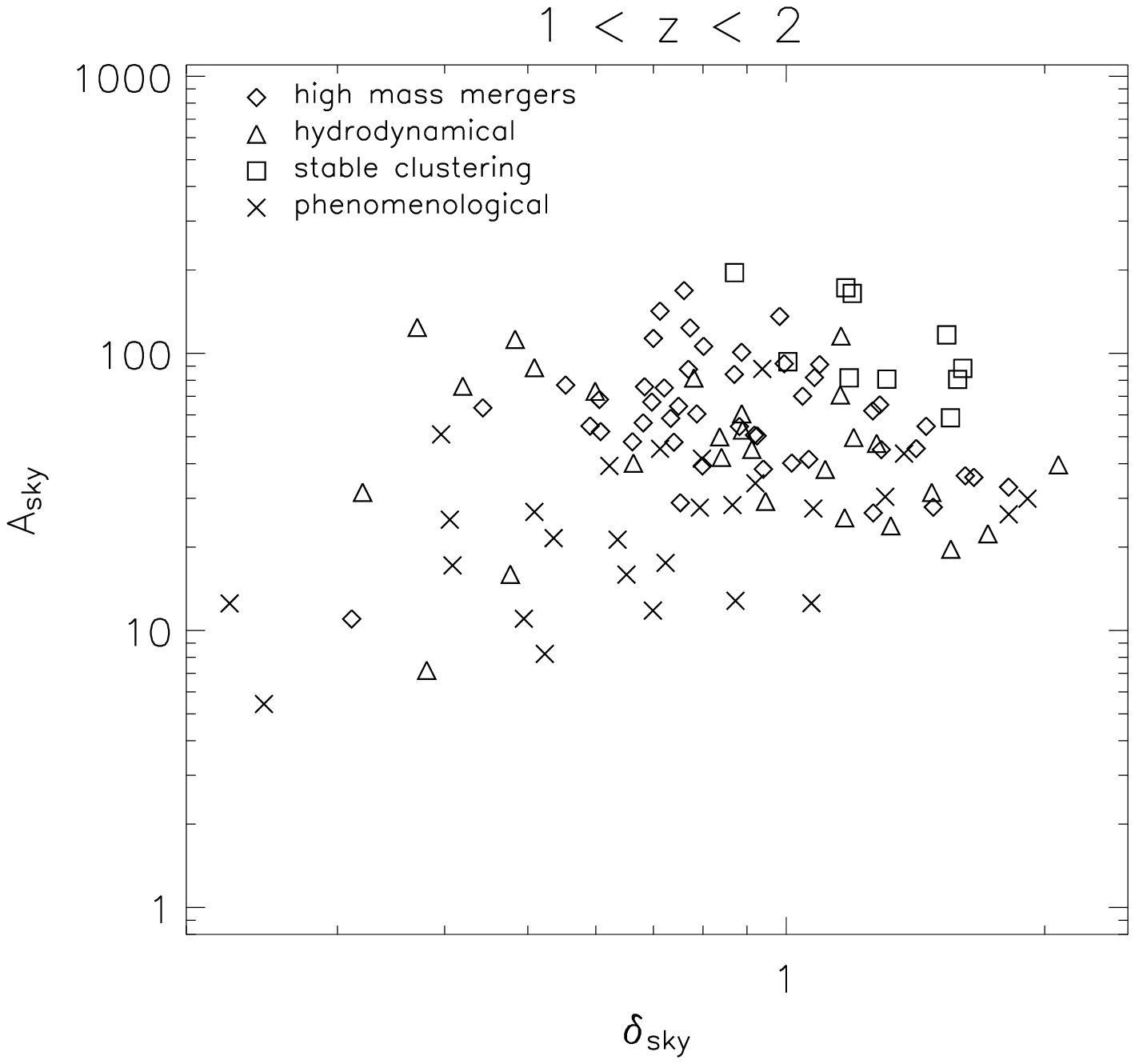,width=8.3cm}} 
{\psfig{file=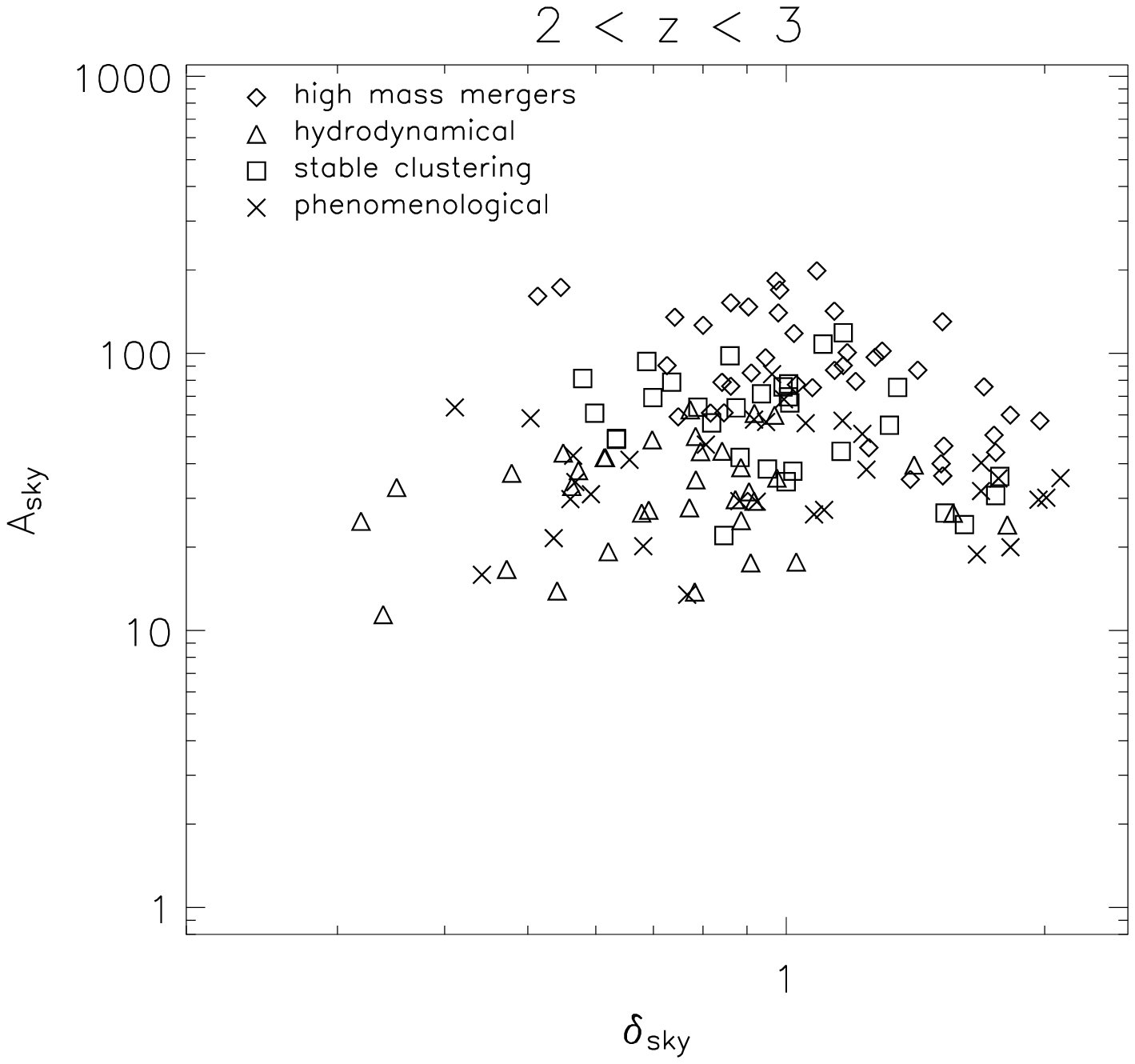,width=8.3cm}} 
{\psfig{file=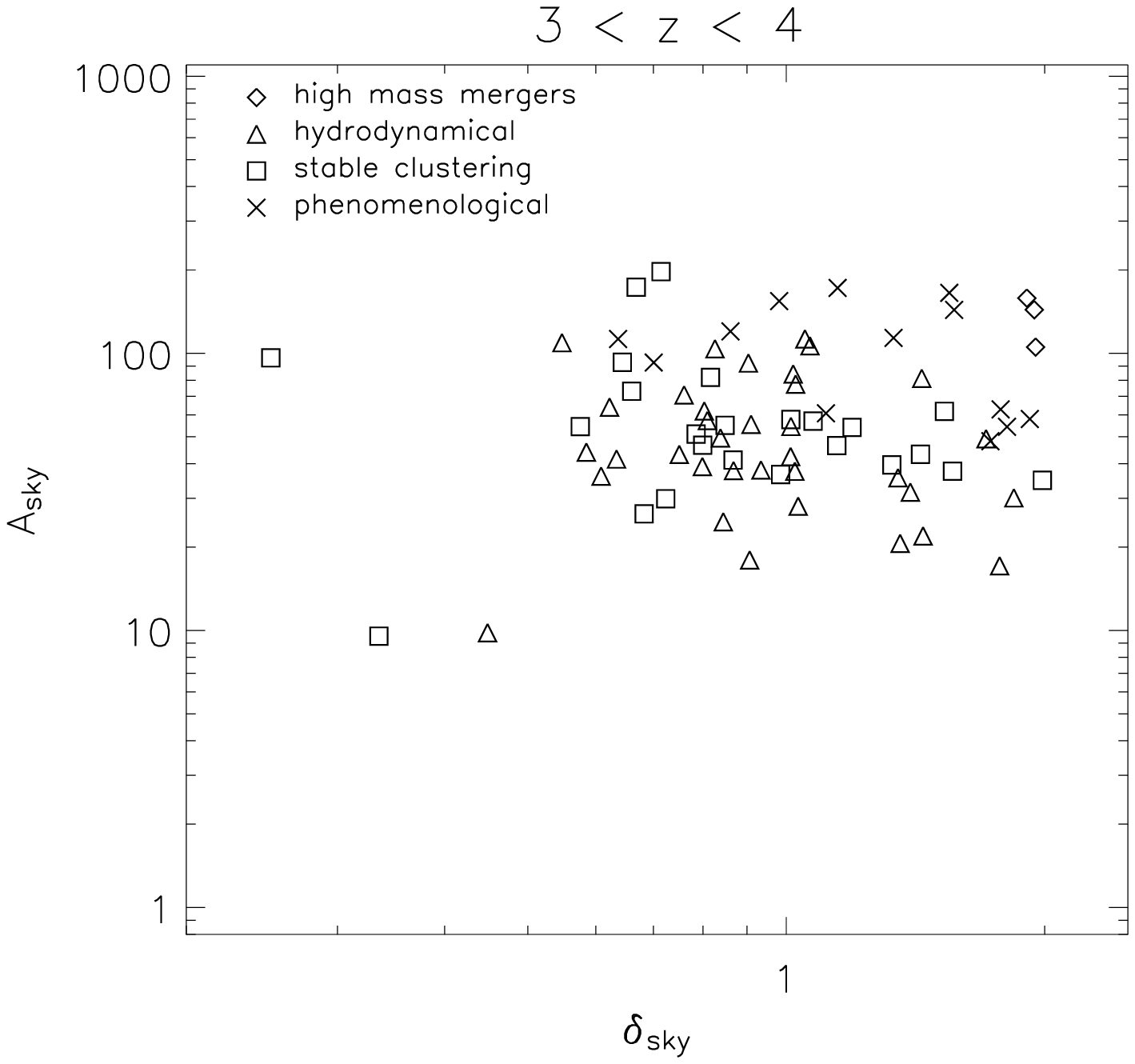,width=8.3cm}} 
\caption{{\bf Figure 6b.} As Fig.\ 6a, but for three different redshift bins.
Again only parameters from `good' fits are shown, i.e. those with
a $\chi^2$ probability larger than 0.1.}
\vskip -0.5cm
\endfigure 

\beginfigure{8}
{\psfig{file=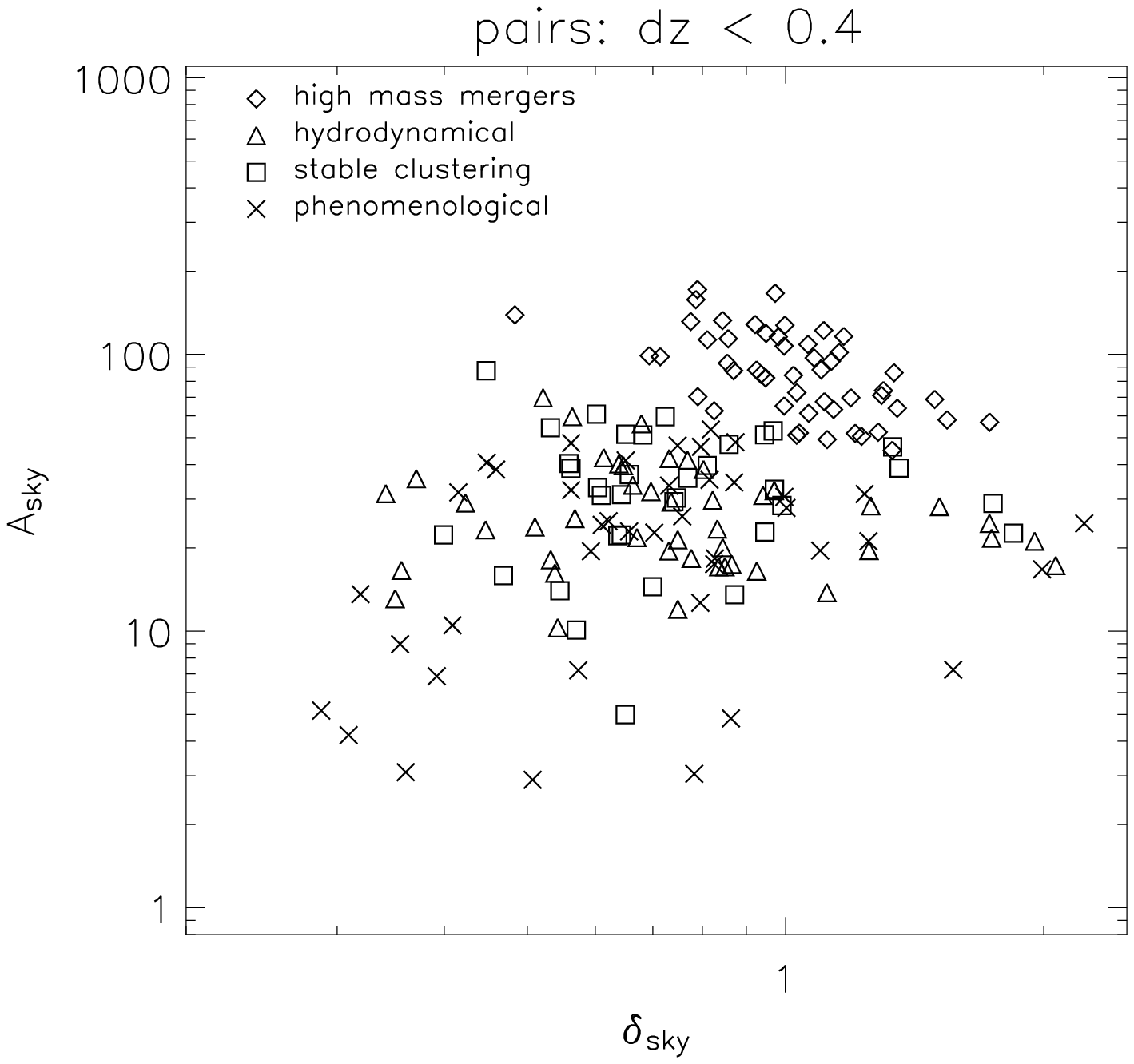,width=8.3cm}} 
\caption{{\bf Figure 6c.} As Fig.\ 6a, but for close redshift pairs with
$\delta z<0.4$.
Again only parameters from `good' fits are shown, i.e. those fits that have
a $\chi^2$ probability larger than 0.1.}
\vskip 0.5cm
\endfigure 

\beginfigure{9}
{\psfig{file=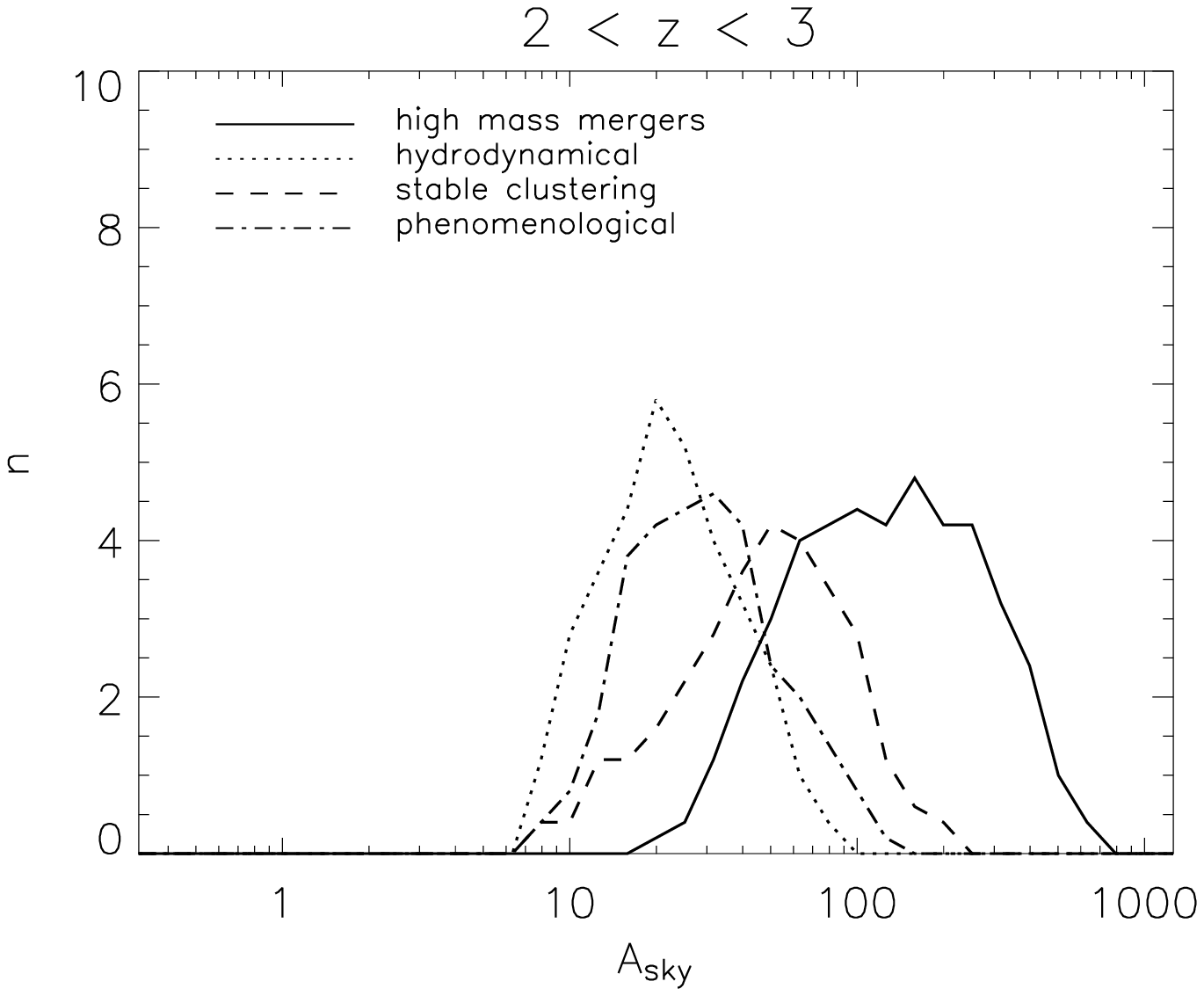,width=9cm}}
{\psfig{file=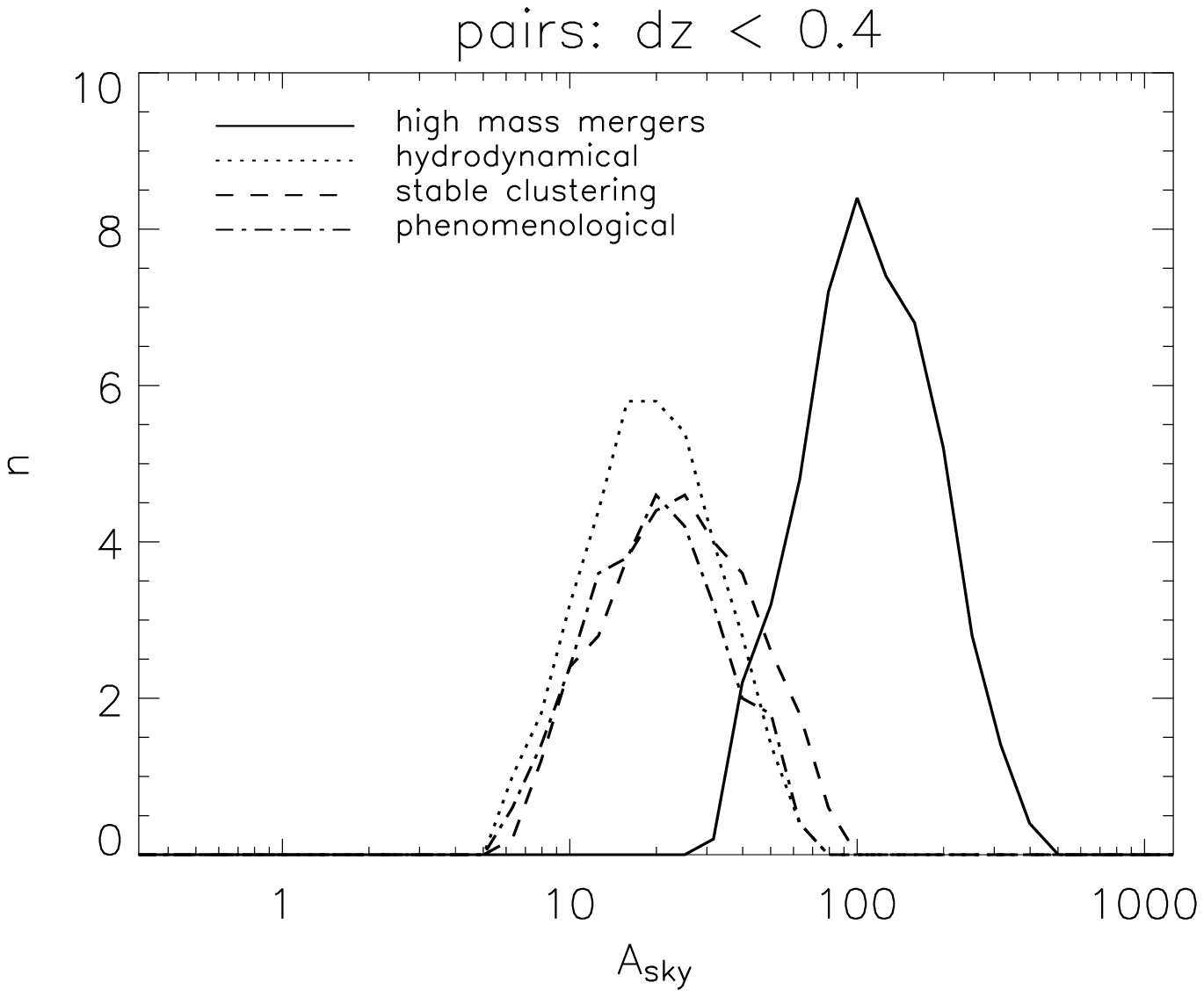,width=9cm}}
\caption{{\bf Figure 7.} Distribution of clustering amplitude $A_{\rm sky}$
over fifty realisation for each model, for the redshift bin $2<z<3$, 
where each model has a sufficient number of sources available (top panel),
and for close redshift pairs with $\delta z<0.4$ (bottom panel).}
\endfigure 

\beginfigure{10}
{\psfig{file=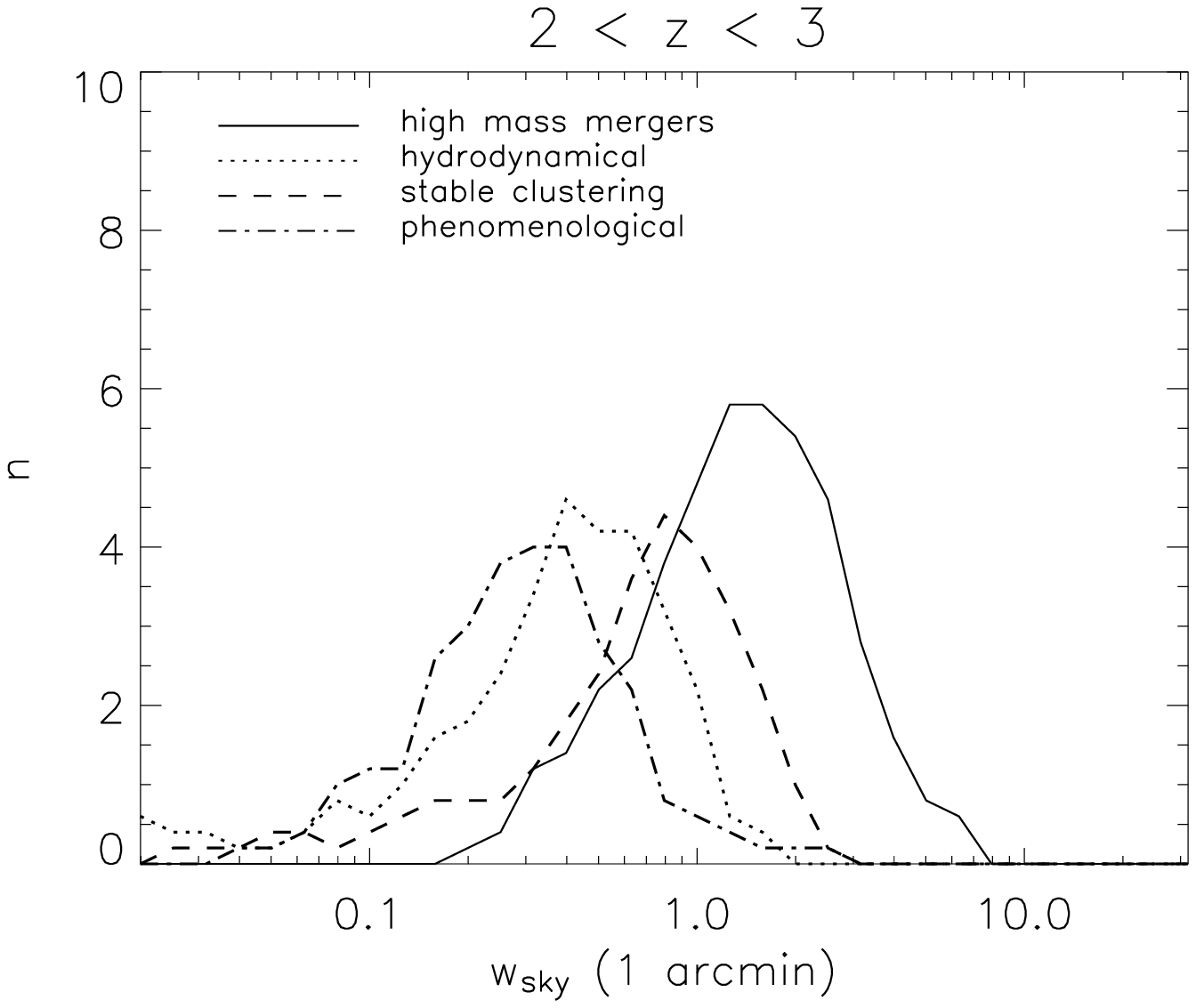,width=9cm}}
{\psfig{file=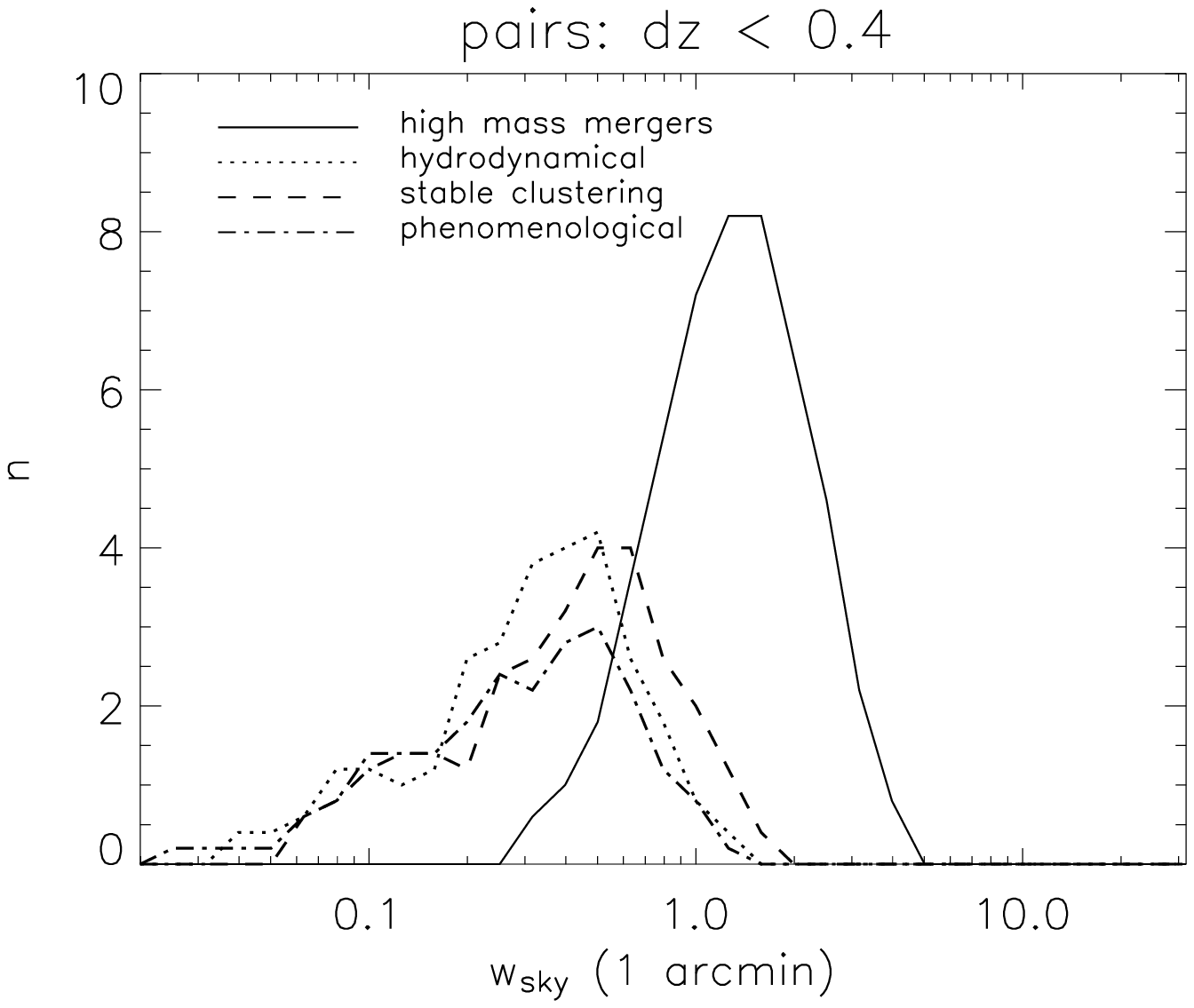,width=9cm}}
\caption{{\bf Figure 8.} Distribution of the angular correlation measure
$\bar w(1')$ over fifty realisation for each model, again for the redshift
bin $2<z<3$ (top panel), and the close redshift pairs with $\delta z<0.4$
(bottom panel).}
\endfigure 

\subsubsection{Examples of mock angular correlation functions}

In Figs.\ 3, 4 and 5 we show examples of $w(\theta)$ and $\bar w(\theta)$
for realisations of all models, for
the case where we have no redshift information (Fig.\ 3) and for the case
where we have (Fig.\ 4 and 5). The {\it same} realisations have been used for
all three figures. Round symbols show the angular correlation function
$w(\theta)$, where open symbols indicate negative values, and error bars
indicate Poisson errors. The best fit power-law is shown as a solid line,
{\it if} a fit was possible. If not, the line is simply omitted.
The sky-averaged angular correlation function $\bar w(\theta)$ is shown 
using stars, and the best-fitting power-law is shown as a dashed
line (again, if a fit was successful).

Before continuing, please note that these examples are by no means
meant to be representative, they are merely shown to demonstrate the
difference between using $w(\theta)$ and $\bar w(\theta)$ for fitting, and
to show the effect of the additional redshift information. The examples
should {\it not} be used to compare the differences in clustering
strength between the models; this will be covered in the next section.

First focusing on Fig.\ 3, we see that the correlation functions
for the complete line-of-sight are noisy, and for two of the
models a fit to $w(\theta)$ fails completely. The figure demonstrates
the use of sky-averaging, as $\bar w(\theta)$, plotted using stars,
is better behaved.
This is perhaps best illustrated in the first panel, but also in the third
panel, where it has the direct consequence that a fit to $\bar w(\theta)$ is
possible where a fit to $w(\theta)$ failed (third panel of Fig. 3).
In general, for most realisations a simple $\chi^2$ fit to $\bar w(\theta)$
turns out to be easier than a fit to the angular correlation function itself. 
This is very helpful for our purpose of comparing models, as it is important
that fitting is possible for a large number of realisations, which needs
to be largely automatic.

A stronger clustering signal is expected for sources selected in a
redshift range, as the signal is less polluted by uncorrelated sources
at very different redshifts. Indeed, Fig.\ 4 shows that for all
realisations the selection of sources in the redshift interval $2<z<3$ boosts
the clustering signal.
Even though the errors are larger due to the small number of sources,
the stronger signal means that a fit is possible in all cases shown, even
for $w(\theta)$ itself, although the sky-averaged correlation function
is to be preferred nevertheless.  However, a fraction of realizations
still produce unacceptable fits.

An alternative to redshift intervals is to only count galaxies that are
paired in redshift space, e.g. have $| z_i - z_j | < 0.4$, as described in 
Section 4.2.1. The result for the same realizations as used for Figs. 3 and 4
is shown in Fig. 5, and again much better results are obtained as compared
to the estimates without any redshift information (Fig. 3). The binned data
looks cleaner than those for the redshift intervals (Fig. 4), which is due to
more galaxies being used in the $DD$ counts. The fitting is therefore somewhat
more reliable, as demonstrated by the small difference between fits to
$w(\theta)$ and $\bar w(\theta)$.

\subsubsection{Distribution over fitting parameters}

So far we have considered single realisations for each model, which should
really be treated as {\it examples} of how the final \shades\ dataset might
appear. In order to be able to make a quantitative comparison possible
between the actual final \shades\ dataset and all models, we need to find
the probability distribution over the fitting parameters for each model
given the survey constraints (area, flux limit, etc.). 

We therefore produced 50 realizations for each model, and fitted a
power-law correlation function to all of these. The resulting
amplitudes $A_{\rm sky}$ and slopes $\delta_{\rm sky}$ of the
sky-averaged correlation function $\bar w(\theta)$ are shown as
scatter plots in Fig.\ 6a, which shows the case where no redshift
information is available, and in Figs.\ 6b and 6c, where we are able
to split up the sample in redshift intervals (three of these are
shown), or select pairs of galaxies in redshift space.

In the case of no redshift information, a significant fraction of the
mock fields do not produce a correlation function that can be fitted
by a power-law, and the number of estimates in Fig.\ 6a is therefore
less than 50 for each model.  However, for each model still more than
half the realization allow a good fit so, crudely speaking, one would
expect that at least one of the two observed \shades\ fields should
produce a good fit.  All models spread out over a relatively large
region of parameter space, and seem to overlap with each other for
most of that region.  This merely reflects the fact that whatever
intrinsic correlation exists in the underlying sub-mm population is
weakened by projection, which produces this large spread in fitting
parameters and the significant overlap between the models. Only the
high-mass merger model shows a larger clustering amplitude overall,
which also results in a larger fraction of the realizations producing
good fits, and a smaller spread in the slope $\delta$.

Let us now use the redshift information to split up the mock samples into
redshift intervals, which should show stronger clustering.
Three intervals are shown in Fig.\ 6b, where
the $2<z<3$ case is the most relevant one as it has the most sources for all
models. The other two intervals both have at least one model where the number
of sources is significantly lacking, which means that only the remaining
models can reasonably be compared. The first thing to notice in all three
panels of Fig.\ 6b is that the clouds of fitted parameter pairs start to
separate out somewhat, reducing the overlap between the models.

Various interesting effects can be seen for the different models.
The stable clustering model (open squares) shows fairly strong
clustering in the $1<z<2$ redshift interval, but because of the low
number of sources in this redshift range (see Fig.\ 2), few of the
realisations actually produce a good fit. The high mass
merger model (open diamonds) shows relatively strong clustering
for $1<z<3$, but of course lack numbers in the highest redshift interval.
The phenomenological model (crosses) shows strongest clustering in that same
$3<z<4$ interval, but this is also somewhat troubled by low source counts.
It also shows the weakest clustering for the lowest redshift interval.
The clustering strength of the hydrodynamical model (open triangles) is
virtually independent of redshift.

In Fig.\ 6c we show the results for the close pairs, i.e. galaxies
with $|\delta z|<0.4$. The points scatter in a similar fashion to the
$2<z<3$ interval, with some differences, and the separating of model point
clouds is comparable between models, except for the simple merger model,
which can quite clearly be distinguished. This diagram should thus be
a good test for high-mass merging versus the other models considered in
this paper.

If we concentrate on the $2<z<3$ interval, which allows the cleanest
comparison between the four models considered here, and on the close pairs,
we do see that
the clouds of points are overlapping significantly, but the distributions
are elongated somewhat along the $\delta$ axis, and have different
mean clustering amplitudes $A$. 
Also, the mean of the distribution is near $\delta=0.8$,
observationally found for a range of galaxy types.

This leads us to finally consider the traditional one-parameter fit to the
data, assuming $\delta=0.8$. This produces a single clustering amplitude
$A_{\rm sky}$ for each mock field, and a distribution over $A_{\rm sky}$
for each model. These distributions
are plotted in Fig.\ 7, for the redshift interval $2<z<3$ (top panel),
and for the close pairs. Interestingly, for this redshift interval, all
distributions are different, and although there is significant overlap, the
final \shades\ dataset will distinguish between the models, especially in
combination with the different
redshift distributions (see Fig.\ 2). For the close pairs, the result of Fig. 6c
is made more apparent, in that the high-mass merger model is clearly different
from the rest of the models, which show almost identical distributions.

Another measure of clustering is the sky-averaged angular correlation
function as it was originally intended: just as a
measurement. Therefore we also plot, in Fig.\ 8, the distribution over
a particular $\bar w(\theta_i)$, which we choose to be $\bar w(1')$,
i.e.  the sky-averaged correlation function within 1 arcmin, for the
same redshift selected data as used for Fig. 7.  The result is a
similar, although the distributions overlap more than those for
$A_{\rm sky}$ (as seen in Fig.\ 7). However, the major advantage with
the measure $\bar w(1')$, or one at a different angle, is that we do
not need to assume a model for the form of the correlation function.

\section{Conclusions}

One of the primary science drivers for the \shades\ project is to place
strong constraints on galaxy formation models by observations of luminous
sub-mm galaxies in the high redshift Universe. In order to achieve this, it
is worth considering how models are best constrained by the data, and
examine the range of possible predictions. With this aim in mind, we have
presented 4 different models of the sub-mm galaxy population, selected to be
widely varying in concept, without worrying about every
aspect of each model. We avoid any strong assumption about the nature or
redshift distribution of the sub-mm population.
Given the uncertainties in what is known about
the sub-mm population, we want to keep open a range of models, even those
that can be questioned in some of their aspects.


For each model, the redshift distribution and clustering properties of
the sub-mm population were predicted for \shades, and
50 realisations were produced, each comprising around 150 sources. Thus,
25 mock \shades\ datasets have been produced for each model.
These simulated \shades\ catalogues were used to investigate the
{\it ability} of the clustering statistics of the final dataset to
constrain the various models collected here. Direct and sky-averaged
estimators of the correlation function have been considered and their
relative merits discussed. We have argued that power-law fits are best
performed on the sky-averaged angular correlation function, and that a
relatively good fit is possible in most cases for this measure. The full
covariance matrix was calculated and diagonalized resulting in a
$\chi^2$ statistic that was used to fit the power-law model to the data
using the Levenberg-Marquardt method as implemented by Press et al. (1988).

All models predict sufficiently strong clustering, so that we expect
to detect clustering within the \scuba\ population when \shades\ is
complete. Although cosmic variance remains a concern, we can certainly
quantify the probability of a given model to produce the observed
dataset, and this is expected to reject some of the models included in
this paper. However, the aim of this paper is not to constrain the
models; this will be done when the full \shades\ dataset is available.
In fact, we simply have observed basic trends between models by
reducing each measured correlation function to two power-law parameters:
its slope and amplitude respectively.

The observed redshift distribution will provide an complementary strong
test of models, even with relatively coarse photometric redshift
information. In fact, the combination of clustering and redshift data
offers the best discriminator between the different models that we have
considered: models with similar redshift distributions have different
clustering strengths, while models with similar clustering properties
have different redshift distributions.

Recently, Blain et al. (2004) considered the clustering of sub-mm sources
in a number of relatively small fields for data with follow-up
spectroscopic redshifts. An approach was adopted that selected galaxy
pairs based only on radial positions, and did not use any angular
information. Pairs of sub-mm sources with $\Delta v$ = 1200 km s$^{-1}$,
equivalent to separations of order 5 Mpc (comoving) at $z=2.5$, were
counted, and compared with the integrated model $\xi(r)$.  For a larger
survey where there is significant angular information, this method is not
optimal, so for \shades\ we prefer to fully exploit the angular
information, with the restricted photometric redshift information as a
subsample selection tool.

The Blain et al. (2004) method is equivalent to the standard method of
calculating $\xi(r)$ by pair counting (using DD/RR-1), but performing this
in single bin with $r<R_{\rm max}$. While angular clustering measurements
ignore radial information, this "radial" method instead ignores angular
information in a similar way. The method should therefore include the
equivalent of an integral constraint. Also, the quoted errors are derived
from Poisson errors (on the pair counts), whereas the errors are expected
to be larger than Poissonian for this estimator (Landy \& Szalay 1993).
Further limitations of the method employed by Blain et al. (2004) are
discussed by Adelberger (2004). Given these issues, it is probably too
early to use the Blain et al. clustering results to distinguish between
models.

The primary conclusion from our analysis is that, with the area
coverage (0.5 square degrees) and the expected number of sources (200-400),
and particularly with the expected photometric redshift information
($\Delta z \sim \pm 0.3$), \shades\ is capable of
distinguishing between widely varying scenarios for the
production of the bright sub-mm population.

\section*{ACKNOWLEDGEMENTS} 

This research was supported in part by the Austrian Science Foundation FWF 
under grant P15868, by two PPARC rolling grants, and by the HPC Europa program
funded through the European Commission contract number RII3-CT-2003-506079.
We thank Duncan Farrah and Tom Babbedge for many useful comments and suggestions.
DHH and IA are partly supported by CONACyT grants 39953-F and 39548-F.

%
%

\end

%% file: mn.tex
%
%
%
%

\catcode `\@=11 

\def\@version{1.6}
\def\@verdate{18th September 1995}

%
%


\newif\ifprod@font

\ifx\@typeface\undefined
  \def\@typeface{Comp. Modern}\prod@fontfalse
\else
  \prod@fonttrue 
\fi

\def\newfam{\alloc@8\fam\chardef\sixt@@n} 

\ifprod@font
\font\fiverm=mtr10 at 5pt
\font\fivebf=mtbx10 at 5pt
\font\fiveit=mtti10 at 5pt
\font\fivesl=mtsl10 at 5pt
\font\fivett=cmtt8 at 5pt     \hyphenchar\fivett=-1
\font\fivecsc=mtcsc10 at 5pt
\font\fivesf=mtss10 at 5pt
\font\fivei=mtmi10 at 5pt      \skewchar\fivei='177
\font\fivesy=mtsy10 at 5pt     \skewchar\fivesy='60

\font\sixrm=mtr10 at 6pt
\font\sixbf=mtbx10 at 6pt
\font\sixit=mtti10 at 6pt
\font\sixsl=mtsl10 at 6pt
\font\sixtt=cmtt8 at 6pt      \hyphenchar\sixtt=-1
\font\sixcsc=mtcsc10 at 6pt
\font\sixsf=mtss10 at 6pt
\font\sixi=mtmi10 at 6pt       \skewchar\sixi='177
\font\sixsy=mtsy10 at 6pt      \skewchar\sixsy='60

\font\sevenrm=mtr10 at 7pt
\font\sevenbf=mtbx10 at 7pt
\font\sevenit=mtti10 at 7pt
\font\sevensl=mtsl10 at 7pt
\font\seventt=cmtt8 at 7pt     \hyphenchar\seventt=-1
\font\sevencsc=mtcsc10 at 7pt
\font\sevensf=mtss10 at 7pt
\font\seveni=mtmi10 at 7pt      \skewchar\seveni='177
\font\sevensy=mtsy10 at 7pt     \skewchar\sevensy='60

\font\eightrm=mtr10 at 8pt
\font\eightbf=mtbx10 at 8pt
\font\eightit=mtti10 at 8pt
\font\eighti=mtmi10 at 8pt      \skewchar\eighti='177
\font\eightsy=mtsy10 at 8pt     \skewchar\eightsy='60
\font\eightsl=mtsl10 at 8pt
\font\eighttt=cmtt8             \hyphenchar\eighttt=-1
\font\eightcsc=mtcsc10 at 8pt
\font\eightsf=mtss10 at 8pt

\font\ninerm=mtr10 at 9pt
\font\ninebf=mtbx10 at 9pt
\font\nineit=mtti10 at 9pt
\font\ninei=mtmi10 at 9pt      \skewchar\ninei='177
\font\ninesy=mtsy10 at 9pt     \skewchar\ninesy='60
\font\ninesl=mtsl10 at 9pt
\font\ninett=cmtt9             \hyphenchar\ninett=-1
\font\ninecsc=mtcsc10 at 9pt
\font\ninesf=mtss10 at 9pt

\font\tenrm=mtr10
\font\tenbf=mtbx10
\font\tenit=mtti10
\font\teni=mtmi10		\skewchar\teni='177
\font\tensy=mtsy10		\skewchar\tensy='60
\font\tenex=cmex10
\font\tensl=mtsl10
\font\tentt=cmtt10		\hyphenchar\tentt=-1
\font\tencsc=mtcsc10
\font\tensf=mtss10

\font\elevenrm=mtr10 at 11pt
\font\elevenbf=mtbx10 at 11pt
\font\elevenit=mtti10 at 11pt
\font\eleveni=mtmi10 at 11pt      \skewchar\eleveni='177
\font\elevensy=mtsy10 at 11pt     \skewchar\elevensy='60
\font\elevensl=mtsl10 at 11pt
\font\eleventt=cmtt10 at 11pt     \hyphenchar\eleventt=-1
\font\elevencsc=mtcsc10 at 11pt
\font\elevensf=mtss10 at 11pt

\font\twelverm=mtr10 at 12pt
\font\twelvebf=mtbx10 at 12pt
\font\twelveit=mtti10 at 12pt
\font\twelvesl=mtsl10 at 12pt
\font\twelvett=cmtt12             \hyphenchar\twelvett=-1
\font\twelvecsc=mtcsc10 at 12pt
\font\twelvesf=mtss10 at 12pt
\font\twelvei=mtmi10 at 12pt      \skewchar\twelvei='177
\font\twelvesy=mtsy10 at 12pt     \skewchar\twelvesy='60

\font\fourteenrm=mtr10 at 14pt
\font\fourteenbf=mtbx10 at 14pt
\font\fourteenit=mtti10 at 14pt
\font\fourteeni=mtmi10 at 14pt      \skewchar\fourteeni='177
\font\fourteensy=mtsy10 at 14pt     \skewchar\fourteensy='60
\font\fourteensl=mtsl10 at 14pt
\font\fourteentt=cmtt12 at 14pt     \hyphenchar\fourteentt=-1
\font\fourteencsc=mtcsc10 at 14pt
\font\fourteensf=mtss10 at 14pt

\font\seventeenrm=mtr10 at 17pt
\font\seventeenbf=mtbx10 at 17pt
\font\seventeenit=mtti10 at 17pt
\font\seventeeni=mtmi10 at 17pt      \skewchar\seventeeni='177
\font\seventeensy=mtsy10 at 17pt     \skewchar\seventeensy='60
\font\seventeensl=mtsl10 at 17pt
\font\seventeentt=cmtt12 at 17pt     \hyphenchar\seventeentt=-1
\font\seventeencsc=mtcsc10 at 17pt
\font\seventeensf=mtss10 at 17pt
\else
\font\fiverm=cmr5
\font\fivei=cmmi5             \skewchar\fivei='177
\font\fivesy=cmsy5            \skewchar\fivesy='60
\font\fivebf=cmbx5

\font\sixrm=cmr6
\font\sixi=cmmi6             \skewchar\sixi='177
\font\sixsy=cmsy6            \skewchar\sixsy='60
\font\sixbf=cmbx6

\font\sevenrm=cmr7
\font\sevenit=cmti7
\font\seveni=cmmi7             \skewchar\seveni='177
\font\sevensy=cmsy7            \skewchar\sevensy='60
\font\sevenbf=cmbx7

\font\eightrm=cmr8
\font\eightbf=cmbx8
\font\eightit=cmti8
\font\eighti=cmmi8			\skewchar\eighti='177
\font\eightsy=cmsy8			\skewchar\eightsy='60
\font\eightsl=cmsl8
\font\eighttt=cmtt8			\hyphenchar\eighttt=-1
\font\eightcsc=cmcsc10 at 8pt
\font\eightsf=cmss8

\font\ninerm=cmr9
\font\ninebf=cmbx9
\font\nineit=cmti9
\font\ninei=cmmi9			\skewchar\ninei='177
\font\ninesy=cmsy9			\skewchar\ninesy='60
\font\ninesl=cmsl9
\font\ninett=cmtt9			\hyphenchar\ninett=-1
\font\ninecsc=cmcsc10 at 9pt
\font\ninesf=cmss9

\font\tenrm=cmr10
\font\tenbf=cmbx10
\font\tenit=cmti10
\font\teni=cmmi10		\skewchar\teni='177
\font\tensy=cmsy10		\skewchar\tensy='60
\font\tenex=cmex10
\font\tensl=cmsl10
\font\tentt=cmtt10		\hyphenchar\tentt=-1
\font\tencsc=cmcsc10
\font\tensf=cmss10

\font\elevenrm=cmr10 scaled \magstephalf
\font\elevenbf=cmbx10 scaled \magstephalf
\font\elevenit=cmti10 scaled \magstephalf
\font\eleveni=cmmi10 scaled \magstephalf	\skewchar\eleveni='177
\font\elevensy=cmsy10 scaled \magstephalf	\skewchar\elevensy='60
\font\elevensl=cmsl10 scaled \magstephalf
\font\eleventt=cmtt10 scaled \magstephalf	\hyphenchar\eleventt=-1
\font\elevencsc=cmcsc10 scaled \magstephalf
\font\elevensf=cmss10 scaled \magstephalf

\font\twelverm=cmr10 scaled \magstep1
\font\twelvebf=cmbx10 scaled \magstep1
\font\twelvei=cmmi10 scaled \magstep1      \skewchar\twelvei='177
\font\twelvesy=cmsy10 scaled \magstep1     \skewchar\twelvesy='60

\font\fourteenrm=cmr10 scaled \magstep2
\font\fourteenbf=cmbx10 scaled \magstep2
\font\fourteenit=cmti10 scaled \magstep2
\font\fourteeni=cmmi10 scaled \magstep2		\skewchar\fourteeni='177
\font\fourteensy=cmsy10 scaled \magstep2	\skewchar\fourteensy='60
\font\fourteensl=cmsl10 scaled \magstep2
\font\fourteentt=cmtt10 scaled \magstep2	\hyphenchar\fourteentt=-1
\font\fourteencsc=cmcsc10 scaled \magstep2
\font\fourteensf=cmss10 scaled \magstep2

\font\seventeenrm=cmr10 scaled \magstep3
\font\seventeenbf=cmbx10 scaled \magstep3
\font\seventeenit=cmti10 scaled \magstep3
\font\seventeeni=cmmi10 scaled \magstep3	\skewchar\seventeeni='177
\font\seventeensy=cmsy10 scaled \magstep3	\skewchar\seventeensy='60
\font\seventeensl=cmsl10 scaled \magstep3
\font\seventeentt=cmtt10 scaled \magstep3	\hyphenchar\seventeentt=-1
\font\seventeencsc=cmcsc10 scaled \magstep3
\font\seventeensf=cmss10 scaled \magstep3
\fi

\def\hexnumber#1{\ifcase#1 0\or1\or2\or3\or4\or5\or6\or7\or8\or9\or
  A\or B\or C\or D\or E\or F\fi}

\def\makestrut{%
  \setbox\strutbox=\hbox{%
    \vrule height.7\baselineskip depth.3\baselineskip width \z@}%
}

\def\baselinestretch{1}
\newskip\tmp@bls

\def\b@ls#1{
  \tmp@bls=#1\relax
  \baselineskip=#1\relax\makestrut
  \normalbaselineskip=\baselinestretch\tmp@bls
  \normalbaselines
}

\def\nostb@ls#1{
  \normalbaselineskip=#1\relax
  \normalbaselines
  \makestrut
}

%

\newfam\scfam  
\newfam\sffam  

\def\mit{\fam\@ne}
\def\cal{\fam\tw@}
\def\em{\ifdim\fontdimen1\font>\z@ \rm\else\it\fi}

\textfont3=\tenex
\scriptfont3=\tenex
\scriptscriptfont3=\tenex

\setbox0=\hbox{\tenex B} \p@renwd=\wd0 

\def\eightpoint{
  \def\rm{\fam0\eightrm}%
  \textfont0=\eightrm \scriptfont0=\sixrm \scriptscriptfont0=\fiverm%
  \textfont1=\eighti  \scriptfont1=\sixi  \scriptscriptfont1=\fivei%
  \textfont2=\eightsy \scriptfont2=\sixsy \scriptscriptfont2=\fivesy%
  \textfont\itfam=\eightit\def\it{\fam\itfam\eightit}%
  \ifprod@font
    \scriptfont\itfam=\sixit
      \scriptscriptfont\itfam=\fiveit
  \else
    \scriptfont\itfam=\eightit
      \scriptscriptfont\itfam=\eightit
  \fi
  \textfont\bffam=\eightbf%
    \scriptfont\bffam=\sixbf%
      \scriptscriptfont\bffam=\fivebf%
  \def\bf{\fam\bffam\eightbf}%
  \textfont\slfam=\eightsl\def\sl{\fam\slfam\eightsl}%
  \ifprod@font
    \scriptfont\slfam=\sixsl
      \scriptscriptfont\slfam=\fivesl
  \else
    \scriptfont\slfam=\eightsl
      \scriptscriptfont\slfam=\eightsl
  \fi
  \textfont\ttfam=\eighttt\def\tt{\fam\ttfam\eighttt}%
  \ifprod@font
    \scriptfont\ttfam=\sixtt
      \scriptscriptfont\ttfam=\fivett
  \else
    \scriptfont\ttfam=\eighttt
      \scriptscriptfont\ttfam=\eighttt
  \fi
  \textfont\scfam=\eightcsc\def\sc{\fam\scfam\eightcsc}%
  \ifprod@font
    \scriptfont\scfam=\sixcsc
      \scriptscriptfont\scfam=\fivecsc
  \else
    \scriptfont\scfam=\eightcsc
      \scriptscriptfont\scfam=\eightcsc
  \fi
  \textfont\sffam=\eightsf\def\sf{\fam\sffam\eightsf}%
  \ifprod@font
    \scriptfont\sffam=\sixsf
      \scriptscriptfont\sffam=\fivesf
  \else
    \scriptfont\sffam=\eightsf
      \scriptscriptfont\sffam=\eightsf
  \fi
  \def\oldstyle{\fam\@ne\eighti}%
  \b@ls{10pt}\rm\@viiipt%
}
\def\@viiipt{}

\def\ninepoint{
  \def\rm{\fam0\ninerm}%
  \textfont0=\ninerm \scriptfont0=\sixrm \scriptscriptfont0=\fiverm%
  \textfont1=\ninei  \scriptfont1=\sixi  \scriptscriptfont1=\fivei%
  \textfont2=\ninesy \scriptfont2=\sixsy \scriptscriptfont2=\fivesy%
  \textfont\itfam=\nineit\def\it{\fam\itfam\nineit}%
  \ifprod@font
    \scriptfont\itfam=\sixit
      \scriptscriptfont\itfam=\fiveit
  \else
    \scriptfont\itfam=\nineit
      \scriptscriptfont\itfam=\nineit
  \fi
  \textfont\bffam=\ninebf%
    \scriptfont\bffam=\sixbf%
      \scriptscriptfont\bffam=\fivebf%
  \def\bf{\fam\bffam\ninebf}%
  \textfont\slfam=\ninesl\def\sl{\fam\slfam\ninesl}%
  \ifprod@font
    \scriptfont\slfam=\sixsl
      \scriptscriptfont\slfam=\fivesl
  \else
    \scriptfont\slfam=\ninesl
      \scriptscriptfont\slfam=\ninesl
  \fi
  \textfont\ttfam=\ninett\def\tt{\fam\ttfam\ninett}%
  \ifprod@font
    \scriptfont\ttfam=\sixtt
      \scriptscriptfont\ttfam=\fivett
  \else
    \scriptfont\ttfam=\ninett
      \scriptscriptfont\ttfam=\ninett
  \fi
  \textfont\scfam=\ninecsc\def\sc{\fam\scfam\ninecsc}%
  \ifprod@font
    \scriptfont\scfam=\sixcsc
      \scriptscriptfont\scfam=\fivecsc
  \else
    \scriptfont\scfam=\ninecsc
      \scriptscriptfont\scfam=\ninecsc
  \fi
  \textfont\sffam=\ninesf\def\sf{\fam\sffam\ninesf}%
  \ifprod@font
    \scriptfont\sffam=\sixsf
      \scriptscriptfont\sffam=\fivesf
  \else
    \scriptfont\sffam=\ninesf
      \scriptscriptfont\sffam=\ninesf
  \fi
  \def\oldstyle{\fam\@ne\ninei}%
  \b@ls{\TextLeading plus \Feathering}\rm\@ixpt%
}
\def\@ixpt{}

\def\tenpoint{
  \def\rm{\fam0\tenrm}%
  \textfont0=\tenrm \scriptfont0=\sevenrm \scriptscriptfont0=\fiverm%
  \textfont1=\teni  \scriptfont1=\seveni  \scriptscriptfont1=\fivei%
  \textfont2=\tensy \scriptfont2=\sevensy \scriptscriptfont2=\fivesy%
  \textfont\itfam=\tenit\def\it{\fam\itfam\tenit}%
  \ifprod@font
    \scriptfont\itfam=\sevenit
      \scriptscriptfont\itfam=\fiveit
  \else
    \scriptfont\itfam=\tenit
      \scriptscriptfont\itfam=\tenit
  \fi
  \textfont\bffam=\tenbf%
    \scriptfont\bffam=\sevenbf%
      \scriptscriptfont\bffam=\fivebf%
  \def\bf{\fam\bffam\tenbf}%
  \textfont\slfam=\tensl\def\sl{\fam\slfam\tensl}%
  \ifprod@font
    \scriptfont\slfam=\sevensl
      \scriptscriptfont\slfam=\fivesl
  \else
    \scriptfont\slfam=\tensl
      \scriptscriptfont\slfam=\tensl
  \fi
  \textfont\ttfam=\tentt\def\tt{\fam\ttfam\tentt}%
  \ifprod@font
    \scriptfont\ttfam=\seventt
      \scriptscriptfont\ttfam=\fivett
  \else
    \scriptfont\ttfam=\tentt
      \scriptscriptfont\ttfam=\tentt
  \fi
  \textfont\scfam=\tencsc\def\sc{\fam\scfam\tencsc}%
  \ifprod@font
    \scriptfont\scfam=\sevencsc
      \scriptscriptfont\scfam=\fivecsc
  \else
    \scriptfont\scfam=\tencsc
      \scriptscriptfont\scfam=\tencsc
  \fi
  \textfont\sffam=\tensf\def\sf{\fam\sffam\tensf}%
  \ifprod@font
    \scriptfont\sffam=\sevensf
      \scriptscriptfont\sffam=\fivesf
  \else
    \scriptfont\sffam=\tensf
      \scriptscriptfont\sffam=\tensf
  \fi
  \def\oldstyle{\fam\@ne\teni}%
  \b@ls{11pt}\rm\@xpt%
}
\def\@xpt{}

\def\elevenpoint{
  \def\rm{\fam0\elevenrm}%
  \textfont0=\elevenrm \scriptfont0=\eightrm \scriptscriptfont0=\sixrm%
  \textfont1=\eleveni  \scriptfont1=\eighti  \scriptscriptfont1=\sixi%
  \textfont2=\elevensy \scriptfont2=\eightsy \scriptscriptfont2=\sixsy%
  \textfont\itfam=\elevenit\def\it{\fam\itfam\elevenit}%
  \ifprod@font
    \scriptfont\itfam=\eightit
      \scriptscriptfont\itfam=\sixit
  \else
    \scriptfont\itfam=\elevenit
      \scriptscriptfont\itfam=\elevenit
  \fi
  \textfont\bffam=\elevenbf%
    \scriptfont\bffam=\eightbf%
      \scriptscriptfont\bffam=\sixbf%
  \def\bf{\fam\bffam\elevenbf}%
  \textfont\slfam=\elevensl\def\sl{\fam\slfam\elevensl}%
  \ifprod@font
    \scriptfont\slfam=\eightsl
      \scriptscriptfont\slfam=\sixsl
  \else
    \scriptfont\slfam=\elevensl
      \scriptscriptfont\slfam=\elevensl
  \fi
  \textfont\ttfam=\eleventt\def\tt{\fam\ttfam\eleventt}%
  \ifprod@font
    \scriptfont\ttfam=\eighttt
      \scriptscriptfont\ttfam=\sixtt
  \else
    \scriptfont\ttfam=\eleventt
      \scriptscriptfont\ttfam=\eleventt
  \fi
  \textfont\scfam=\elevencsc\def\sc{\fam\scfam\elevencsc}%
  \ifprod@font
    \scriptfont\scfam=\eightcsc
      \scriptscriptfont\scfam=\sixcsc
  \else
    \scriptfont\scfam=\elevencsc
      \scriptscriptfont\scfam=\elevencsc
  \fi
  \textfont\sffam=\elevensf\def\sf{\fam\sffam\elevensf}%
  \ifprod@font
    \scriptfont\sffam=\eightsf
      \scriptscriptfont\sffam=\sixsf
  \else
    \scriptfont\sffam=\elevensf
      \scriptscriptfont\sffam=\elevensf
  \fi
  \def\oldstyle{\fam\@ne\eleveni}%
  \b@ls{13pt}\rm\@xipt%
}
\def\@xipt{}

\def\fourteenpoint{
  \def\rm{\fam0\fourteenrm}%
  \textfont0\fourteenrm  \scriptfont0\tenrm  \scriptscriptfont0\sevenrm%
  \textfont1\fourteeni   \scriptfont1\teni   \scriptscriptfont1\seveni%
  \textfont2\fourteensy  \scriptfont2\tensy  \scriptscriptfont2\sevensy%
  \textfont\itfam=\fourteenit\def\it{\fam\itfam\fourteenit}%
  \ifprod@font
    \scriptfont\itfam=\tenit
      \scriptscriptfont\itfam=\sevenit
  \else
    \scriptfont\itfam=\fourteenit
      \scriptscriptfont\itfam=\fourteenit
  \fi
  \textfont\bffam=\fourteenbf%
    \scriptfont\bffam=\tenbf%
      \scriptscriptfont\bffam=\sevenbf%
  \def\bf{\fam\bffam\fourteenbf}%
  \textfont\slfam=\fourteensl\def\sl{\fam\slfam\fourteensl}%
  \ifprod@font
    \scriptfont\slfam=\tensl
      \scriptscriptfont\slfam=\sevensl
  \else
    \scriptfont\slfam=\fourteensl
      \scriptscriptfont\slfam=\fourteensl
  \fi
  \textfont\ttfam=\fourteentt\def\tt{\fam\ttfam\fourteentt}%
  \ifprod@font
    \scriptfont\ttfam=\tentt
      \scriptscriptfont\ttfam=\seventt
  \else
    \scriptfont\ttfam=\fourteentt
      \scriptscriptfont\ttfam=\fourteentt
  \fi
  \textfont\scfam=\fourteencsc\def\sc{\fam\scfam\fourteencsc}%
  \ifprod@font
    \scriptfont\scfam=\tencsc
      \scriptscriptfont\scfam=\sevencsc
  \else
    \scriptfont\scfam=\fourteencsc
      \scriptscriptfont\scfam=\fourteencsc
  \fi
  \textfont\sffam=\fourteensf\def\sf{\fam\sffam\fourteensf}%
  \ifprod@font
    \scriptfont\sffam=\tensf
      \scriptscriptfont\sffam=\sevensf
  \else
    \scriptfont\sffam=\fourteensf
      \scriptscriptfont\sffam=\fourteensf
  \fi
  \def\oldstyle{\fam\@ne\fourteeni}%
  \b@ls{17pt}\rm\@xivpt%
}
\def\@xivpt{}

\def\seventeenpoint{
  \def\rm{\fam0\seventeenrm}%
  \textfont0\seventeenrm  \scriptfont0\twelverm  \scriptscriptfont0\tenrm%
  \textfont1\seventeeni   \scriptfont1\twelvei   \scriptscriptfont1\teni%
  \textfont2\seventeensy  \scriptfont2\twelvesy  \scriptscriptfont2\tensy%
  \textfont\itfam=\seventeenit\def\it{\fam\itfam\seventeenit}%
  \ifprod@font
    \scriptfont\itfam=\twelveit
      \scriptscriptfont\itfam=\tenit
  \else
    \scriptfont\itfam=\seventeenit
      \scriptscriptfont\itfam=\seventeenit
  \fi
  \textfont\bffam=\seventeenbf%
    \scriptfont\bffam=\twelvebf%
      \scriptscriptfont\bffam=\tenbf%
  \def\bf{\fam\bffam\seventeenbf}%
  \textfont\slfam=\seventeensl\def\sl{\fam\slfam\seventeensl}%
  \ifprod@font
    \scriptfont\slfam=\twelvesl
      \scriptscriptfont\slfam=\tensl
  \else
    \scriptfont\slfam=\seventeensl
      \scriptscriptfont\slfam=\seventeensl
  \fi
  \textfont\ttfam=\seventeentt\def\tt{\fam\ttfam\seventeentt}%
  \ifprod@font
    \scriptfont\ttfam=\twelvett
      \scriptscriptfont\ttfam=\tentt
  \else
    \scriptfont\ttfam=\seventeentt
      \scriptscriptfont\ttfam=\seventeentt
  \fi
  \textfont\scfam=\seventeencsc\def\sc{\fam\scfam\seventeencsc}%
  \ifprod@font
    \scriptfont\scfam=\twelvecsc
      \scriptscriptfont\scfam=\tencsc
  \else
    \scriptfont\scfam=\seventeencsc
      \scriptscriptfont\scfam=\seventeencsc
  \fi
  \textfont\sffam=\seventeensf\def\sf{\fam\sffam\seventeensf}%
  \ifprod@font
    \scriptfont\sffam=\twelvesf
      \scriptscriptfont\sffam=\tensf
  \else
    \scriptfont\sffam=\seventeensf
      \scriptscriptfont\sffam=\seventeensf
  \fi
  \def\oldstyle{\fam\@ne\seventeeni}%
  \b@ls{20pt}\rm\@xviipt%
}
\def\@xviipt{}

\lineskip=1pt      \normallineskip=\lineskip
\lineskiplimit=\z@ \normallineskiplimit=\lineskiplimit


\def\loadboldmathnames{%
  \def\balpha{{\bmath{\alpha}}}%
  \def\bbeta{{\bmath{\beta}}}%
  \def\bgamma{{\bmath{\gamma}}}%
  \def\bdelta{{\bmath{\delta}}}%
  \def\bepsilon{{\bmath{\epsilon}}}%
  \def\bzeta{{\bmath{\zeta}}}%
  \def\boldeta{{\bmath{\eta}}}%
  \def\btheta{{\bmath{\theta}}}%
  \def\biota{{\bmath{\iota}}}%
  \def\bkappa{{\bmath{\kappa}}}%
  \def\blambda{{\bmath{\lambda}}}%
  \def\bmu{{\bmath{\mu}}}%
  \def\bnu{{\bmath{\nu}}}%
  \def\bxi{{\bmath{\xi}}}%
  \def\bpi{{\bmath{\pi}}}%
  \def\brho{{\bmath{\rho}}}%
  \def\bsigma{{\bmath{\sigma}}}%
  \def\btau{{\bmath{\tau}}}%
  \def\bupsilon{{\bmath{\upsilon}}}%
  \def\bphi{{\bmath{\phi}}}%
  \def\bchi{{\bmath{\chi}}}%
  \def\bpsi{{\bmath{\psi}}}%
  \def\bomega{{\bmath{\omega}}}%
  \def\bvarepsilon{{\bmath{\varepsilon}}}%
  \def\bvartheta{{\bmath{\vartheta}}}%
  \def\bvarpi{{\bmath{\varpi}}}%
  \def\bvarrho{{\bmath{\varrho}}}%
  \def\bvarsigma{{\bmath{\varsigma}}}%
  \def\bvarphi{{\bmath{\varphi}}}%
  \def\baleph{{\bmath{\aleph}}}%
  \def\bimath{{\bmath{\imath}}}%
  \def\bjmath{{\bmath{\jmath}}}%
  \def\bell{{\bmath{\ell}}}%
  \def\bwp{{\bmath{\wp}}}%
  \def\bRe{{\bmath{\Re}}}%
  \def\bIm{{\bmath{\Im}}}%
  \def\bpartial{{\bmath{\partial}}}%
  \def\binfty{{\bmath{\infty}}}%
  \def\bprime{{\bmath{\prime}}}%
  \def\bemptyset{{\bmath{\emptyset}}}%
  \def\bnabla{{\bmath{\nabla}}}%
  \def\btop{{\bmath{\top}}}%
  \def\bbot{{\bmath{\bot}}}%
  \def\btriangle{{\bmath{\triangle}}}%
  \def\bforall{{\bmath{\forall}}}%
  \def\bexists{{\bmath{\exists}}}%
  \def\bneg{{\bmath{\neg}}}%
  \def\bflat{{\bmath{\flat}}}%
  \def\bnatural{{\bmath{\natural}}}%
  \def\bsharp{{\bmath{\sharp}}}%
  \def\bclubsuit{{\bmath{\clubsuit}}}%
  \def\bdiamondsuit{{\bmath{\diamondsuit}}}%
  \def\bheartsuit{{\bmath{\heartsuit}}}%
  \def\bspadesuit{{\bmath{\spadesuit}}}%
  \def\bsmallint{{\bmath{\smallint}}}%
  \def\btriangleleft{{\bmath{\triangleleft}}}%
  \def\btriangleright{{\bmath{\triangleright}}}%
  \def\bbigtriangleup{{\bmath{\bigtriangleup}}}%
  \def\bbigtriangledown{{\bmath{\bigtriangledown}}}%
  \def\bwedge{{\bmath{\wedge}}}%
  \def\bvee{{\bmath{\vee}}}%
  \def\bcap{{\bmath{\cap}}}%
  \def\bcup{{\bmath{\cup}}}%
  \def\bddagger{{\bmath{\ddagger}}}%
  \def\bdagger{{\bmath{\dagger}}}%
  \def\bsqcap{{\bmath{\sqcap}}}%
  \def\bsqcup{{\bmath{\sqcup}}}%
  \def\buplus{{\bmath{\uplus}}}%
  \def\bamalg{{\bmath{\amalg}}}%
  \def\bdiamond{{\bmath{\diamond}}}%
  \def\bbullet{{\bmath{\bullet}}}%
  \def\bwr{{\bmath{\wr}}}%
  \def\bdiv{{\bmath{\div}}}%
  \def\bodot{{\bmath{\odot}}}%
  \def\boslash{{\bmath{\oslash}}}%
  \def\botimes{{\bmath{\otimes}}}%
  \def\bominus{{\bmath{\ominus}}}%
  \def\boplus{{\bmath{\oplus}}}%
  \def\bmp{{\bmath{\mp}}}%
  \def\bpm{{\bmath{\pm}}}%
  \def\bcirc{{\bmath{\circ}}}%
  \def\bbigcirc{{\bmath{\bigcirc}}}%
  \def\bsetminus{{\bmath{\setminus}}}%
  \def\bcdot{{\bmath{\cdot}}}%
  \def\bast{{\bmath{\ast}}}%
  \def\btimes{{\bmath{\times}}}%
  \def\bstar{{\bmath{\star}}}%
  \def\bpropto{{\bmath{\propto}}}%
  \def\bsqsubseteq{{\bmath{\sqsubseteq}}}%
  \def\bsqsupseteq{{\bmath{\sqsupseteq}}}%
  \def\bparallel{{\bmath{\parallel}}}%
  \def\bmid{{\bmath{\mid}}}%
  \def\bdashv{{\bmath{\dashv}}}%
  \def\bvdash{{\bmath{\vdash}}}%
  \def\bnearrow{{\bmath{\nearrow}}}%
  \def\bsearrow{{\bmath{\searrow}}}%
  \def\bnwarrow{{\bmath{\nwarrow}}}%
  \def\bswarrow{{\bmath{\swarrow}}}%
  \def\bLeftrightarrow{{\bmath{\Leftrightarrow}}}%
  \def\bLeftarrow{{\bmath{\Leftarrow}}}%
  \def\bRightarrow{{\bmath{\Rightarrow}}}%
  \def\bleq{{\bmath{\leq}}}%
  \def\bgeq{{\bmath{\geq}}}%
  \def\bsucc{{\bmath{\succ}}}%
  \def\bprec{{\bmath{\prec}}}%
  \def\bapprox{{\bmath{\approx}}}%
  \def\bsucceq{{\bmath{\succeq}}}%
  \def\bpreceq{{\bmath{\preceq}}}%
  \def\bsupset{{\bmath{\supset}}}%
  \def\bsubset{{\bmath{\subset}}}%
  \def\bsupseteq{{\bmath{\supseteq}}}%
  \def\bsubseteq{{\bmath{\subseteq}}}%
  \def\bin{{\bmath{\in}}}%
  \def\bni{{\bmath{\ni}}}%
  \def\bgg{{\bmath{\gg}}}%
  \def\bll{{\bmath{\ll}}}%
  \def\bnot{{\bmath{\not}}}%
  \def\bleftrightarrow{{\bmath{\leftrightarrow}}}%
  \def\bleftarrow{{\bmath{\leftarrow}}}%
  \def\brightarrow{{\bmath{\rightarrow}}}%
  \def\bmapstochar{{\bmath{\mapstochar}}}%
  \def\bsim{{\bmath{\sim}}}%
  \def\bsimeq{{\bmath{\simeq}}}%
  \def\bperp{{\bmath{\perp}}}%
  \def\bequiv{{\bmath{\equiv}}}%
  \def\basymp{{\bmath{\asymp}}}%
  \def\bsmile{{\bmath{\smile}}}%
  \def\bfrown{{\bmath{\frown}}}%
  \def\bleftharpoonup{{\bmath{\leftharpoonup}}}%
  \def\bleftharpoondown{{\bmath{\leftharpoondown}}}%
  \def\brightharpoonup{{\bmath{\rightharpoonup}}}%
  \def\brightharpoondown{{\bmath{\rightharpoondown}}}%
  \def\blhook{{\bmath{\lhook}}}%
  \def\brhook{{\bmath{\rhook}}}%
  \def\bldotp{{\bmath{\ldotp}}}%
  \def\bcdotp{{\bmath{\cdotp}}}%
}

\def\,{\relax\ifmmode \mskip\thinmuskip\else \thinspace\fi}
\let\protect=\relax

\long\def\@ifundefined#1#2#3{\expandafter\ifx\csname
  #1\endcsname\relax#2\else#3\fi}




\newtoks\math@groups \math@groups={}
\def\addtom@thgroup#1#2{#1\expandafter{\the#1#2}} 



\def\addtosizeh@ok#1#2#3#4{%
  \expandafter\def\csname @#1pt\endcsname{%
    \def\s@ze{#2}\def\ss@ze{#3}\def\sss@ze{#4}\the\math@groups%
  }%
}



\let\resetsizehook=\addtosizeh@ok


\ifprod@font
  \addtosizeh@ok{viii} {8} {6}  {5}
  \addtosizeh@ok{ix}   {9} {6}  {5}
  \addtosizeh@ok{x}    {10}{7}  {5}
  \addtosizeh@ok{xi}   {11}{8}  {6}
  \addtosizeh@ok{xiv}  {14}{10} {7}
  \addtosizeh@ok{xvii} {17}{12}{10}
\else
  \addtosizeh@ok{viii} {8}     {6}     {5}
  \addtosizeh@ok{ix}   {9}     {6}     {5}
  \addtosizeh@ok{x}    {10}    {7}     {5}
  \addtosizeh@ok{xi}   {10.95} {8}     {6}
  \addtosizeh@ok{xiv}  {14.4}  {10}    {7}
  \addtosizeh@ok{xvii} {17.28} {12}    {10}
\fi

\def\get@font#1#2#3{%
  \edef\fonts@ze{\romannumeral#3}
  \edef\fontn@me{\fonts@ze#1}
  \@ifundefined{\fontn@me}%
    {
     \global\expandafter\font\csname \fontn@me\endcsname=#2 at #3pt}%
    {}%
}

\def\ass@tfont#1#2{%
  \xdef\fam@name{\csname #1\endcsname}%
  \xdef\font@name{\csname #2\endcsname}%
  \let\textfont@name\font@name
  \textfont\fam@name\textfont@name
}

\def\ass@sfont#1#2{%
  \xdef\fam@name{\csname #1\endcsname}%
  \xdef\font@name{\csname #2\endcsname}%
  \let\textfont@name\font@name
  \scriptfont\fam@name\textfont@name
}

\def\ass@ssfont#1#2{%
  \xdef\fam@name{\csname #1\endcsname}%
  \xdef\font@name{\csname #2\endcsname}%
  \let\textfont@name\font@name
  \scriptscriptfont\fam@name\textfont@name
}


\def\NewSymbolFont#1#2{%
  \expandafter\ifx\csname sym#1fam\endcsname\relax 
    \expandafter\newfam\csname sym#1fam\endcsname
    \expandafter\edef\csname sym#1fam\endcsname{\the\allocationnumber}%
    \addtom@thgroup\math@groups{%
      \get@font{#1}{#2}{\s@ze}%
      \ass@tfont{sym#1fam}{\fontn@me}%
      \get@font{#1}{#2}{\ss@ze}%
      \ass@sfont{sym#1fam}{\fontn@me}%
      \get@font{#1}{#2}{\sss@ze}%
      \ass@ssfont{sym#1fam}{\fontn@me}%
    }%
  \else
    \errmessage{Family `#1' already defined}%
  \fi
}


\def\NewMathSymbol#1#2#3#4{%
  \edef\f@mly{\expandafter\hexnumber{\csname sym#3fam\endcsname}}%
  \mathchardef#1="#2\f@mly#4\relax
}


\newif\ifd@f

\def\NewMathDelimiter#1#2#3#4#5#6{%
  \d@ftrue
  \expandafter\ifx\csname sym#3fam\endcsname\relax
    \d@ffalse \errmessage{Family `#3' is not defined}%
  \fi
  \expandafter\ifx\csname sym#5fam\endcsname\relax
    \d@ffalse \errmessage{Family `#5' is not defined}%
  \fi
  \ifd@f
    \edef\f@mly{\expandafter\hexnumber{\csname sym#3fam\endcsname}}%
    \edef\f@mlytw@{\expandafter\hexnumber{\csname sym#5fam\endcsname}}%
    \xdef#1{\delimiter"#2\f@mly #4\f@mlytw@ #6\relax}%
  \fi
}


\def\setboxz@h{\setbox\z@\hbox}
\def\wdz@{\wd\z@}
\def\boxz@{\box\z@}
\def\setbox@ne{\setbox\@ne}
\def\wd@ne{\wd\@ne}

\def\math@atom#1#2{%
   \binrel@{#1}\binrel@@{#2}}
\def\binrel@#1{\setboxz@h{\thinmuskip0mu
  \medmuskip\m@ne mu\thickmuskip\@ne mu$#1\m@th$}%
 \setbox@ne\hbox{\thinmuskip0mu\medmuskip\m@ne mu\thickmuskip
  \@ne mu${}#1{}\m@th$}%
 \setbox\tw@\hbox{\hskip\wd@ne\hskip-\wdz@}}
\def\binrel@@#1{\ifdim\wd2<\z@\mathbin{#1}\else\ifdim\wd\tw@>\z@
 \mathrel{#1}\else{#1}\fi\fi}

\def\m@thit{1}

\def\set@skchar#1{\global\expandafter\skewchar
  \csname\fontn@me\endcsname=#1\relax}

\def\NewMathAlphabet#1#2#3{%
  \def\tst{#3}%
  \ifx\tst\empty\else 
    \expandafter\gdef\csname #1@sc\endcsname{}
  \fi
  \expandafter\def\csname #1\endcsname{
    \protect\csname @#1\endcsname}%
  \expandafter\def\csname @#1\endcsname##1{
    {%
    \begingroup
      \get@font{#1}{#2}{\s@ze}%
      \@ifundefined{#1@sc}{}{\set@skchar{#3}}%
      \ass@tfont{m@thit}{\fontn@me}%
      \get@font{#1}{#2}{\ss@ze}%
      \@ifundefined{#1@sc}{}{\set@skchar{#3}}%
      \ass@sfont{m@thit}{\fontn@me}%
      \get@font{#1}{#2}{\sss@ze}%
      \@ifundefined{#1@sc}{}{\set@skchar{#3}}%
      \ass@ssfont{m@thit}{\fontn@me}%
      \math@atom{##1}{%
      \mathchoice%
        {\hbox{$\m@th\displaystyle##1$}}%
        {\hbox{$\m@th\textstyle##1$}}%
        {\hbox{$\m@th\scriptstyle##1$}}%
        {\hbox{$\m@th\scriptscriptstyle##1$}}}%
    \endgroup
    }%
  }%
}


\newif\iffirstta  \firsttatrue

\def\set@hchar#1{\global\expandafter\hyphenchar
  \csname\fontn@me\endcsname=#1\relax}

\def\NewTextAlphabet#1#2#3{%
  \iffirstta
    \global\firsttafalse
    \newfam\scratchfam
    \edef\scrt@fam{\the\allocationnumber}%
  \fi
  \def\tst{#3}%
  \ifx\tst\empty\else 
    \expandafter\gdef\csname #1@hc\endcsname{}
  \fi
  \expandafter\def\csname #1\endcsname{
    \protect\csname t@#1\endcsname}%
  \long\expandafter\def\csname t@#1\endcsname##1{
    \ifmmode
      \typeout{Warning: do not use \expandafter\string\csname #1\endcsname
        \space in math mode}\fi%
    {%
      \get@font{#1}{#2}{\s@ze}\let\t@xtfnt=\fontn@me\relax
      \@ifundefined{#1@hc}{}{\set@hchar{#3}}%
      \ass@tfont{scrt@fam}{\fontn@me}%
      \get@font{#1}{#2}{\ss@ze}%
      \@ifundefined{#1@hc}{}{\set@hchar{#3}}%
      \ass@sfont{scrt@fam}{\fontn@me}%
      \get@font{#1}{#2}{\sss@ze}%
      \@ifundefined{#1@hc}{}{\set@hchar{#3}}%
      \ass@ssfont{scrt@fam}{\fontn@me}%
      \fam\scratchfam\csname\t@xtfnt\endcsname
    ##1%
    }%
  }%
  \expandafter\def\csname #1shape
    \endcsname{\protect\csname @#1shape\endcsname}%
  \expandafter\def\csname @#1shape\endcsname{
    \ifmmode
      \typeout{Warning: do not use \expandafter\string\csname
        #1shape\endcsname \space in math mode}\fi
      \get@font{#1}{#2}{\s@ze}\let\t@xtfnt=\fontn@me\relax
      \@ifundefined{#1@hc}{}{\set@hchar{#3}}%
      \ass@tfont{scrt@fam}{\fontn@me}%
      \get@font{#1}{#2}{\ss@ze}%
      \@ifundefined{#1@hc}{}{\set@hchar{#3}}%
      \ass@sfont{scrt@fam}{\fontn@me}%
      \get@font{#1}{#2}{\sss@ze}%
      \@ifundefined{#1@hc}{}{\set@hchar{#3}}%
      \ass@ssfont{scrt@fam}{\fontn@me}%
      \fam\scratchfam\csname\t@xtfnt\endcsname
  }%
}


\ifprod@font
  \def\math@itfnt{mtmib10}
  \def\math@syfnt{mtbsy10}
\else
  \def\math@itfnt{cmmib10}
  \def\math@syfnt{cmbsy10}
\fi

\def\m@thsy{2}

\def\bmath{\protect\@bmath}
\def\@bmath#1{%
  {%
  \begingroup
    \get@font{mthit}{\math@itfnt}{\s@ze}\set@skchar{'177}%
    \ass@tfont{m@thit}{\fontn@me}%
    \get@font{mthit}{\math@itfnt}{\ss@ze}\set@skchar{'177}%
    \ass@sfont{m@thit}{\fontn@me}%
    \get@font{mthit}{\math@itfnt}{\sss@ze}\set@skchar{'177}%
    \ass@ssfont{m@thit}{\fontn@me}%
    \get@font{mthsy}{\math@syfnt}{\s@ze}\set@skchar{'60}%
    \ass@tfont{m@thsy}{\fontn@me}%
    \get@font{mthsy}{\math@syfnt}{\ss@ze}\set@skchar{'60}%
    \ass@sfont{m@thsy}{\fontn@me}%
    \get@font{mthsy}{\math@syfnt}{\sss@ze}\set@skchar{'60}%
    \ass@ssfont{m@thsy}{\fontn@me}%
    \math@atom{#1}{%
    \mathchoice%
      {\hbox{$\m@th\displaystyle#1$}}%
      {\hbox{$\m@th\textstyle#1$}}%
      {\hbox{$\m@th\scriptstyle#1$}}%
      {\hbox{$\m@th\scriptscriptstyle#1$}}}%
  \endgroup
  }%
}



\def\diameter{{\ifmmode\mathchoice
{\ooalign{\hfil\hbox{$\displaystyle/$}\hfil\crcr
{\hbox{$\displaystyle\mathchar"20D$}}}}
{\ooalign{\hfil\hbox{$\textstyle/$}\hfil\crcr
{\hbox{$\textstyle\mathchar"20D$}}}}
{\ooalign{\hfil\hbox{$\scriptstyle/$}\hfil\crcr
{\hbox{$\scriptstyle\mathchar"20D$}}}}
{\ooalign{\hfil\hbox{$\scriptscriptstyle/$}\hfil\crcr
{\hbox{$\scriptscriptstyle\mathchar"20D$}}}}
\else{\ooalign{\hfil/\hfil\crcr\mathhexbox20D}}%
\fi}}

\def\sq{\ifmmode\squareforqed\else{\unskip\nobreak\hfil
\penalty50\hskip1em\null\nobreak\hfil\squareforqed
\parfillskip=0pt\finalhyphendemerits=0\endgraf}\fi}
\def\squareforqed{\hbox{\rlap{$\sqcap$}$\sqcup$}}


\def\bbbc{{\mathchoice {\setbox0=\hbox{$\displaystyle\rm C$}\hbox{\hbox
to0pt{\kern0.4\wd0\vrule height0.9\ht0\hss}\box0}}
{\setbox0=\hbox{$\textstyle\rm C$}\hbox{\hbox
to0pt{\kern0.4\wd0\vrule height0.9\ht0\hss}\box0}}
{\setbox0=\hbox{$\scriptstyle\rm C$}\hbox{\hbox
to0pt{\kern0.4\wd0\vrule height0.9\ht0\hss}\box0}}
{\setbox0=\hbox{$\scriptscriptstyle\rm C$}\hbox{\hbox
to0pt{\kern0.4\wd0\vrule height0.9\ht0\hss}\box0}}}}
\def\bbbq{{\mathchoice {\setbox0=\hbox{$\displaystyle\rm
Q$}\hbox{\raise
0.15\ht0\hbox to0pt{\kern0.4\wd0\vrule height0.8\ht0\hss}\box0}}
{\setbox0=\hbox{$\textstyle\rm Q$}\hbox{\raise
0.15\ht0\hbox to0pt{\kern0.4\wd0\vrule height0.8\ht0\hss}\box0}}
{\setbox0=\hbox{$\scriptstyle\rm Q$}\hbox{\raise
0.15\ht0\hbox to0pt{\kern0.4\wd0\vrule height0.7\ht0\hss}\box0}}
{\setbox0=\hbox{$\scriptscriptstyle\rm Q$}\hbox{\raise
0.15\ht0\hbox to0pt{\kern0.4\wd0\vrule height0.7\ht0\hss}\box0}}}}
\def\bbbt{{\mathchoice {\setbox0=\hbox{$\displaystyle\rm
T$}\hbox{\hbox to0pt{\kern0.3\wd0\vrule height0.9\ht0\hss}\box0}}
{\setbox0=\hbox{$\textstyle\rm T$}\hbox{\hbox
to0pt{\kern0.3\wd0\vrule height0.9\ht0\hss}\box0}}
{\setbox0=\hbox{$\scriptstyle\rm T$}\hbox{\hbox
to0pt{\kern0.3\wd0\vrule height0.9\ht0\hss}\box0}}
{\setbox0=\hbox{$\scriptscriptstyle\rm T$}\hbox{\hbox
to0pt{\kern0.3\wd0\vrule height0.9\ht0\hss}\box0}}}}
\def\bbbs{{\mathchoice
{\setbox0=\hbox{$\displaystyle     \rm S$}\hbox{\raise0.5\ht0\hbox
to0pt{\kern0.35\wd0\vrule height0.45\ht0\hss}\hbox
to0pt{\kern0.55\wd0\vrule height0.5\ht0\hss}\box0}}
{\setbox0=\hbox{$\textstyle        \rm S$}\hbox{\raise0.5\ht0\hbox
to0pt{\kern0.35\wd0\vrule height0.45\ht0\hss}\hbox
to0pt{\kern0.55\wd0\vrule height0.5\ht0\hss}\box0}}
{\setbox0=\hbox{$\scriptstyle      \rm S$}\hbox{\raise0.5\ht0\hbox
to0pt{\kern0.35\wd0\vrule height0.45\ht0\hss}\raise0.05\ht0\hbox
to0pt{\kern0.5\wd0\vrule height0.45\ht0\hss}\box0}}
{\setbox0=\hbox{$\scriptscriptstyle\rm S$}\hbox{\raise0.5\ht0\hbox
to0pt{\kern0.4\wd0\vrule height0.45\ht0\hss}\raise0.05\ht0\hbox
to0pt{\kern0.55\wd0\vrule height0.45\ht0\hss}\box0}}}}
\def\bbbz{{\mathchoice {\hbox{$\sf\textstyle Z\kern-0.4em Z$}}
{\hbox{$\sf\textstyle Z\kern-0.4em Z$}}
{\hbox{$\sf\scriptstyle Z\kern-0.3em Z$}}
{\hbox{$\sf\scriptscriptstyle Z\kern-0.2em Z$}}}}


\def\Nulle{0} 
\def\Afe{1}   
\def\Hae{2}   
\def\Hbe{3}   
\def\Hce{4}   
\def\Hde{5}   


\newcount\LastMac       \LastMac=\Nulle

\newskip\half      \half=5.5pt plus 1.5pt minus 2.25pt
\newskip\one       \one=11pt plus 3pt minus 5.5pt
\newskip\onehalf   \onehalf=16.5pt plus 5.5pt minus 8.25pt
\newskip\two       \two=22pt plus 5.5pt minus 11pt

\def\Half{\addvspace{\half}}
\def\One{\addvspace{\one}}
\def\OneHalf{\addvspace{\onehalf}}
\def\Two{\addvspace{\two}}

\def\Raggedright{
  \rightskip=\z@ plus \hsize\relax
}

\def\Fullout{
  \rightskip=\z@\relax
}

\def\Hang#1#2{
  \hangindent=#1%
  \hangafter=#2\relax
}


\newif\ifsp@page
\def\pagestyle#1{\csname ps@#1\endcsname}
\def\thispagestyle#1{\global\sp@pagetrue\gdef\sp@type{#1}}

\def\ps@titlepage{%
  \def\@oddhead{\eightpoint\noindent \the\CatchLine
    \ifprod@font\else\qquad Printed\ \today\qquad
      (MN plain \TeX\ macros\ v\@version)\fi \hfil}%
  \let\@evenhead=\@oddhead
  \def\@oddfoot{\eightpoint\copyright\ \@pubyear\ RAS\hfil}%
  \def\@evenfoot{\hfil\eightpoint\noindent\copyright\ \@pubyear\ RAS}%
}

\def\ps@headings{%
  \def\@oddhead{\elevenpoint\it\noindent
    \hfill\the\RightHeader\hskip1.5em\rm\folio}%
  \def\@evenhead{\elevenpoint\noindent
    \folio\hskip1.5em\it\the\LeftHeader\hfill}%
  \def\@oddfoot{\eightpoint\noindent\copyright\ \@pubyear\ RAS,
    MNRAS {\bf \@volume}, \@pagerange\hfil}%
  \def\@evenfoot{\hfil\eightpoint\copyright\ \@pubyear\ RAS,
    MNRAS {\bf \@volume}, \@pagerange}%
}

\def\ps@plate{%
  \def\@oddhead{\eightpoint\noindent\plt@cap\hfil}%
  \def\@evenhead{\eightpoint\noindent\plt@cap\hfil}%
  \def\@oddfoot{\eightpoint\noindent\copyright\ \@pubyear\ RAS,
    MNRAS {\bf \@volume}, \@pagerange\hfil}%
  \def\@evenfoot{\hfil\eightpoint\copyright\ \@pubyear\ RAS,
    MNRAS {\bf \@volume}, \@pagerange}%
}



\def\title#1{
  \bgroup
    \vbox to 8pt{\vss}%
    \seventeenpoint
    \Raggedright
    \noindent \strut{\bf #1}\par
  \egroup
}

\def\author#1{
  \bgroup
    \ifnum\LastMac=\Afe \OneHalf\else \vskip 21pt\fi
    \fourteenpoint
    \Raggedright
    \noindent \strut #1\par
    \vskip 3pt%
  \egroup
}

\def\affiliation#1{
  \bgroup
    \vskip -4pt%
    \eightpoint
    \Raggedright
    \noindent \strut {\it #1}\par
  \egroup
  \LastMac=\Afe\relax
}

\def\acceptedline#1{
  \bgroup
    \Two
    \eightpoint
    \Raggedright
    \noindent \strut #1\par
  \egroup
}

\long\def\abstract#1{%
  \bgroup
    \vskip 20pt%
    \leftskip 11pc\rightskip\z@
    \noindent{\ninebf ABSTRACT}\par
    \tenpoint
    \Fullout
    \noindent #1\par
  \egroup
}

\long\def\keywords#1{
  \bgroup
    \Half
    \leftskip 11pc\rightskip\z@
    \tenpoint
    \Fullout
    \noindent\hbox{\bf Key words:}\ #1\par
  \egroup
}


\def\maketitle{%
  \EndOpening
  \ifsinglecol \else \MakePage\fi
}


\def\pageoffset#1#2{\hoffset=#1\relax\voffset=#2\relax}


\def\@nameuse#1{\csname #1\endcsname}
\def\arabic#1{\@arabic{\@nameuse{#1}}}
\def\alph#1{\@alph{\@nameuse{#1}}}
\def\Alph#1{\@Alph{\@nameuse{#1}}}
\def\@arabic#1{\number #1}
\def\@Alph#1{\ifcase#1\or A\or B\or C\or D\else\@Ialph{#1}\fi}
\def\@Ialph#1{\ifcase#1\or \or \or \or \or E\or F\or G\or H\or I\or J\or
   K\or L\or M\or N\or O\or P\or Q\or R\or S\or T\or U\or V\or W\or X\or
   Y\or Z\else\errmessage{Counter out of range}\fi}
\def\@alph#1{\ifcase#1\or a\or b\or c\or d\else\@ialph{#1}\fi}
\def\@ialph#1{\ifcase#1\or \or \or \or \or e\or f\or g\or h\or i\or j\or
   k\or l\or m\or n\or o\or p\or q\or r\or s\or t\or u\or v\or w\or x\or y\or
   z\else\errmessage{Counter out of range}\fi}


\newcount\Eqnno
\newcount\SubEqnno

\def\theeq{\arabic{Eqnno}}
\def\thesubeq{\alph{SubEqnno}}

\def\stepeq{\relax
  \global\SubEqnno \z@
  \global\advance\Eqnno \@ne\relax
  {\rm (\theeq)}%
}

\def\startsubeq{\relax
  \global\SubEqnno \z@
  \global\advance\Eqnno \@ne\relax
  \stepsubeq
}

\def\stepsubeq{\relax
  \global\advance\SubEqnno \@ne\relax
  {\rm (\theeq\thesubeq)}%
}


\newcount\Sec        
\newcount\SecSec
\newcount\SecSecSec

\def\thesection{\arabic{Sec}}
\def\thesubsection{\thesection.\arabic{SecSec}}
\def\thesubsubsection{\thesubsection.\arabic{SecSecSec}}

\Sec=\z@

\def\:{\let\@sptoken= } \:  
\def\:{\@xifnch} \expandafter\def\: {\futurelet\@tempc\@ifnch}

\def\@ifnextchar#1#2#3{%
  \let\@tempMACe #1%
  \def\@tempMACa{#2}%
  \def\@tempMACb{#3}%
  \futurelet \@tempMACc\@ifnch%
}

\def\@ifnch{%
\ifx \@tempMACc \@sptoken%
  \let\@tempMACd\@xifnch%
\else%
  \ifx \@tempMACc \@tempMACe%
    \let\@tempMACd\@tempMACa%
  \else%
    \let\@tempMACd\@tempMACb%
  \fi%
\fi%
\@tempMACd%
}

\def\@ifstar#1#2{\@ifnextchar *{\def\@tempMACa*{#1}\@tempMACa}{#2}}

\newskip\@tempskipb

\def\addvspace#1{%
  \ifvmode\else \endgraf\fi%
  \ifdim\lastskip=\z@%
    \vskip #1\relax%
  \else%
    \@tempskipb#1\relax\@xaddvskip%
  \fi%
}

\def\@xaddvskip{%
  \ifdim\lastskip<\@tempskipb%
    \vskip-\lastskip%
    \vskip\@tempskipb\relax%
  \else%
    \ifdim\@tempskipb<\z@%
      \ifdim\lastskip<\z@ \else%
        \advance\@tempskipb\lastskip%
        \vskip-\lastskip\vskip\@tempskipb%
      \fi%
    \fi%
  \fi%
}

\newskip\@tmpSKIP

\def\addpen#1{%
  \ifvmode
    \if@nobreak
    \else
      \ifdim\lastskip=\z@
        \penalty#1\relax
      \else
        \@tmpSKIP=\lastskip
        \vskip -\lastskip
        \penalty#1\vskip\@tmpSKIP
      \fi
    \fi
  \fi
}

\newcount\@clubpen   \@clubpen=\clubpenalty
\newif\if@nobreak    \@nobreakfalse

\def\@noafterindent{%
  \global\@nobreaktrue
  \everypar{\if@nobreak
              \global\@nobreakfalse
              \clubpenalty \@M
              {\setbox\z@\lastbox}%
              \LastMac=\Nulle\relax%
            \else
              \clubpenalty \@clubpen
              \everypar{}%
            \fi}%
}

\newcount\gds@cbrk   \gds@cbrk=-300

\def\@nohdbrk{\interlinepenalty \@M\relax}

\let\@par=\par
\def\@restorepar{\def\par{\@par}}

\newif\if@endpe   \@endpefalse
 
\def\@doendpe{\@endpetrue \@nobreakfalse \LastMac=\Nulle\relax%
     \def\par{\@restorepar\everypar{}\par\@endpefalse}%
              \everypar{\setbox\z@\lastbox\everypar{}\@endpefalse}%
}

\def\section{\@ifstar{\@ssection}{\@section}}

\def\@section#1{
  \if@nobreak
    \everypar{}%
    \ifnum\LastMac=\Hae \addvspace{\half}\fi
  \else
    \addpen{\gds@cbrk}%
    \addvspace{\two}%
  \fi
  \bgroup
    \ninepoint\bf
    \Raggedright
    \global\advance\Sec \@ne
    \ifappendix
      \global\Eqnno=\z@ \global\SubEqnno=\z@\relax
      \def\ch@ck{#1}%
      \ifx\ch@ck\empty \def\c@lon{}\else\def\c@lon{:}\fi
      \noindent\@nohdbrk APPENDIX\ \thesection\c@lon\hskip 0.5em%
        \uppercase{#1}\par
    \else
      \noindent\@nohdbrk\thesection\hskip 1pc \uppercase{#1}\par
    \fi
    \global\SecSec=\z@
  \egroup
  \nobreak
  \vskip\half
  \nobreak
  \@noafterindent
  \LastMac=\Hae\relax
}

\def\@ssection#1{
  \if@nobreak
    \everypar{}%
    \ifnum\LastMac=\Hae \addvspace{\half}\fi
  \else
    \addpen{\gds@cbrk}%
    \addvspace{\two}%
  \fi
  \bgroup
    \ninepoint\bf
    \Raggedright
    \noindent\@nohdbrk\uppercase{#1}\par
  \egroup
  \nobreak
  \vskip\half
  \nobreak
  \@noafterindent
  \LastMac=\Hae\relax
}

\def\subsection{\@ifstar{\@ssubsection}{\@subsection}}

\def\@subsection#1{
  \if@nobreak
    \everypar{}%
    \ifnum\LastMac=\Hae \addvspace{1pt plus 1pt minus .5pt}\fi
  \else
    \addpen{\gds@cbrk}%
    \addvspace{\onehalf}%
  \fi
  \bgroup
    \ninepoint\bf
    \Raggedright
    \global\advance\SecSec \@ne
    \noindent\@nohdbrk\thesubsection \hskip 1pc\relax #1\par
    \global\SecSecSec=\z@
  \egroup
  \nobreak
  \vskip\half
  \nobreak
  \@noafterindent
  \LastMac=\Hbe\relax
}

\def\@ssubsection#1{
  \if@nobreak
    \everypar{}%
    \ifnum\LastMac=\Hae \addvspace{1pt plus 1pt minus .5pt}\fi
  \else
    \addpen{\gds@cbrk}%
    \addvspace{\onehalf}%
  \fi
  \bgroup
    \ninepoint\bf
    \Raggedright
    \noindent\@nohdbrk #1\par
  \egroup
  \nobreak
  \vskip\half
  \nobreak
  \@noafterindent
  \LastMac=\Hbe\relax
}

\def\subsubsection{\@ifstar{\@ssubsubsection}{\@subsubsection}}

\def\@subsubsection#1{
  \if@nobreak
    \everypar{}%
    \ifnum\LastMac=\Hbe \addvspace{1pt plus 1pt minus .5pt}\fi
  \else
    \addpen{\gds@cbrk}%
    \addvspace{\onehalf}%
  \fi
  \bgroup
    \ninepoint\it
    \Raggedright
    \global\advance\SecSecSec \@ne
    \noindent\@nohdbrk\thesubsubsection \hskip 1pc\relax #1\par
  \egroup
  \nobreak
  \vskip\half
  \nobreak
  \@noafterindent
  \LastMac=\Hce\relax
}

\def\@ssubsubsection#1{
  \if@nobreak
    \everypar{}%
    \ifnum\LastMac=\Hbe \addvspace{1pt plus 1pt minus .5pt}\fi
  \else
    \addpen{\gds@cbrk}%
    \addvspace{\onehalf}%
  \fi
  \bgroup
    \ninepoint\it
    \Raggedright
    \noindent\@nohdbrk #1\par
  \egroup
  \nobreak
  \vskip\half
  \nobreak
  \@noafterindent
  \LastMac=\Hce\relax
}

\def\paragraph#1{
  \if@nobreak
    \everypar{}%
  \else
    \addpen{\gds@cbrk}%
    \addvspace{\one}%
  \fi%
  \bgroup%
    \ninepoint\it
    \noindent #1\ \nobreak%
  \egroup
  \LastMac=\Hde\relax
  \ignorespaces
}


\newif\ifappendix

\def\appendix{%
  \global\appendixtrue
  \def\thesection{\Alph{Sec}}%
  \def\thesubsection{\thesection\arabic{SecSec}}%
  \def\theeq{\thesection\arabic{Eqnno}}%
  \Sec=\z@ \SecSec=\z@ \SecSecSec=\z@ \Eqnno=\z@ \SubEqnno=\z@\relax
}




\def\beginlist{%
  \par\if@nobreak \else\addvspace{\half}\fi%
  \bgroup%
    \ninepoint
    \let\item=\list@item%
}

\def\list@item{%
  \par\noindent\hskip 1em\relax%
  \ignorespaces%
}

\def\endlist{\par\egroup\addvspace{\half}\@doendpe}


\def\beginrefs{%
  \par
  \bgroup
    \eightpoint
    \Fullout
    \let\bibitem=\bib@item
}

\def\bib@item{%
  \par\parindent=1.5em\Hang{1.5em}{1}%
  \everypar={\Hang{1.5em}{1}\ignorespaces}%
  \noindent\ignorespaces
}

\def\endrefs{\par\egroup\@doendpe}


\newtoks\CatchLine

\def\@journal{Mon.\ Not.\ R.\ Astron.\ Soc.\ }  
\def\@pubyear{1994}        
\def\@pagerange{000--000}  
\def\@volume{000}          
\def\@microfiche{}         %

\def\pubyear#1{\gdef\@pubyear{#1}\@makecatchline}
\def\pagerange#1{\gdef\@pagerange{#1}\@makecatchline}
\def\volume#1{\gdef\@volume{#1}\@makecatchline}
\def\microfiche#1{\gdef\@microfiche{and Microfiche\ #1}\@makecatchline}

\def\@makecatchline{%
  \global\CatchLine{%
    {\rm \@journal {\bf \@volume},\ \@pagerange\ (\@pubyear)\ \@microfiche}}%
}

\@makecatchline 

\newtoks\LeftHeader
\def\shortauthor#1{
  \global\LeftHeader{#1}%
}

\newtoks\RightHeader
\def\shorttitle#1{
  \global\RightHeader{#1}%
}

\def\PageHead{
  \begingroup
    \ifsp@page
      \csname ps@\sp@type\endcsname
    \fi
    \ifodd\pageno
      \let\the@head=\@oddhead
    \else
      \let\the@head=\@evenhead
    \fi
    \vbox to \z@{\vskip-22.5\p@%
      \hbox to \PageWidth{\vbox to8.5\p@{}%
        \the@head
      }%
    \vss}%
  \endgroup
  \nointerlineskip
}

\gdef\PageFoot{%
  \nointerlineskip%
  \begingroup
  \ifsp@page
    \csname ps@\sp@type\endcsname
    \global\sp@pagefalse
  \fi
  \vbox to 22pt{\vfil%
    \hbox to \PageWidth{%
      \eightpoint\strut\noindent
      \ifodd\pageno
        \@oddfoot
      \else
        \@evenfoot
      \fi
    }%
  }%
  \endgroup
}

\def\today{%
  \number\day\space
  \ifcase\month\or January\or February\or March\or April\or May\or June\or
    July\or August\or September\or October\or November\or December\fi
  \space\number\year%
}

\def\authorcomment#1{%
  \gdef\PageFoot{%
    \nointerlineskip%
    \vbox to 20pt{\vfil%
      \hbox to \PageWidth{\elevenpoint\noindent \hfil #1 \hfil}}%
  }%
}


\newif\ifplate@page
\newbox\plt@box

\def\beginplatepage{%
  \let\plate=\plate@head
  \let\caption=\fig@caption
  \global\setbox\plt@box=\vbox\bgroup
  \TEMPDIMEN=\PageWidth 
  \hsize=\PageWidth\relax
}

\def\endplatepage{\par\egroup\global\plate@pagetrue}
\def\plate@head#1{\gdef\plt@cap{#1}}


\def\letters{%
  \gdef\folio{\ifnum\pageno<\z@ L\romannumeral-\pageno
    \else L\number\pageno \fi}%
}


\newdimen\mathindent

\global\mathindent=\z@
\global\everydisplay{\global\@dspwd=\displaywidth\displaysetup}


\def\@displaylines#1{
  {}$\displ@y\hbox{\vbox{\halign{$\@lign\hfil\displaystyle##\hfil$\crcr
  #1\crcr}}}${}%
}

\def\@eqalign#1{\null\vcenter{\openup\jot\m@th
  \ialign{\strut\hfil$\displaystyle{##}$&$\displaystyle{{}##}$\hfil
      \crcr#1\crcr}}%
}

\def\@eqalignno#1{
  \global\advance\@dspwd by -\mathindent%
  {}$\displ@y\hbox{\vbox{\halign to\@dspwd%
  {\hfil$\@lign\displaystyle{##}$\tabskip\z@skip
  &$\@lign\displaystyle{{}##}$\hfil\tabskip\centering
  &\llap{$\@lign##$}\tabskip\z@skip\crcr
  #1\crcr}}}${}%
}


\global\let\displaylines=\@displaylines
\global\let\eqalign=\@eqalign
\global\let\eqalignno=\@eqalignno
\global\let\leqalignno=\@eqalignno

\newdimen\@dspwd   \@dspwd=\z@
\newif\if@eqno
\newif\if@leqno
\newtoks\@eqn
\newtoks\@eq

\def\displaysetup#1$${\displaytest#1\eqno\eqno\displaytest}

\def\displaytest#1\eqno#2\eqno#3\displaytest{%
 \if!#3!\ldisplaytest#1\leqno\leqno\ldisplaytest
 \else\@eqnotrue\@leqnofalse\@eqn={#2}\@eq={#1}\fi
 \generaldisplay$$}

\def\ldisplaytest#1\leqno#2\leqno#3\ldisplaytest{%
\@eq={#1}%
 \if!#3!\@eqnofalse\else\@eqnotrue\@leqnotrue
  \@eqn={#2}\fi}

\def\generaldisplay{%
  \if@eqno
    \if@leqno
      \hbox to \displaywidth{\noindent
        \rlap{$\displaystyle\the\@eqn$}%
        \hskip\mathindent$\displaystyle\the\@eq$\hfil}%
    \else
      \hbox to \displaywidth{\noindent
        \hskip\mathindent
        $\displaystyle\the\@eq$\hfil$\displaystyle\the\@eqn$}%
    \fi
  \else
    \hbox to \displaywidth{\noindent
      \hskip\mathindent$\displaystyle\the\@eq$\hfil}%
  \fi
}


\def\@notice{%
  \par\Two%
  \noindent{\b@ls{11pt}\ninerm This paper has been produced using the
    Royal Astronomical Society/Blackwell Science \TeX\ macros.\par}%
}

\outer\def\bye{\@notice\par\vfill\supereject\end}


\def\start@mess{%
  Monthly notices of the RAS journal style (\@typeface)\space
    v\@version,\space \@verdate.%
}

\everyjob{\Warn{\start@mess}}



\newif\if@debug \@debugfalse  

\def\Print#1{\if@debug\immediate\write16{#1}\else \fi}
\def\Warn#1{\immediate\write16{#1}}
\def\wlog#1{}

\newcount\Iteration 

\def\Single{0} \def\Double{1}                 
\def\Figure{0} \def\Table{1}                  

\def\InStack{0}  
\def\InZoneA{1}
\def\InZoneB{2}
\def\InZoneC{3}

\newcount\TEMPCOUNT 
\newdimen\TEMPDIMEN 
\newbox\TEMPBOX     
\newbox\VOIDBOX     

\newcount\LengthOfStack 
\newcount\MaxItems      
\newcount\StackPointer
\newcount\Point         
\newcount\NextFigure    
\newcount\NextTable     
\newcount\NextItem      

\newcount\StatusStack   
\newcount\NumStack      
\newcount\TypeStack     
\newcount\SpanStack     
\newcount\BoxStack      

\newcount\ItemSTATUS    
\newcount\ItemNUMBER    
\newcount\ItemTYPE      
\newcount\ItemSPAN      
\newbox\ItemBOX         
\newdimen\ItemSIZE      

\newdimen\PageHeight    
\newdimen\TextLeading   
\newdimen\Feathering    
\newcount\LinesPerPage  
\newdimen\ColumnWidth   
\newdimen\ColumnGap     
\newdimen\PageWidth     
\newdimen\BodgeHeight   
\newcount\Leading       

\newdimen\ZoneBSize  
\newdimen\TextSize   
\newbox\ZoneABOX     
\newbox\ZoneBBOX     
\newbox\ZoneCBOX     

\newif\ifFirstSingleItem
\newif\ifFirstZoneA
\newif\ifMakePageInComplete
\newif\ifMoreFigures \MoreFiguresfalse 
\newif\ifMoreTables  \MoreTablesfalse  

\newif\ifFigInZoneB 
\newif\ifFigInZoneC 
\newif\ifTabInZoneB 
\newif\ifTabInZoneC

\newif\ifZoneAFullPage

\newbox\MidBOX    
\newbox\LeftBOX
\newbox\RightBOX
\newbox\PageBOX   

\newif\ifLeftCOL  
\LeftCOLtrue

\newdimen\ZoneBAdjust

\newcount\ItemFits
\def\Yes{1}
\def\No{2}


\MaxItems=15
\NextFigure=\z@        
\NextTable=\@ne

\BodgeHeight=6pt
\TextLeading=11pt    
\Leading=11
\Feathering=\z@      
\LinesPerPage=61     
\topskip=\TextLeading
\ColumnWidth=20pc    
\ColumnGap=2pc       

\newskip\ItemSepamount  
\ItemSepamount=\TextLeading plus \TextLeading minus 4pt

\parskip=\z@ plus .1pt
\parindent=18pt
\widowpenalty=\z@
\clubpenalty=10000
\tolerance=1500
\hbadness=1500
\abovedisplayskip=6pt plus 2pt minus 1pt
\belowdisplayskip=6pt plus 2pt minus 1pt
\abovedisplayshortskip=6pt plus 2pt minus 1pt
\belowdisplayshortskip=6pt plus 2pt minus 1pt

\frenchspacing

\ninepoint 

\PageHeight=682pt
\PageWidth=2\ColumnWidth
\advance\PageWidth by \ColumnGap

\pagestyle{headings}




\newcount\DUMMY \StatusStack=\allocationnumber
\newcount\DUMMY \newcount\DUMMY \newcount\DUMMY 
\newcount\DUMMY \newcount\DUMMY \newcount\DUMMY 
\newcount\DUMMY \newcount\DUMMY \newcount\DUMMY
\newcount\DUMMY \newcount\DUMMY \newcount\DUMMY 
\newcount\DUMMY \newcount\DUMMY \newcount\DUMMY

\newcount\DUMMY \NumStack=\allocationnumber
\newcount\DUMMY \newcount\DUMMY \newcount\DUMMY 
\newcount\DUMMY \newcount\DUMMY \newcount\DUMMY 
\newcount\DUMMY \newcount\DUMMY \newcount\DUMMY 
\newcount\DUMMY \newcount\DUMMY \newcount\DUMMY 
\newcount\DUMMY \newcount\DUMMY \newcount\DUMMY

\newcount\DUMMY \TypeStack=\allocationnumber
\newcount\DUMMY \newcount\DUMMY \newcount\DUMMY 
\newcount\DUMMY \newcount\DUMMY \newcount\DUMMY 
\newcount\DUMMY \newcount\DUMMY \newcount\DUMMY 
\newcount\DUMMY \newcount\DUMMY \newcount\DUMMY 
\newcount\DUMMY \newcount\DUMMY \newcount\DUMMY

\newcount\DUMMY \SpanStack=\allocationnumber
\newcount\DUMMY \newcount\DUMMY \newcount\DUMMY 
\newcount\DUMMY \newcount\DUMMY \newcount\DUMMY 
\newcount\DUMMY \newcount\DUMMY \newcount\DUMMY 
\newcount\DUMMY \newcount\DUMMY \newcount\DUMMY 
\newcount\DUMMY \newcount\DUMMY \newcount\DUMMY

\newbox\DUMMY   \BoxStack=\allocationnumber
\newbox\DUMMY   \newbox\DUMMY \newbox\DUMMY 
\newbox\DUMMY   \newbox\DUMMY \newbox\DUMMY 
\newbox\DUMMY   \newbox\DUMMY \newbox\DUMMY 
\newbox\DUMMY   \newbox\DUMMY \newbox\DUMMY 
\newbox\DUMMY   \newbox\DUMMY \newbox\DUMMY

\def\wlog{\immediate\write\m@ne}


\def\GetItemAll#1{%
 \GetItemSTATUS{#1}
 \GetItemNUMBER{#1}
 \GetItemTYPE{#1}
 \GetItemSPAN{#1}
 \GetItemBOX{#1}
}

\def\GetItemSTATUS#1{%
 \Point=\StatusStack
 \advance\Point by #1
 \global\ItemSTATUS=\count\Point
}

\def\GetItemNUMBER#1{%
 \Point=\NumStack
 \advance\Point by #1
 \global\ItemNUMBER=\count\Point
}

\def\GetItemTYPE#1{%
 \Point=\TypeStack
 \advance\Point by #1
 \global\ItemTYPE=\count\Point
}

\def\GetItemSPAN#1{%
 \Point\SpanStack
 \advance\Point by #1
 \global\ItemSPAN=\count\Point
}

\def\GetItemBOX#1{%
 \Point=\BoxStack
 \advance\Point by #1
 \global\setbox\ItemBOX=\vbox{\copy\Point}
 \global\ItemSIZE=\ht\ItemBOX
 \global\advance\ItemSIZE by \dp\ItemBOX
 \TEMPCOUNT=\ItemSIZE
 \divide\TEMPCOUNT by \Leading
 \divide\TEMPCOUNT by 65536
 \advance\TEMPCOUNT \@ne
 \ItemSIZE=\TEMPCOUNT pt
 \global\multiply\ItemSIZE by \Leading
}


\def\JoinStack{%
 \ifnum\LengthOfStack=\MaxItems 
  \Warn{WARNING: Stack is full...some items will be lost!}
 \else
  \Point=\StatusStack
  \advance\Point by \LengthOfStack
  \global\count\Point=\ItemSTATUS
  \Point=\NumStack
  \advance\Point by \LengthOfStack
  \global\count\Point=\ItemNUMBER
  \Point=\TypeStack
  \advance\Point by \LengthOfStack
  \global\count\Point=\ItemTYPE
  \Point\SpanStack
  \advance\Point by \LengthOfStack
  \global\count\Point=\ItemSPAN
  \Point=\BoxStack
  \advance\Point by \LengthOfStack
  \global\setbox\Point=\vbox{\copy\ItemBOX}
  \global\advance\LengthOfStack \@ne
  \ifnum\ItemTYPE=\Figure 
   \global\MoreFigurestrue
  \else
   \global\MoreTablestrue
  \fi
 \fi
}


\def\LeaveStack#1{%
 {\Iteration=#1
 \loop
 \ifnum\Iteration<\LengthOfStack
  \advance\Iteration \@ne
  \GetItemSTATUS{\Iteration}
   \advance\Point by \m@ne
   \global\count\Point=\ItemSTATUS
  \GetItemNUMBER{\Iteration}
   \advance\Point by \m@ne
   \global\count\Point=\ItemNUMBER
  \GetItemTYPE{\Iteration}
   \advance\Point by \m@ne
   \global\count\Point=\ItemTYPE
  \GetItemSPAN{\Iteration}
   \advance\Point by \m@ne
   \global\count\Point=\ItemSPAN
  \GetItemBOX{\Iteration}
   \advance\Point by \m@ne
   \global\setbox\Point=\vbox{\copy\ItemBOX}
 \repeat}
 \global\advance\LengthOfStack by \m@ne
}


\newif\ifStackNotClean

\def\CleanStack{%
 \StackNotCleantrue
 {\Iteration=\z@
  \loop
   \ifStackNotClean
    \GetItemSTATUS{\Iteration}
    \ifnum\ItemSTATUS=\InStack
     \advance\Iteration \@ne
     \else
      \LeaveStack{\Iteration}
    \fi
   \ifnum\LengthOfStack<\Iteration
    \StackNotCleanfalse
   \fi
 \repeat}
}


\def\FindItem#1#2{%
 \global\StackPointer=\m@ne 
 {\Iteration=\z@
  \loop
  \ifnum\Iteration<\LengthOfStack
   \GetItemSTATUS{\Iteration}
   \ifnum\ItemSTATUS=\InStack
    \GetItemTYPE{\Iteration}
    \ifnum\ItemTYPE=#1
     \GetItemNUMBER{\Iteration}
     \ifnum\ItemNUMBER=#2
      \global\StackPointer=\Iteration
      \Iteration=\LengthOfStack 
     \fi
    \fi
   \fi
  \advance\Iteration \@ne
 \repeat}
}


\def\FindNext{%
 \global\StackPointer=\m@ne 
 {\Iteration=\z@
  \loop
  \ifnum\Iteration<\LengthOfStack
   \GetItemSTATUS{\Iteration}
   \ifnum\ItemSTATUS=\InStack
    \GetItemTYPE{\Iteration}
   \ifnum\ItemTYPE=\Figure
    \ifMoreFigures
      \global\NextItem=\Figure
      \global\StackPointer=\Iteration
      \Iteration=\LengthOfStack 
    \fi
   \fi
   \ifnum\ItemTYPE=\Table
    \ifMoreTables
      \global\NextItem=\Table
      \global\StackPointer=\Iteration
      \Iteration=\LengthOfStack 
    \fi
   \fi
  \fi
  \advance\Iteration \@ne
 \repeat}
}


\def\ChangeStatus#1#2{%
 \Point=\StatusStack
 \advance\Point by #1
 \global\count\Point=#2
}



\def\Zone{\InZoneA}

\ZoneBAdjust=\z@

\def\MakePage{
 \global\ZoneBSize=\PageHeight
 \global\TextSize=\ZoneBSize
 \global\ZoneAFullPagefalse
 \global\topskip=\TextLeading
 \MakePageInCompletetrue
 \MoreFigurestrue
 \MoreTablestrue
 \FigInZoneBfalse
 \FigInZoneCfalse
 \TabInZoneBfalse
 \TabInZoneCfalse
 \global\FirstSingleItemtrue
 \global\FirstZoneAtrue
 \global\setbox\ZoneABOX=\box\VOIDBOX
 \global\setbox\ZoneBBOX=\box\VOIDBOX
 \global\setbox\ZoneCBOX=\box\VOIDBOX
 \loop
  \ifMakePageInComplete
 \FindNext
 \ifnum\StackPointer=\m@ne
  \NextItem=\m@ne
  \MoreFiguresfalse
  \MoreTablesfalse
 \fi
 \ifnum\NextItem=\Figure
   \FindItem{\Figure}{\NextFigure}
   \ifnum\StackPointer=\m@ne \global\MoreFiguresfalse
   \else
    \GetItemSPAN{\StackPointer}
    \ifnum\ItemSPAN=\Single \def\Zone{\InZoneB}\relax
     \ifFigInZoneC \global\MoreFiguresfalse\fi
    \else
     \def\Zone{\InZoneA}
     \ifFigInZoneB \def\Zone{\InZoneC}\fi
    \fi
   \fi
   \ifMoreFigures\Print{}\FigureItems\fi
 \fi
\ifnum\NextItem=\Table
   \FindItem{\Table}{\NextTable}
   \ifnum\StackPointer=\m@ne \global\MoreTablesfalse
   \else
    \GetItemSPAN{\StackPointer}
    \ifnum\ItemSPAN=\Single\relax
     \ifTabInZoneC \global\MoreTablesfalse\fi
    \else
     \def\Zone{\InZoneA}
     \ifTabInZoneB \def\Zone{\InZoneC}\fi
    \fi
   \fi
   \ifMoreTables\Print{}\TableItems\fi
 \fi
   \MakePageInCompletefalse 
   \ifMoreFigures\MakePageInCompletetrue\fi
   \ifMoreTables\MakePageInCompletetrue\fi
 \repeat
 \ifZoneAFullPage
  \global\TextSize=\z@
  \global\ZoneBSize=\z@
  \global\vsize=\z@\relax
  \global\topskip=\z@\relax
  \vbox to \z@{\vss}
  \eject
 \else
 \global\advance\ZoneBSize by -\ZoneBAdjust
 \global\vsize=\ZoneBSize
 \global\hsize=\ColumnWidth
 \global\ZoneBAdjust=\z@
 \ifdim\TextSize<23pt
 \Warn{}
 \Warn{* Making column fall short: TextSize=\the\TextSize *}
 \vskip-\lastskip\eject\fi
 \fi
}

\def\MakeRightCol{
 \global\TextSize=\ZoneBSize
 \MakePageInCompletetrue
 \MoreFigurestrue
 \MoreTablestrue
 \global\FirstSingleItemtrue
 \global\setbox\ZoneBBOX=\box\VOIDBOX
 \def\Zone{\InZoneB}
 \loop
  \ifMakePageInComplete
 \FindNext
 \ifnum\StackPointer=\m@ne
  \NextItem=\m@ne
  \MoreFiguresfalse
  \MoreTablesfalse
 \fi
 \ifnum\NextItem=\Figure
   \FindItem{\Figure}{\NextFigure}
   \ifnum\StackPointer=\m@ne \MoreFiguresfalse
   \else
    \GetItemSPAN{\StackPointer}
    \ifnum\ItemSPAN=\Double\relax
     \MoreFiguresfalse\fi
   \fi
   \ifMoreFigures\Print{}\FigureItems\fi
 \fi
 \ifnum\NextItem=\Table
   \FindItem{\Table}{\NextTable}
   \ifnum\StackPointer=\m@ne \MoreTablesfalse
   \else
    \GetItemSPAN{\StackPointer}
    \ifnum\ItemSPAN=\Double\relax
     \MoreTablesfalse\fi
   \fi
   \ifMoreTables\Print{}\TableItems\fi
 \fi
   \MakePageInCompletefalse 
   \ifMoreFigures\MakePageInCompletetrue\fi
   \ifMoreTables\MakePageInCompletetrue\fi
 \repeat
 \ifZoneAFullPage
  \global\TextSize=\z@
  \global\ZoneBSize=\z@
  \global\vsize=\z@\relax
  \global\topskip=\z@\relax
  \vbox to \z@{\vss}
  \eject
 \else
 \global\vsize=\ZoneBSize
 \global\hsize=\ColumnWidth
 \ifdim\TextSize<23pt
 \Warn{}
 \Warn{* Making column fall short: TextSize=\the\TextSize *}
 \vskip-\lastskip\eject\fi
\fi
}

\def\FigureItems{
 \Print{Considering...}
 \ShowItem{\StackPointer}
 \GetItemBOX{\StackPointer} 
 \GetItemSPAN{\StackPointer}
  \CheckFitInZone 
  \ifnum\ItemFits=\Yes
   \ifnum\ItemSPAN=\Single
     \ChangeStatus{\StackPointer}{\InZoneB} 
     \global\FigInZoneBtrue
     \ifFirstSingleItem
      \hbox{}\vskip-\BodgeHeight
     \global\advance\ItemSIZE by \TextLeading
     \fi
     \unvbox\ItemBOX\ItemSep
     \global\FirstSingleItemfalse
     \global\advance\TextSize by -\ItemSIZE
     \global\advance\TextSize by -\TextLeading
   \else
    \ifFirstZoneA
     \global\advance\ItemSIZE by \TextLeading
     \global\FirstZoneAfalse\fi
    \global\advance\TextSize by -\ItemSIZE
    \global\advance\TextSize by -\TextLeading
    \global\advance\ZoneBSize by -\ItemSIZE
    \global\advance\ZoneBSize by -\TextLeading
    \ifFigInZoneB\relax
     \else
     \ifdim\TextSize<3\TextLeading
     \global\ZoneAFullPagetrue
     \fi
    \fi
    \ChangeStatus{\StackPointer}{\Zone}
    \ifnum\Zone=\InZoneC \global\FigInZoneCtrue\fi
  \fi
   \Print{TextSize=\the\TextSize}
   \Print{ZoneBSize=\the\ZoneBSize}
  \global\advance\NextFigure \@ne
   \Print{This figure has been placed.}
  \else
   \Print{No space available for this figure...holding over.}
   \Print{}
   \global\MoreFiguresfalse
  \fi
}

\def\TableItems{
 \Print{Considering...}
 \ShowItem{\StackPointer}
 \GetItemBOX{\StackPointer} 
 \GetItemSPAN{\StackPointer}
  \CheckFitInZone 
  \ifnum\ItemFits=\Yes
   \ifnum\ItemSPAN=\Single
    \ChangeStatus{\StackPointer}{\InZoneB}
     \global\TabInZoneBtrue
     \ifFirstSingleItem
      \hbox{}\vskip-\BodgeHeight
     \global\advance\ItemSIZE by \TextLeading
     \fi
     \unvbox\ItemBOX\ItemSep
     \global\FirstSingleItemfalse
     \global\advance\TextSize by -\ItemSIZE
     \global\advance\TextSize by -\TextLeading
   \else
    \ifFirstZoneA
    \global\advance\ItemSIZE by \TextLeading
    \global\FirstZoneAfalse\fi
    \global\advance\TextSize by -\ItemSIZE
    \global\advance\TextSize by -\TextLeading
    \global\advance\ZoneBSize by -\ItemSIZE
    \global\advance\ZoneBSize by -\TextLeading
    \ifFigInZoneB\relax
     \else
     \ifdim\TextSize<3\TextLeading
     \global\ZoneAFullPagetrue
     \fi
    \fi
    \ChangeStatus{\StackPointer}{\Zone}
    \ifnum\Zone=\InZoneC \global\TabInZoneCtrue\fi
   \fi
  \global\advance\NextTable \@ne
   \Print{This table has been placed.}
  \else
  \Print{No space available for this table...holding over.}
   \Print{}
   \global\MoreTablesfalse
  \fi
}


\def\CheckFitInZone{%
{\advance\TextSize by -\ItemSIZE
 \advance\TextSize by -\TextLeading
 \ifFirstSingleItem
  \advance\TextSize by \TextLeading
 \fi
 \ifnum\Zone=\InZoneA\relax
  \else \advance\TextSize by -\ZoneBAdjust
 \fi
 \ifdim\TextSize<3\TextLeading \global\ItemFits=\No
 \else \global\ItemFits=\Yes\fi}
}

\def\BeginOpening{%
  \ninepoint
  \thispagestyle{titlepage}%
  \global\setbox\ItemBOX=\vbox\bgroup%
    \hsize=\PageWidth%
    \hrule height \z@
    \ifsinglecol\vskip 6pt\fi 
}

\let\begintopmatter=\BeginOpening  

\def\EndOpening{%
  \One
  \egroup
  \ifsinglecol
    \box\ItemBOX%
    \vskip\TextLeading plus 2\TextLeading
    \@noafterindent
  \else
    \ItemNUMBER=\z@%
    \ItemTYPE=\Figure
    \ItemSPAN=\Double
    \ItemSTATUS=\InStack
    \JoinStack
  \fi
}


\newif\if@here  \@herefalse

\def\no@float{\global\@heretrue}
\let\nofloat=\relax 

\def\beginfigure{%
  \@ifstar{\global\@dfloattrue \@bfigure}{\global\@dfloatfalse \@bfigure}%
}

\def\@bfigure#1{%
  \par
  \if@dfloat
    \ItemSPAN=\Double
    \TEMPDIMEN=\PageWidth
  \else
    \ItemSPAN=\Single
    \TEMPDIMEN=\ColumnWidth
  \fi
  \ifsinglecol
    \TEMPDIMEN=\PageWidth
  \else
    \ItemSTATUS=\InStack
    \ItemNUMBER=#1%
    \ItemTYPE=\Figure
  \fi
  \bgroup
    \hsize=\TEMPDIMEN
    \global\setbox\ItemBOX=\vbox\bgroup
      \eightpoint\nostb@ls{10pt}%
      \let\caption=\fig@caption
      \ifsinglecol \let\nofloat=\no@float\fi
}

\def\fig@caption#1{%
  \vskip 5.5pt plus 6pt%
  \bgroup 
    \eightpoint\nostb@ls{10pt}%
    \setbox\TEMPBOX=\hbox{#1}%
    \ifdim\wd\TEMPBOX>\TEMPDIMEN
      \noindent \unhbox\TEMPBOX\par
    \else
      \hbox to \hsize{\hfil\unhbox\TEMPBOX\hfil}%
    \fi
  \egroup
}

\def\endfigure{%
  \par\egroup 
  \egroup
  \ifsinglecol
    \if@here \midinsert\global\@herefalse\else \topinsert\fi
      \unvbox\ItemBOX
    \endinsert
  \else
    \JoinStack
    \Print{Processing source for figure \the\ItemNUMBER}%
  \fi
}


\newbox\tab@cap@box
\def\tab@caption#1{\global\setbox\tab@cap@box=\hbox{#1\par}}

\newtoks\tab@txt@toks
\long\def\tab@txt#1{\global\tab@txt@toks={#1}\global\table@txttrue}

\newif\iftable@txt  \table@txtfalse
\newif\if@dfloat    \@dfloatfalse

\def\begintable{%
  \@ifstar{\global\@dfloattrue \@btable}{\global\@dfloatfalse \@btable}%
}

\def\@btable#1{%
  \par
  \if@dfloat
    \ItemSPAN=\Double
    \TEMPDIMEN=\PageWidth
  \else
    \ItemSPAN=\Single
    \TEMPDIMEN=\ColumnWidth
  \fi
  \ifsinglecol
    \TEMPDIMEN=\PageWidth
  \else
    \ItemSTATUS=\InStack
    \ItemNUMBER=#1%
    \ItemTYPE=\Table
  \fi
  \bgroup
    \eightpoint\nostb@ls{10pt}%
    \global\setbox\ItemBOX=\vbox\bgroup
      \let\caption=\tab@caption
      \let\tabletext=\tab@txt
      \ifsinglecol \let\nofloat=\no@float\fi
}

\def\endtable{%
  \par\egroup 
  \egroup
  \setbox\TEMPBOX=\hbox to \TEMPDIMEN{%
    \eightpoint\nostb@ls{10pt}%
    \hss
    \vbox{%
      \hsize=\wd\ItemBOX
      \ifvoid\tab@cap@box
      \else
        \noindent\unhbox\tab@cap@box
        \vskip 5.5pt plus 6pt%
      \fi
      \box\ItemBOX
      \iftable@txt
        \vskip 10pt%
        \noindent\the\tab@txt@toks
        \global\table@txtfalse
      \fi
    }%
    \hss
  }%
  \ifsinglecol
    \if@here \midinsert\global\@herefalse\else \topinsert\fi
      \box\TEMPBOX
    \endinsert
  \else
    \global\setbox\ItemBOX=\box\TEMPBOX
    \JoinStack
    \Print{Processing source for table \the\ItemNUMBER}%
  \fi
}

\def\UnloadZoneA{%
\FirstZoneAtrue
 \Iteration=\z@
  \loop
   \ifnum\Iteration<\LengthOfStack
    \GetItemSTATUS{\Iteration}
    \ifnum\ItemSTATUS=\InZoneA
     \GetItemBOX{\Iteration}
     \ifFirstZoneA \vbox to \BodgeHeight{\vfil}%
     \FirstZoneAfalse\fi
     \unvbox\ItemBOX\ItemSep
     \LeaveStack{\Iteration}
     \else
     \advance\Iteration \@ne
   \fi
 \repeat
}

\def\UnloadZoneC{%
\Iteration=\z@
  \loop
   \ifnum\Iteration<\LengthOfStack
    \GetItemSTATUS{\Iteration}
    \ifnum\ItemSTATUS=\InZoneC
     \GetItemBOX{\Iteration}
     \ItemSep\unvbox\ItemBOX
     \LeaveStack{\Iteration}
     \else
     \advance\Iteration \@ne
   \fi
 \repeat
}


\def\ShowItem#1{
  {\GetItemAll{#1}
  \Print{\the#1:
  {TYPE=\ifnum\ItemTYPE=\Figure Figure\else Table\fi}
  {NUMBER=\the\ItemNUMBER}
  {SPAN=\ifnum\ItemSPAN=\Single Single\else Double\fi}
  {SIZE=\the\ItemSIZE}}}
}

\def\ShowStack{%
 \Print{}
 \Print{LengthOfStack = \the\LengthOfStack}
 \ifnum\LengthOfStack=\z@ \Print{Stack is empty}\fi
 \Iteration=\z@
 \loop
 \ifnum\Iteration<\LengthOfStack
  \ShowItem{\Iteration}
  \advance\Iteration \@ne
 \repeat
}

\def\B#1#2{%
\hbox{\vrule\kern-0.4pt\vbox to #2{%
\hrule width #1\vfill\hrule}\kern-0.4pt\vrule}
}


\newif\ifsinglecol   \singlecolfalse

\def\onecolumn{%
  \global\output={\singlecoloutput}%
  \global\hsize=\PageWidth
  \global\vsize=\PageHeight
  \global\ColumnWidth=\hsize
  \global\TextLeading=12pt
  \global\Leading=12
  \global\singlecoltrue
  \global\let\onecolumn=\relax
  \global\let\footnote=\sing@footnote
  \global\let\vfootnote=\sing@vfootnote
  \ninepoint 
  \message{(Single column)}%
}

\def\singlecoloutput{%
  \shipout\vbox{\PageHead\vbox to \PageHeight{\pagebody\vss}\PageFoot}%
  \advancepageno
  \ifplate@page
    \shipout\vbox{%
      \sp@pagetrue
      \def\sp@type{plate}%
      \global\plate@pagefalse
      \PageHead\vbox to \PageHeight{\unvbox\plt@box\vfil}\PageFoot%
    }%
    \message{[plate]}%
    \advancepageno
  \fi
  \ifnum\outputpenalty>-\@MM \else\dosupereject\fi%
}

\def\ItemSep{\vskip\ItemSepamount\relax}

\def\ItemSepbreak{\par\ifdim\lastskip<\ItemSepamount
  \removelastskip\penalty-200\ItemSep\fi%
}


\let\@@endinsert=\endinsert 

\def\endinsert{\egroup 
  \if@mid \dimen@\ht\z@ \advance\dimen@\dp\z@ \advance\dimen@12\p@
    \advance\dimen@\pagetotal \advance\dimen@-\pageshrink
    \ifdim\dimen@>\pagegoal\@midfalse\p@gefalse\fi\fi
  \if@mid \ItemSep\box\z@\ItemSepbreak
  \else\insert\topins{\penalty100 
    \splittopskip\z@skip
    \splitmaxdepth\maxdimen \floatingpenalty\z@
    \ifp@ge \dimen@\dp\z@
    \vbox to\vsize{\unvbox\z@\kern-\dimen@}
    \else \box\z@\nobreak\ItemSep\fi}\fi\endgroup%
}


\def\gobbleone#1{}
\def\gobbletwo#1#2{}
\let\footnote=\gobbletwo 
\let\vfootnote=\gobbleone

\def\sing@footnote#1{\let\@sf\empty 
  \ifhmode\edef\@sf{\spacefactor\the\spacefactor}\/\fi
  \hbox{$^{\hbox{\eightpoint #1}}$}\@sf\sing@vfootnote{#1}%
}

\def\sing@vfootnote#1{\insert\footins\bgroup\eightpoint\b@ls{9pt}%
  \interlinepenalty\interfootnotelinepenalty
  \splittopskip\ht\strutbox 
  \splitmaxdepth\dp\strutbox \floatingpenalty\@MM
  \leftskip\z@skip \rightskip\z@skip \spaceskip\z@skip \xspaceskip\z@skip
  \noindent $^{\scriptstyle\hbox{#1}}$\hskip 4pt%
    \footstrut\futurelet\next\fo@t%
}

\def\footnoterule{\kern-3\p@ \hrule height \z@ \kern 3\p@}

\skip\footins=19.5pt plus 12pt minus 1pt
\count\footins=1000
\dimen\footins=\maxdimen

\def\note#1#2{%
  \let\@sf=\empty \ifhmode\edef\@sf{\spacefactor\the\spacefactor}\/\fi
  #1\insert\footins\bgroup
    \eightpoint\b@ls{10pt}\rm
    \interlinepenalty\interfootnotelinepenalty
    \splitmaxdepth\dp\strutbox \floatingpenalty\@MM
    \leftskip\z@skip \rightskip\z@skip \spaceskip\z@skip \xspaceskip\z@skip
    \noindent\footstrut #1$\,$#2\strut\par
  \egroup
  \@sf\relax}


\def\landscape{%
  \global\TEMPDIMEN=\PageWidth
  \global\PageWidth=\PageHeight
  \global\PageHeight=\TEMPDIMEN
  \global\let\landscape=\relax
  \onecolumn
  \message{(landscape)}%
  \raggedbottom
}


\output{%
  \ifLeftCOL
    \global\setbox\LeftBOX=\vbox to \ZoneBSize{\box255\unvbox\ZoneBBOX
      \ifvoid\footins\else
        \vskip\skip\footins\unvbox\footins\fi
    }%
    \global\LeftCOLfalse
    \MakeRightCol
  \else
    \setbox\RightBOX=\vbox to \ZoneBSize{\box255\unvbox\ZoneBBOX
      \ifvoid\footins\else
        \vskip\skip\footins\unvbox\footins\fi
    }%
    \setbox\MidBOX=\hbox{\box\LeftBOX\hskip\ColumnGap\box\RightBOX}%
    \setbox\PageBOX=\vbox to \PageHeight{%
      \UnloadZoneA\box\MidBOX\UnloadZoneC}%
    \shipout\vbox{\PageHead\vbox to \PageHeight{\box\PageBOX\vss}\PageFoot}%
    \advancepageno
    \ifplate@page
      \shipout\vbox{%
        \sp@pagetrue
        \def\sp@type{plate}%
        \global\plate@pagefalse
        \PageHead\vbox to \PageHeight{\unvbox\plt@box\vfil}\PageFoot%
      }%
      \message{[plate]}%
      \advancepageno
    \fi
    \global\LeftCOLtrue
    \CleanStack
    \MakePage
  \fi
}


\Warn{\start@mess}

\newif\ifCUPmtplainloaded 
\ifprod@font
  \global\CUPmtplainloadedtrue
\fi


\catcode `\@=12 

